\documentclass[useAMS,usenatbib,usegraphicx]{mn2e}

\usepackage[T1]{fontenc}
\usepackage{graphicx}
\usepackage{journals}
\usepackage{subfig}
\usepackage{color}
\usepackage{times}
\usepackage{paralist}
\usepackage{booktabs}

\definecolor{vert}{RGB}{49,148,31}
\definecolor{grey}{rgb}{0.4,0.6,0.6}
\definecolor{darkgreen}{rgb}{0.,0.7,0.}

\title[MAGGIE group finder]{MAGGIE\@: Models and Algorithms for Galaxy Groups,
Interlopers and Environment}
\author[Manuel Duarte and Gary A. Mamon ]{Manuel Duarte\thanks{E-mail:
    acanthostega@hotmail.fr} and Gary A. Mamon\thanks{E-mail: gam@iap.fr} \\
Institut d'Astrophysique de Paris (UMR 7095: CNRS \&
UPMC, Sorbonne Universit\'es), F--75014 Paris, France }

\begin{document}
\date{Accepted 2015 August 3. Received 2015 August 3; in original form
  2014 December 10; \textcolor{red}{Consolidated version with Erratum}
  \textcolor{darkgreen}{\& additional corrections}}
\pagerange{\pageref{firstpage}--\pageref{lastpage}} \pubyear{2013}
\maketitle
\label{firstpage}
\begin{abstract}
Combining our knowledge of halo structure and internal kinematics from
cosmological dark
matter simulations and the distribution of halo interlopers in projected
phase space measured
in cosmological galaxy simulations, we develop {\sc maggie}, a prior- and
halo-based, probabilistic, abundance matching (AM) grouping algorithm for
doubly complete subsamples (in distance
and luminosity) of flux-limited samples. We test {\sc maggie-l} and 
{\sc maggie-m} (in
which group
masses are derived from AM applied to the group luminosities and stellar
masses, respectively) on groups of at least three galaxies extracted from a
mock Sloan Digital Sky Survey Legacy redshift survey, incorporating realistic
observational errors on galaxy luminosities and
stellar masses. In comparison with the optimal Friends-of-Friends (FoF) group
finder, groups
extracted with {\sc maggie} are much less likely to be secondary fragments of true
groups; in primary fragments, its galaxy memberships (relative to the
virial sphere of the real-space group) 
are much more complete and usually more reliable, and its masses are much
less biased and
usually with less scatter, as are its group luminosities and stellar masses
(computed in {\sc maggie}
using the membership probabilities as weights). FoF outperforms {\sc maggie} only
for high-mass clusters: for the reliability of the galaxy population and the
dispersion of its total mass. In comparison with our implementation of the
Yang et al. group finder, {\sc maggie} reaches much higher completeness and
slightly lower group fragmentation and dispersion on group total  
masses, luminosities and stellar masses, but slightly greater bias in the
latter two and lower reliabilities. {\sc maggie} should therefore lead to sharper
trends of environmental effects on galaxies
and more accurate mass/orbit modelling.
\end{abstract}

\section{Introduction}

In the hierarchical growth of structure in the Universe, dominated by gravity
(and dark energy), matter flows from low to high density regions. To first
order, galaxies, which form in small dark matter
haloes, follow this evolution and cluster into galaxy systems called
\emph{clusters} or  \emph{groups}, depending on their mass (clusters are
often defined with masses within the virial radius greater than $10^{14} \rm M_\odot$).

The properties of galaxies within these systems (hereafter denoted
\emph{groups} for simplicity), now attached to dark matter \emph{subhaloes},
are likely to be modified by the peculiar environment of their parent
groups. Many physical processes should indeed alter galaxy properties in
groups: the high galaxy density in groups will lead to galaxy interactions and
possibly mergers; the deeper gravitational potentials of the more massive
groups will produce higher velocity dispersions for the galaxy population,
favouring rapid flybys over mergers (e.g., \citealp{Mamon92}); the tides from the group
potential will prevent outer gas from accreting onto galaxy disks
\citep{LTC80}; the diffuse intra-group gas will exert ram pressure on the
galaxy's gas \citep{GG72} and either compress it, enhancing star formation
(e.g., \citealp{KBS84}), or when the pressure gets very high it will expel
the gas (\citeauthor{GG72}), decreasing subsequent star formation.

Galaxy groups and clusters thus represent an ideal laboratory to test the
environmental effects on galaxies in models of galaxy formation and
evolution. Groups and clusters are also an important tool to probe
cosmological parameters, such as the dark energy parameter \citep{Wang+98}.
Moreover, clusters have been recently used to test a major prediction of
general relativity, with the recent discovery of weak but significant signs of
gravitational redshifts \citep{Wojtak+11}.

Since the early discovery of morphological segregation of galaxies in
clusters \citep{Shapley26,Hubble36,vandenBergh60}, i.e.\ where inner regions
of clusters preferentially contain elliptical galaxies, whose red colours are
indicative of old stellar populations, it has been clear that the efficiency
with which stars form within galaxies must depend on their environment. In
other words, high density environments act to \emph{quench} star formation in
galaxies. More
specifically, the specific star formation rate (SSFR) of galaxies (star
formation rate divided by stellar mass) is likely to be a function of two
separate environmental parameters: the \emph{global environment}
characterized by the total mass of their group, and the \emph{local environment}
that measures the position of the galaxy within its group.

\cite{Peng+10} studied the dependence of SSFR with stellar mass and
environment, where they quantified the latter by the distance to the 5th
nearest neighbour. They found that, at low stellar mass, the SSFR varied more with the
density of the environment, while at high mass, the environmental effects
are small and the SSFR anti-correlates with stellar mass. Unfortunately, the
use of an environment tracer such as the 5th nearest neighbour produces a mix
between the global and local environments. In contrast to
\citeauthor{Peng+10}, \cite{Weinmann+06} and
\cite{vonderLinden+10} both considered indicators of both
the global and local environments. 
\citeauthor{Weinmann+06}
found that the fraction of late-type 
satellite galaxies appears more anti-correlated with group mass than with
stellar mass, while \citeauthor{vonderLinden+10} found that high-mass galaxies also
show some moderate dependence of SSFR with the relative distance to the
group/cluster centre (albeit limited to low clustercentric radii).

These possible disagreements highlight the importance of properly measuring the global
and local environments.
Unfortunately, a clean characterization of the \emph{real space} environment
from the \emph{redshift space} observed distribution of galaxies  is
difficult since the \emph{redshift distortions} \citep{Jackson+72} caused by
the velocity dispersion of the galaxy group distorts the group into
elongated structures pointing towards the observer, i.e. Fingers-of-God
\citep{Tully+78}. Moreover, because of redshift distortions, real space
groups can be merged into single groups in redshift space. Conversely,
grouping algorithms inevitably lead to some fragmentation of real-space
groups, so that the secondary fragments do not represent bona-fide groups of
their own (although some may represent subgroups of real-space groups).
Finally, even without group merging and fragmentation, a group finder may
miss some of the real-space group galaxies, leading to incomplete galaxy
membership; 
or conversely
may include additionally galaxies that lie outside of the 
virial sphere  of the real space group, producing unreliable galaxy membership.
One then wonders to what extent the strength of environmental effects on
galaxy properties may be washed out by the imperfect extraction  of the global
and local environments by the
grouping algorithm (group finder). 

Many galaxy group catalogues have already been published, usually following the
first publications of data from galaxy surveys. First attempts were made with
visual selections based on well-defined criteria
(\citealp{Abell+58,Zwicky+61} for clusters and \citealp{Rose77,Hickson82} for
compact groups). The first automated (and simple) algorithm has been the
percolation `\emph{Friends-of-Friends}' (FoF) method, first introduced by
\citet{TG76} and \citet{Huchra+82}, in which groups are built by collections
of galaxies linked, two-by-two, by their proximity. Redshift distortions are
taken into account by the use of two different linking lengths, along the
line-of-sight (LOS) and transverse directions. There is a fairly wide range
of pairs of linking lengths used in the literature. In a previous study
(\citealp{DM+14a}, hereafter Paper~I), we have analysed several mock SDSS
samples of galaxies to optimize the pairs of linking lengths for minimal
group fragmentation and merging, maximal galaxy completeness and reliability,
and maximal group mass accuracy (see also \citealp{Eke+04},
\citealp{Berlind+06}, and \citealp{Robotham+11}), and compared their optimal
linking lengths with those of ten previous implementations of the FoF
algorithm.
Another fairly non-parametric grouping algorithm is to
partition redshift space into Voronoi cells, constructed from Delaunay
triangulation, providing local galaxy 
number densities that are  inversely proportional to the volumes of
the Voronoi cells (\citealp{Marinoni+02},
see also \citealp{Gerke+05}).

Building on our recently gained knowledge from cosmological $N$-body
simulations, grouping algorithms have begun to appear, where priors on galaxy
group properties are incorporated to improve their extraction from galaxy
redshift surveys.  In pioneering studies, \citet{Yang+05,Yang+07} developed
an iterative method halo-based group finder 
that uses a density contrast
criterion in \emph{projected phase space} (PPS, i.e.\ projected radius and
LOS velocity dispersion), assuming a \citeauthor*{NFW96} (\citeyear{NFW96},
hereafter NFW) surface density profile and a Maxwellian LOS velocity
distribution, both in reasonably good agreement with what is found in the
group- and cluster-mass haloes of dissipationless cosmological
simulations. In the \citet{Yang+05,Yang+07} group finder, the group masses, hence virial radii, are determined by
\emph{abundance matching} (AM, first introduced by \citealp{mh02}), 
which assumes a one-to-one correspondence between group luminosity or stellar
mass and its total (halo) mass to 
match the cumulative distribution functions (CDFs) of the 
cosmic halo mass function (HMF) and the
group luminosity or stellar mass function (measured, here, in the previous iteration
of the algorithm).  AM between
groups and haloes has also been introduced by \citet{MunozCuartas+12} in
their FoF algorithm that links haloes rather than galaxies: they consider halo
virial radii in the transverse direction and maximum circular velocity in the
LOS direction, combining the two links in an ellipsoidal fashion.

But galaxy surveys come with observational problems that are difficult to
handle: surveys suffer from edge effects and from bright (saturated) stars
masking regions, and those
 with photometric redshifts have large and sometimes
catastrophic redshift errors. Probabilistic methods appear to be a promising way to deal with
these aspects. For example,~\cite{Liu+08} designed a probabilistic FoF
method for surveys with photometric redshifts,
\cite{AWB12} incorporated priors on the galaxy luminosity function, while
\cite{Rykoff+14} assumed a prior 
on the existence of the Red Sequence (see also \citealp{GY00}).
\citet{DominguezRomero+12} have recently adapted the \citeauthor{Yang+07} group finder
into a probabilistic algorithm:
they initially assign haloes to single galaxies, and use AM
like \citeauthor{Yang+07} to assign group masses and radii.
But \citeauthor{DominguezRomero+12} end their algorithm with a hard assignment of
galaxies to their groups.

These studies can be improved in several respects:
\vspace{-0.5\baselineskip}

\begin{enumerate}
    \item In their prediction of the density in PPS,~\cite{Yang+05,Yang+07}
        and~\cite{DominguezRomero+12} assume that the LOS velocity
        dispersion is independent of projected radius, while cosmological
        $N$-body simulations (starting with \citealp{CL96})
        indicate a convex profile in log-log. One can easily predict this
        LOS velocity dispersion profile  (see \citealp{ML+05} for a
        single integral expression) by solving
        the Jeans equation of
        local dynamical equilibrium, adopting the velocity anisotropy
        profile of the particles in the  haloes of $\Lambda$CDM cosmological
        simulations (hereafter $\Lambda$CDM haloes).

    \item~\cite{Yang+05,Yang+07} and~\cite{DominguezRomero+12} assume
        that the LOS velocity distribution is Maxwellian, whereas the
        velocity anisotropy alters this Gaussianity \citep{Merritt87}, hence one
        can do better and predict its precise shape from the
        three-dimensional velocity distribution \citep*{MBB13}.

    \item Rather than use a threshold in the PPS density as proposed by~\cite{Yang+05,Yang+07}
        and~\cite{DominguezRomero+12}, one can take
        advantage of our knowledge of the distribution of galaxies in PPS
        for two terms:
            a) the galaxies within the virial sphere of the parent
                real-space group (hereafter the \emph{halo} term);
            b) the galaxies that are in the virial cone but outside the
                virial sphere (hereafter the \emph{interloper} term).
        The interloper PPS density was quantified by~\cite*{MBM10} using a
        cosmological simulation, and it turns out to be fairly independent of
        halo mass (for cluster-mass haloes).
Comparing the PPS densities from the halo and interloper terms
        yields a probability of membership. There is no need to perform a
        hard assignment of galaxies to groups in the end as was done by
        \citet{DominguezRomero+12}: group
        properties are easily obtained using the membership probabilities as
        weights.

    \item~\cite{Yang+07} employ a complicated and imprecise scheme (see their
      fig.~4) to
        estimate how the luminosity incompleteness varies with redshift in
        their flux-limited sample. Errors in the luminosity incompleteness
        will propagate, among other places, to the AM
        technique they use to infer group masses. However, the issue can be entirely avoided by
        restricting the group finder to subsamples that are doubly complete in
        both distance and luminosity.  Admittedly, such samples are, at
        best, less than one-third the size of the parent flux-limited samples (see
        \citealp{Tempel+14} for the SDSS).
        However, the very large sizes of the samples from recent or ongoing galaxy
        spectroscopic surveys (250$\,$000 for the Two Degree Field Galaxy
        Redshift Survey [2dFGRS, \citealp{Colless+01}], 125$\,$000 for the
        Six Degree Field Galaxy Survey [6dFGS, \citealp{Jones+09}],
        700$\,$000 for the primary spectroscopic sample of the  Sloan
        Digital Sky Survey [SDSS, \citealp{Abazajian+09}], 300$\,$000 for
        the ongoing Galaxy and Mass Assembly survey [GAMA,
        \citealp{Hopkins+13}]) lead to substantial sizes for  doubly complete
        subsamples, which can be used for studies of environmental effects on galaxies. 
        Moreover, it is wiser to study environmental effects on a group
        catalogue derived from a doubly-complete galaxy subsample, rather
        than start with a group catalogue derived from a flux-limited
        subsample and then cut it into a doubly complete subsample of groups
        to study environmental effects. For
        example, 
        \cite{Tempel+14} have recently produced publicly available FoF group catalogues
        that they had run on doubly-complete SDSS galaxy subsamples.

\item \cite{Yang+05, Yang+07}, \cite{MunozCuartas+12} and \cite{DominguezRomero+12}
  wisely test their grouping algorithms using mocks. However, their adopted
  definitions for
  purity and contamination take values above (and below) unity, while we
  prefer a measure of the reliability that is restricted to values between
  zero and unity (see
  Sect.~\ref{testproceedures} below). 
  Moreover, these mocks should include observational errors (on galaxy
  luminosities and stellar masses), and while this is briefly mentioned by
  \citet{Yang+05}, it is not clear what level of errors were considered by
  them and \citet{Yang+07}, while observational errors were not mentioned by
  \citeauthor{DominguezRomero+12}. 
\end{enumerate}

In this work, we present a new probabilistic grouping algorithm, {\sf Models
  and Algorithms for Galaxy Groups, Interlopers and Environment}, a.k.a. {\sc
  maggie}. The galaxy membership of {\sc maggie} groups is determined
probabilistically, combining the distribution of interlopers in PPS measured
by \cite{MBM10} with a realistic model for the distribution in PPS of halo
members, while the group masses are determined by AM in an
iterative fashion, as in \cite{Yang+05,Yang+07}.

We present {\sc maggie} in Sect.~\ref{sec:maggie},
and our mocks and testing procedure are described in Sect.~\ref{sec:tests}.
We compare, in Sect.~\ref{sec:results}, the optimal FoF group finder with
two implementations of {\sc maggie}, on their ability to recover physical properties
and galaxy membership of real-space groups, we discuss our findings in
Sect.~\ref{sec:discus} and summarize our results in Sect.~\ref{sec:maggie_discussion_conclusion}.

\section{maggie}
\label{sec:maggie}

We present here a complete description of the different steps of {\sc
maggie}\@. We start with a basic description of the algorithm, and then we
explain how we take into account the edges of the galaxy sample.

\subsection{Basic Group Finder}
\label{algo}

We assume that we have a galaxy sample that is doubly complete in distance
and luminosity, with positions on the sky (right
ascension and declination), redshifts, as well as  apparent
magnitudes in a given waveband and/or stellar masses. 
This is the minimum required data set. 

{\sc maggie} groups are built around the most luminous galaxy ({\sc maggie-l}) 
or the most massive in stars ({\sc maggie-m}).
This galaxy is assumed to be the \emph{central galaxy} and at rest relative to the group.
Although the most massive group
        galaxies can be offset and not at rest with the group
        (e.g., \citealp{Skibba+11}),
we prefer this definition to the barycentre,
since the galaxy number density profiles in
        clusters are known to be less cuspy when clusters are centered on
        their barycentres \citep{BeersTonry86}, and indeed most analyses
        adopt the central galaxy as the position of the group centre.

{\sc maggie} then builds groups with the following iterative method. 
Steps 2 to 4 are similar to those of \cite{Yang+05,Yang+07}.

\renewcommand{\theenumi}{(\arabic{enumi})}
\renewcommand{\theenumii}{(\arabic{enumi}\alph{enumii})}
\renewcommand{\theenumiii}{(\arabic{enumi}\alph{enumii}\roman{enumiii})}

\begin{enumerate}
    \item \emph{Sort galaxies by decreasing stellar mass and loop over potential groups}\label{galaxyloop}

        We loop over the potential group central galaxies, sorted by
        decreasing galaxy stellar mass ({\sc maggie-m}) or luminosity ({\sc
        maggie-l}), performing the following steps:\\

            \item \emph{Group total masses}

                \begin{enumerate}
                    \item \emph{Initial group total masses}
                      \label{pass1}
                        On first pass, we determine the mass of each
                        group, either by adopting
                         group masses  $M = 300 \,L_r$ ({\sc
                        maggie-l}) or using the relation between halo mass
                        and central galaxy stellar mass ({\sc maggie-m})
                        that~\cite*{BCW+10} derived from AM
                        (basically matching the halo mass and central galaxy
                        stellar mass CDFs). This initial choice has no effect on the
                        final outcome (see Sect.~\ref{sec:prior_relation}).

                    \item \emph{Group total masses on subsequent iterations}
                      \label{passnext}
                        On subsequent passes, we determine the group mass by
                        performing our own AM between our
                        group luminosity ({\sc maggie-l}) or group stellar
                        mass ({\sc maggie-m}) function (determined in the
                        previous pass) and a chosen HMF:
                        \begin{eqnarray}
                            \label{eq:AM-l}
                            N(>L_{\mathrm{group}})&=&N(>M)
                                \qquad (\hbox{\sc maggie-l}) \\
                            \label{eq:AM-m}
                            N(>m_{\mathrm{group}})&=&N(>M)
                                \qquad (\hbox{\sc maggie-m}) \ ,
                        \end{eqnarray}
                        where $M$ is the group (halo) total mass, while
                        $L_{\rm group}$ and $m_{\rm group}$ and represent the
                        group luminosity and stellar mass,
                        respectively.\footnote{We denote stellar
                          masses as $m$ and total (group/halo) masses
                          (including dark matter and gas) as $M$.}
                        (In this AM, we
                          must assume that the group has the same central
                          galaxy as in the previous iteration, which is true
                          for the great majority of groups.)
                        The
                        cumulative mass functions are considered for the
                        comoving volume of the subsample, i.e. 
                        \begin{equation} 
                        N(>M) = \int_{z_{ \rm min }}^{z_{ \rm max }} \left ({{\rm d}V\over {\rm
                            d}z}\right)\,{\rm d}z \int_M^\infty
                        f(M',z)\, {\rm d}M' \ ,
                        \label{eq:NofM}
                        \end{equation} 
                        where $f(M)$ is the
                        differential HMF.
                        Numerically solving equation~(\ref{eq:AM-l})
                        or (\ref{eq:AM-m}) together with equation~(\ref{eq:NofM}) 
                        provides the group total mass as a function of the
                        group luminosity or stellar mass.  
                        Practical details on the HMF are provided in
                        Sect.~\ref{sec:hmf_test}.\footnote{In this work, the
                        cumulative HMF $N(>M)$ is derived
                        by a maximum likelihood estimate of the
                        parameters of the analytical differential HMF 
                        of \cite{Tinker+08} to the list of halo
                        masses (within the sphere of radius $r_{200}$) of the 
                        the Millennium-II simulation from which our mocks
                        were built. A
                        theoretical HMF can be chosen when
                        working on real data. 
                        Also, while it would be preferable to fit an
                        analytical form to the group 
                        luminosity function or stellar mass function, to
                        avoid shot noise and cosmic variance, such a fit is
                        difficult with a  single or
                        double \citealp{Schechter76} function. We therefore use the
                        raw list of luminosities or stellar masses for the
                        AM. We solve
                          equation~(\ref{eq:AM-l}) or
                          equation~(\ref{eq:AM-m}) by performing linear
                          interpolation  (in log-log space) of the cumulative
                          HMF (which is not analytical despite the analytical
                          nature of the differential HMF).}
                        
                \end{enumerate}

            \item \emph{Group radii}

                We estimate the group radius from the group mass, using
                \begin{equation}
                    r_{\rm 200} =
                        {\left[
                            {G \,M_{\rm 200} \over 100\,H^2(z)}
                        \right]}^{1/3} \ ,
                    \label{rad200}
                \end{equation}
                where $r_{200}$ is our proxy for the virial radius and is
                the radius of the sphere (hereafter, \emph{virial sphere})
                centered on the position of the central galaxy and whose mean density is 200
                times the \emph{critical density} of the Universe, $\rho_{\rm crit}
                = 3 H^2(z)/(8\,\pi G)$, while $M_{200}$ is the mass within
                the virial sphere. The Hubble constant, in
                equation~(\ref{rad200}), for a flat Universe, is
                \begin{equation}
                    H(z) = H_0\,
                        \sqrt{\Omega_{\rm m}{(1+z)}^3 + 1-\Omega_{\rm m}} \ ,
                \end{equation}
                where $\Omega_{\rm m}$ is the cosmological density parameter
                at $z=0$.
                The factor 100 in equation~(\ref{rad200}) is really
                $\Delta/2$ for the overdensity relative to critical of
                $\Delta=200$.\footnote{Note that \cite{Yang+07} use virial
                  radii corresponding to mean densities equal to 180 times
                  the \emph{mean density} of the Universe, which for their
                  assumed $\Omega_{\rm m}=0.238$ 
                  corresponds to 43 times the critical density of the
                  Universe. For typical NFW density profiles,
                  the virial radius used by \citeauthor{Yang+07} is roughly
                  $1.86\,(c/10)^{-0.052}$ times the radius $r_{200}$ used here.}\\

            \item \emph{Coordinates in projected phase space}

                The projected separation $R$ (hereafter, \emph{projected
                  radius}) 
                and LOS velocity $v$ of
                a galaxy relative to a central group galaxy (assumed at rest
                in the group) are written with the
                standard cosmological formulae:
                \begin{eqnarray}
                R &=& {\theta \, d_{\rm ang}(z_{\rm group})} \ ,\\
                v &=& c\,{\left(z-z_{\rm group}\right)
                    \over 1+z_{\rm group} \ ,} \label{vofz}
                \end{eqnarray}
                where $\theta$ is the angular separation, $c$ is
                the speed of light,
                \[
                    d_{\rm ang}(z) =
                    {c\over 1+z}\,\int {{\rm d}z' \over  H(z')}
                \]
                is the cosmological angular distance (for a flat Universe),
                and $z_{\rm group}$ is the redshift of the central group
                galaxy.\footnote{In this article, all instances of the
                symbol $v$ represent line-of-sight velocities of galaxies
            relative to the group.}\\

            \item \emph{Membership probability}

                In {\sc maggie}, galaxies are not assigned to groups, but are
                provided with probabilities that they belong to a given group,
                i.e. to the virial sphere of the real-space group.

                The probability that a galaxy lies within the virial sphere
                of the real space group is necessarily zero if the galaxy is
                outside the \emph{virial cone} (circumscribing the virial
                sphere). Inside, the virial cone, the probability is obtained
                by comparing the predicted densities in PPS of the \emph{halo}
                members (galaxies within the virial sphere) and the
                \emph{interlopers} (galaxies within the virial cone, but
                outside the virial sphere). This can be written
                \begin{equation}
                    p(R,v) =
                    \left \{
                    \begin{array}{ll}
                        \displaystyle {g_{\rm h}(R,v)
                        \over g_{\rm h}(R,v) + g_{\rm i}(R,v)}
                        & \qquad R \leq r_{200} \\
                        & \\
                        0 & \qquad R > r_{200}
                    \end{array}
                    \right.\label{prob1}
                \end{equation}
                where $g_{\rm h}$ and $g_{\rm i}$ are the densities in PPS
                of the halo members and
                interlopers, respectively. In practice, since the computation of the first
                expression of equation~(\ref{prob1}) is limited to the
                galaxies within the virial cone, there are few galaxy
                distances to compute around each group centre.

                \begin{enumerate}
                    \item \emph{Halo density in projected phase space}

                        Given a galaxy number density profile $\nu(r)$, the
                        density of halo particles in PPS is (following
                        \citealp{MBB13}, replacing infinities by the virial
                        radius):
                        \begin{equation}
                            g_{\rm h}(R,v) = \Sigma_{\rm sph}(R)\,
                            \left\langle h(v|R,r)\right\rangle_{\rm LOS-sph} \ ,
                            \label{ghalodef}
                        \end{equation}
                        where $\Sigma_{\rm sph}$ is the surface density of
                        the galaxies limited to the virial sphere:
                        \begin{equation}
                            \Sigma_{\rm sph}(R) = 2\,\int_R^{r_{200}} \nu(r) \,{r\,{\rm d}r\over
                              \sqrt{r^2-R^2}} \ ,
                            \label{Sigmasph}
                        \end{equation}
                        while $h(v|R,r)$ is the probability of having a LOS
                        velocity $v$ at the position in space given by
                        ($R,r$), or when taking a LOS coordinate whose
                        origin is at the group centre, at position
                        ($R,z$=$\sqrt{r^2-R^2}$). Combining
                        equations~(\ref{ghalodef}) and (\ref{Sigmasph}) one
                        obtains
                        \begin{equation}
                            g_{\rm h}(R,v) =
                                2\,\int_R^{r_{200}}
                                \nu(r) \,h(v|R,r) \,{r\,{\rm d}r\over
                              \sqrt{r^2-R^2}} \;.
                        \end{equation}
                        Assuming Gaussian (Maxwellian) three-dimensional
                        velocities,\footnote{It is easy to improve this
                            model using the joint $q$-Gaussian (Tsallis)
                            velocity dispersion that~\cite{Beraldo+15} found
                            to represent better the 3D velocity distribution
                        in $\Lambda$CDM haloes.}~\cite{MBB13} have shown that
                        the LOS velocity distribution at position $(R,r)$ is
                        also a Gaussian:
                         \begin{equation}
                            h(v|R,r) = {1\over \sqrt{2\pi\sigma_z^2(R,r)}}\,\exp\left[-{v^2\over
                                2\,\sigma_z^2(R,r)}\right] \ ,
                            \label{hofvz1}
                        \end{equation}
                        with
                        \begin{equation}
                            \sigma_z^2(R,r) = \left (1-\beta(r) {R^2\over r^2} \right)\,\sigma_r^2(r) \ ,
                            \label{sigmaz}
                        \end{equation}
                        where $\beta = 1 - \sigma_\theta^2/\sigma_r^2$ is
                        the velocity anisotropy  (for radial velocity
                        dispersion $\sigma_r$ and one component of the
                        tangential velocity dispersion $\sigma_\theta$). In
                        the presence of measurement errors of the LOS
                        velocity, assumed Gaussian with zero bias and standard
                        deviation $\epsilon(v)$, the new distribution of LOS
                        velocities is the convolution of the zero-error
                        $h(v|R,r)$ of equation~(\ref{hofvz1}) by a Gaussian
                        of standard deviation $\epsilon(v)$. Then,
                        in the
                        expression of $h(v|R,r)$ (eq.~[\ref{hofvz1}]),  
                        the local LOS velocity variance $\sigma_z^2$ in equation~(\ref{sigmaz}) 
                        is replaced by $\sigma_z^2(R,r) + \epsilon^2(v)$.
                        The radial velocity variance $\sigma_r^2$ in
                        equation~(\ref{sigmaz}) is obtained
                        (eq.~[\ref{sigmaofr}] in Appendix~\ref{sec:sigr})  from the
                        stationary spherical Jeans
                        equation of local dynamical equilibrium
                        \begin{equation}
                            {{\rm d} \left (\nu \sigma_r^2 \right) \over {\rm d}r} + 2\,\beta(r) \,{\nu
                            \sigma_r^2\over r} = - \nu {G M(r) \over r^2} \ ,
                            \label{Jeans}
                        \end{equation}
                        where $M(r)$ is our chosen total mass profile.\footnote{The general
  form of the Jeans equation in an
  expanding Universe contains extra terms that do not appear in our
  ``standard'' Jeans equation~(\ref{Jeans}) for the density of the Universe,
  dark energy, streaming motions and
  non-stationarity \citep{Falco+13}. However, the solution of
  equation~(\ref{sigmaofr}) of the ``standard'' Jeans equation is a highly
  accurate solution of the ``general'' Jeans equation for
  $r < 2\,r_{100} \simeq 2.7 r_{200}$ \citep{Falco+13}.} 

                        We assume that the galaxy distribution follows the
                        mass distribution, and assume an NFW model for these
                        two quantities. Denoting $a$ the scale radius of the
                        NFW density profile
                        \[
                            \nu_{\rm NFW}(r) \propto {1\over r\,{(r+a)}^2} \ ,
                        \]
                        (in the NFW model, $a$ happens to be equal to the
                        radius where the logarithmic  slope of the density
                        profile is equal to 
                        $-2$), we define the \emph{concentration parameter}
                        $c_{200} = r_{200}/a$. We adopt the scaling
                        between $r_{200}$ and $M_{200} = M(r_{200})$ from
                        the measurements on $\Lambda$CDM haloes at $z=0$
                        by~\cite{MDvdB+08}. The NFW density profile can then
                        be written
                        \begin{eqnarray}
                            \nu(r)&\!\!=\!\!& {N_{200} \over 4 \pi r_{200}^3}\,\widehat \nu \left ({r\over
                              r_{200}}\right) \ ,\\
                            \widehat \nu(x) &\!\!=\!\!& {1\over \ln(c_{200}+1)-c_{200}/(c_{200}+1)}\,{x^{-1}\over
                              {(x+1/c_{\textcolor{darkgreen}{200}})}^2}\ ,
                            \label{nunfw}
                        \end{eqnarray}
                        where $N_{200}$ is the number of predicted galaxies
                        (above some minimum luminosity or stellar mass)
                        within the virial sphere. We shall see, below, that
                        the normalization $N_{200}$ cancels from
                        equation~(\ref{prob1}). The mass profile of the
                        groups is
                        \begin{equation}
                            M(r) = M_{200} \,{\ln(x+1)-x/(x+1)\over \ln (c_{200}+1)-c_{200}/(c_{200}+1)}
                            \ ,
                        \end{equation}
                        where, again, $x = r/r_{200}$.

                        Finally, we adopt the velocity anisotropy profile
                        that~\cite{ML+05} found to represent well the
                        particles in cluster-mass $\Lambda$CDM haloes
                        \begin{equation}
                            \beta(r) = {1\over 2} \,{r\over r+r_\beta} \ ,
                            \label{betaML}
                        \end{equation}
                        with $r_\beta \simeq r_{200}/c_{200}$ \citep{MBM10}.

                        For our choice of NFW mass model and
                        \citeauthor{ML+05} anisotropy model, the radial
                        velocity variance is given in
                        equation~(\ref{sigmaofrML}) of
                        Appendix~\ref{sec:sigr}.

                    \item \emph{Interloper surface density in projected
                        phase space}\label{sec:gilop}

                        Analyzing the distribution of dark matter particles within a
                        hydrodynamical cosmological $N$-body
                        simulation,~\cite{MBM10} have found that the
                        distribution of interlopers in PPS can be written as
                        a Gaussian of the LOS velocity plus a constant term,
                        where the coefficients of the Gaussian depend on
                        projected
                        radius:
                        \begin{eqnarray}
                            g_{\rm i}(R,v) &=& {N_{200}\over r_{200}^2 v_{200}}\,
                            \widehat g_{\rm i} \left ({R\over
                              r_{200}},{v\over v_{200}}\right) \ ,
                            \label{gilop}\\
                            \widehat g_{\rm i}(X,u) &=&
                            A (X)\,\exp \left [-{1\over 2}\,{u^2 \over
                             \widehat   \sigma_{\rm i}^2(X)} \right] + B  ,
                            \label{gilophat}
                        \end{eqnarray}
                        where
                        \begin{eqnarray}
                            A(X) &=&
                            {\rm dex} \left (-1.061 + 0.364\, X^2 - 0.580\, X^4 \right. \nonumber \\
                            &\mbox{}& \qquad \left. + 0.533\, X^6 \right)
                            \ ,
                            \label{Ailop}
                            \\
                            \widehat \sigma_{\rm i} (X) &\!\!\!\!=\!\!\!\!&
                            0.612 - 0.0653\,X^2 \;, 
                            \label{sigmailop}\\
                            B &=& 0.0075 \ ,
                            \label{Bilop}
                        \end{eqnarray}
                        where cosmic variance fluctuations are 0.11, 0.23 and
                        0.40 dex for $\widehat \sigma_{\rm i}(X)$, $A(X)$, and $B$,
                        respectively \citep{MBM10}.
                        The velocity  $v_{200}$ is the circular velocity at
                        $r_{200}$, i.e. $v_{200} = 10\,H(z)\,r_{200}$.
                        In the presence of velocity measurement errors of
                        dispersion $\epsilon(v)$, one should replace 
                        $\widehat \sigma_{\rm i}^2$ by 
                        $\widehat \sigma_{\rm i}^2+\epsilon^2(v)/v_{200}^2$.

                        Since galaxies are somewhat biased tracers of the
                        dark matter distribution, one needs to re-estimate
                        the functions  $A(X)$,
                        $\sigma_{\rm i}(X)$, as well as $B$ from a mock that is
                        built from the galaxy distribution rather than the 
                        dark matter particles.
                        In Appendix~\ref{sec:giGuo}, 
                        we present our analysis of the $z$=0 output of the SAM of
                        \cite{Guo+11}, deriving
                       \begin{eqnarray}
                            \log_{10} A(X) &=&  -1.092 - \textcolor{red}{0.01922}\,X^3 + 0.1829\,X^6 \ , 
                            \label{AilopGuo}
                            \\
                            \sigma_{\rm i}(X) &=& 0.6695 - 0.1004\, X^2 \ , 
                            \label{sigmailopGuo}\\
                            B &=& 0.0067 \ .
                            \label{BilopGuo}
                        \end{eqnarray}
                        We thus
                        adopt, in this work, the functional fits provided in
                        equations~(\ref{AilopGuo}), 
                        (\ref{sigmailopGuo}), and (\ref{BilopGuo}), 
                        which admittedly are
                        close to those of equations~(\ref{Ailop}),
                        (\ref{sigmailop}), and (\ref{Bilop}).

                        Equations~(\ref{gilop}) -- (\ref{Bilop}) depend
                        little on halo mass in the cluster-mass regime
                        \citep{MBM10}, and we assume here that these
                        equations extend to group masses, in particular for
                        the functional forms
                        (eqs.~[\ref{AilopGuo}]-\ref{BilopGuo}]) that we derived in
                        Appendix~\ref{sec:giGuo} for the \citeauthor{Guo+11}
                        SAM. 

                \end{enumerate}

                We note that the normalization $N_{200}$ appears in both
                $g_{\rm h}$ and $g_{\rm i}$, so it cancels out of the
                probability $p(R,v)$ of equation~(\ref{prob1}).

                In our scheme, central galaxies have $R=0$ and $v=0$, by
                definition, and we set to unity their probability of membership
                (since the NFW central surface density diverges). To
                avoid too much group fragmentation, we do not assign a
                galaxy as a potential central group galaxy  if it has a
                probability $p > p_{\rm cen}$ of belonging to another group
                of greater central galaxy stellar mass (since we proceed
                with groups of decreasing central galaxy stellar masses).
                Here, $p_{\rm cen}$ is a free parameter of {\sc maggie}. If
                $p_{\rm cen}=1$, all galaxies can be group centres (case of
                maximum group fragmentation and no group merging).
                If $p_{\rm cen} = 0$, no
                satellite galaxy of a massive group can be the centre of
                another one (no group fragmentation, but maximal group
                merging). 
                In other words, with $p_{\rm cen}=0$, galaxies
                lying in the virial cone of a massive central galaxy, but far in the
                foreground/background, will be assigned membership
                probabilities to the group around this first galaxy, but will
                not be assigned membership probabilities to potential groups
                around potential central galaxies lying in the same virial cone.
                However, if the central galaxy of the first group was wrongly
                determined, then one can effectively have group
                fragmentation, even with $p_{\rm cen}=0$ (but this occurs
                very rarely).
                Our tests showed that the performance of {\sc maggie} was
                independent of $p_{\rm cen}$ for $0 < p_{\rm cen} < 0.5$, and we
                adopted $p_{\rm cen}=0.001$.\\

            \item \emph{Group global properties}

                The group global properties are obtained by using the galaxy
                membership probabilities as weights, i.e.\ group luminosities
                $L_{\rm group}$ and stellar masses $m_{\rm group}$ are
                obtained with
                \begin{eqnarray}
                    L_{\rm group} &=& \sum_i p(R_i,v_i) \,L_i \;,
                    \label{Lgroup}\\
                    m_{\rm group} &=& \sum_i p(R_i,v_i) \,m_i \;.
                    \label{mgroup}
                \end{eqnarray}
                over all galaxies with $p(R,v) \geq p_{\rm mem}$, where $p_{\rm
                  mem}$ is another free parameter of {\sc maggie}. If $p_{\rm
                  mem}=1$, the group luminosities and stellar masses
                will correspond to the values of the central galaxies, while
                if $p_{\rm mem}=0$, all galaxies within the virial cone will
                be considered when computing the luminosities and stellar
                masses, even those that contribute a tiny
                probability. Clearly, there should be little difference
                between setting $p_{\rm mem}$ = 0.001 or $p_{\rm mem}=0$. But
                physically, galaxies with extremely low $p_{\rm mem}$
                typically correspond to interlopers that are many group standard
                deviations in the foreground or background,\footnote{Groups
                  lying very close to the virial cone also have very low
                  membership probabilities.} in projection and it makes
                little sense to keep them in the group. We thus choose to set
                $p_{\rm mem} = 0.001$.

    \item \emph{Loop convergence}

        We return to step~\ref{galaxyloop}, waiting for convergence when the number of groups
        found on the current pass matches the numbers found in the previous 3
        passes. While the number of groups evolves towards a fixed value, it
        does not converge after 20 passes, hence we stop the
        iteration after the twentieth pass.\footnote{The number of groups
          oscillates around a value, but in an aperiodic fashion.}

\end{enumerate}

Note that the central galaxy is in general the most luminous ({\sc maggie-l})
or the most massive in stars ({\sc maggie-m}). However, there are rare exceptions
where a group may contain a galaxy that is more luminous or massive than its
central, and yet is not the central of another previously found  group,
i.e. a group whose central is  more luminous or massive.

\subsection{Edge effects}

Aside from all-sky surveys, galaxy surveys have edges on the sky. Moreover,
all volume-limited subsamples of galaxy surveys (including all-sky) will
have edges in redshift space. Galaxy groups lying too close to an edge may
be truncated. The grouping algorithm may detect the truncated group without
knowing how much of the group lies beyond the survey edge. There is
therefore no simple recipe to handle survey edges.

For groups lying near a survey edge, following \cite{Yang+07} we generate 700
galaxies (\citeauthor{Yang+07} use 200) following the
NFW profile, using the halo concentration  estimated by the halo mass from
\citet{MDvdB+08}. 
Then, we project galaxies on the celestial sphere and we
estimate the number of galaxies weighted by their probabilities that fall
outside the galaxy survey zone. For this, we also need to generate galaxy
velocities. We assume that the 3D velocity distribution is Maxwellian and that
the velocity  anisotropy is that given by \citet{ML+05}. Next, we compute the
fraction of galaxies that are outside the survey (still weighted by galaxy
probabilities) and then the total stellar mass and luminosity  of the group
are corrected by dividing by  this fraction (see \citealp{Yang+07}).
Admittedly, if a large group is centered just beyond the survey edge, only a
small fraction of this group will intersect our survey mask, so we will
underestimate its virial radius and mass.

\section{Tests of {\sc maggie} on mock catalogues}
\label{sec:tests}

We test {\sc maggie} using realistic mock, doubly complete in distance and
luminosity,  galaxy redshift catalogues, which we had
previously used in Paper~I to optimize the FoF linking lengths. The
construction of the mock catalogues and the description of the tests are
discussed in detail in Paper~I, and are briefly recalled below.

\subsection{Mock galaxy sample}
\label{sec:mock}

We have constructed a mock galaxy catalogue corresponding to the extent on
the sky and depth of the largest contiguous (2.2 sr) region of the 
primary (Legacy) spectroscopic sample of the SDSS. For this, we replicated the galaxy outputs at $z=0$
generated from  the \citet{Guo+11} semi-analytical model (SAM) of galaxy
formation and evolution, which was run on the halo merger trees extracted from the Millennium-II
dissipationless cosmological $N$-body simulation \citep{BoylanKolchin+09}, 
which itself had been  run in a box 
of comoving size $L_{\rm box} = 100\;h^{-1}\mathrm{Mpc}$, with
cosmological parameters
$\Omega_m=0.25$, $\Omega_\Lambda=0.75$, $H_0=73$ and $\sigma_8=0.9$, and
particle mass $1.1\times 10^7 \rm M_\odot$.
Haloes were identified by
applying the Friends-of-Friends (FoF) technique to the  real space particle
data.

In the output of the \citeauthor{Guo+11} SAM, each galaxy is associated to a
halo, making it easy to compare the groups extracted from our algorithm to
the real space groups. \citeauthor{Guo+11} found that the $z$=0 galaxy
luminosity and stellar mass functions agree well with the
corresponding observed functions, making their galaxy catalogue
realistic and useful to test our algorithm on data similar to observations.

The maximal redshift spanned by the simulation box is approximately
${H_0}{L_{\mathrm{box}}}/c\approx0.025$. Simulating the SDSS survey requires
a deeper sample (see Table~\ref{tab:subsamp}). For this, we have juxtaposed
several boxes of the galaxy 
catalogue, applying random translations and rotations in galaxies
coordinates to avoid perspective effects \citep{Blaizot+05}. This produced a
larger \emph{superbox} composed of the replicas of the galaxies in the
computation box. We placed the observer at the middle of one of the sides of
the superbox (see Fig.~1 of Paper~I). Redshifts of the galaxies were computed
using velocities given in the galaxy catalogue and adding the Hubble flow to
it (see Paper~I).

\begin{table}
    \caption{Doubly complete subsamples of the mock SDSS/Legacy
    survey.\label{tab:subsamp}}
    \centering
    \begin{tabular}{lccccr}
        \toprule
        \multicolumn{1}{c}{Subsample} & $z_{\min}$ & $z_{\max}$ & $M_r^{\max}$ & $L_{\min}/L*$ &
        galaxies \\
        \midrule
        Nearby & 0.01 & 0.053 & --19.0 & 0.14 & 72$\,$510 \\
        Distant & 0.01 & 0.102 & --20.5 & 0.56 & 213$\,$546 \\
        \bottomrule
    \end{tabular}
\end{table}

Our mock survey had no holes caused by
saturated stars or bad data. Nevertheless, we allowed for observational errors on galaxy
luminosities and stellar masses.
According to Appendix~\ref{app:errm},
the errors on galaxy stellar masses, determined by comparison of
different stellar mass algorithms on hundreds of thousands of SDSS galaxies,
are roughly 0.2 dex. This value is much more conservative than the value of
0.10 dex \citep{Taylor+11}, and 
0.15 dex \citep{Mendel+14}, but consistent with the 95\% confidence errors of
0.30 and 0.35 dex for blue and red galaxies, as
deduced by \cite{Conroy+09}.
In Appendix~\ref{app:errL}, we estimate the errors on galaxy 
luminosities, taking into account errors in photometry and redshift,
uncertainties on extinction corrections and k-corrections, and neglect of
peculiar velocities. We find that the errors on galaxy luminosities 
 are of order of 0.08 dex at our minimum redshift of $z=0.01$
decreasing to 0.06 dex at our maximum redshift of $z\simeq 0.1$.
In our analysis, we have therefore generated Gaussian errors without bias and
with dispersion of 0.2 dex for log stellar masses and 0.08 dex 
for log luminosities. 

From our flux-limited mock galaxy survey, we constructed several subsamples
that are doubly complete in distance and luminosity. We focus our results on
the two subsamples shown in Table~\ref{tab:subsamp}.

\subsection{Flags}
We flagged all galaxies belonging to real-space FoF groups containing at least
one member that was on the other side of the periodic box (their groups would thus be
split by the transformations of the box). 

We also flagged the galaxies in the extracted groups 
that lie close to the redshift space edges of
the doubly complete subsamples: 
to be very conservative, we flagged all extracted groups lying closer (roughly 2.5 Mpc) to
the angular edges than would be the virial radius of a massive ($\log_{10}
M=15.2$) cluster, and all groups lying closer to the redshift limits than 13
times this distance  (see
\citealp{MBM10}) to 
account for redshift distortions.\footnote{The number of unflagged galaxies
  depends on the group finder and the subsample.}

We ran {\sc maggie} on all galaxies of the mock (flagged or unflagged), and
subsequently flag the groups that contain at least one flagged galaxy with 
$p > p_{\rm mem}$.

\subsection{Testing procedures}
\label{testproceedures}
Following Paper~I, we applied a suite of tests to groups containing no
flagged galaxies to assess the performance of {\sc maggie}, its robustness to
some of the assumptions, and to compare it
to other group finders.
The tests check how well the  sample of \emph{extracted groups} in
redshift space (hereafter, EGs) matches the sample of \emph{true groups} in
real space (hereafter, TGs). The TGs are defined as the set of galaxies that
lie within the virial sphere around the centre of the real space group,
i.e. the position of the most bound particle of the halo \citep{BoylanKolchin+09}.

In an optimal grouping algorithm,  the TGs minimally suffer from
\emph{fragmentation} into several EGs. A fragmented TG contains the central
galaxies (see beginning of Sect.~\ref{algo})
 of several
EGs. 
There should also be 
minimal \emph{merging} of several TGs into a unique EG (the EG contains
the central galaxies of several TGs, each with $p\geq p_{\rm
  mem}$). Following~\cite{Yang+07} and Paper~I, the EGs and 
TGs are linked by their respective central galaxies. When fragmentation
occurs, the primary EG is that 
containing the central galaxy of the parent TG\@. When merging occurs, the primary TG is
that containing the central galaxy of the EG\@.
We refer the reader to fig.~3 of Paper~I for illustrations of group
fragmentation and merging.

Also, in the optimal grouping algorithm, the galaxies of the EG should represent a
maximally  \emph{complete} sample of the parent TG galaxies, and a maximally
\emph{reliable} (pure) sample, i.e.\ with as high as possible fraction of galaxies
that belong to the parent TG (recall that the TG is the set of galaxies
within the virial sphere).

Finally, the optimal grouping algorithm should produce EG luminosities, stellar
masses and total masses as close as
possible to those of the parent TG, i.e.\ with minimal \emph{bias} and
\emph{scatter}. While bias can be corrected for, a measurement with strong
scatter will be \emph{inefficient}.

When TGs are fragmented, it makes little sense to measure the reliability of
the galaxy membership of
the secondary EGs (secondary fragments), and when TGs are merged, it would
similarly not be useful to
measure the completeness of the galaxy membership of a secondary TG\@. And it only makes sense to
compare EG properties with the corresponding TG ones for primary fragments or relative to
primary parent TGs. So all measures of completeness, reliability, as well as
bias and scatter of group luminosity, stellar and total masses  
are limited to the primary EGs.
The reader is 
referred to Paper~I for more details.

Since the galaxy membership of {\sc maggie} groups is probabilistic, some of the statistical
tests must be modified.
In Paper~I, we defined the galaxy reliability as\footnote{Note that our
  \emph{reliability}, which can take values in the range $[0-1]$,
is different from the \emph{purity} used by \cite{Yang+07} and
\cite{DominguezRomero+12},
  defined as TG/EG=$R/C$, 
  and also different from  one minus their \emph{contamination}, 
defined as
  (EG--TG$\cap$EG)/TG=$C(1/R$--$1)$, which both can be greater or smaller than unity.}
\begin{equation}
    R={\frac{\rm TG \cap EG}{\rm EG}}=\frac{N_{i\in \rm TG\cap
    EG}}{N_{i\in \rm EG}} \ ,
\label{Rold}
\end{equation}
where we adopt the notation $N_{i \in {\cal E}}$ to represent the number of
elements in space ${\cal E}$.
For our probabilistic {\sc maggie}  group finder, we modify
equation~(\ref{Rold}) to 
\begin{equation}
    R=\frac{\rm TG \cap EG}{\rm EG}=\frac{\sum_{i\in \rm TG\cap EG}
      p_i}{\sum_{i\in \rm  EG}p_i} \ ,
\label{Rnew}
\end{equation}
where $p_i \equiv p(R_i,v_i)$ is the probability of membership of galaxy $i$
(eq.~[\ref{prob1}]).
The equivalent of equation~(\ref{Rnew}) 
for the completeness would be
\begin{equation}
C={\rm TG \cap EG \over TG} = {\sum_{i \in \rm TG\cap EG} p_i \over N_{i\in \rm
  TG} } \ .
\label{Cprob}
\end{equation}
However, it is inconsistent to consider probabilities in the numerator of
equation~(\ref{Cprob}) and not in its denominator. We therefore adopt instead a
definition based on hard assignments:
\begin{equation}
C={\rm TG \cap EG \over TG} = {N_{i \in \rm TG\cap EG {\rm \  AND \ } p_i> p_{\rm mem}} \over N_{i\in \rm
  TG}}  \ .
\label{Cdef2}
\end{equation}
Since our chosen value of $p_{\rm mem}$ is very small, the definition of
completeness in equation~(\ref{Cdef2}) is very close to the definition of paper~I

For group  luminosities and stellar masses, we use the probabilities as
in equations~(\ref{Lgroup}) and (\ref{mgroup}), respectively. 

Finally, we did not  measure group merging in this work.
The logical way of estimating group merging is to request that two TG
centrals are members of the same EG.
But with a probabilistic method such as
{\sc maggie}, a given galaxy may be a member of several EGs (with different
membership probabilities in each, all with $p>p_{\rm cen}$). 
So it is not clear a group merger occurs
when one of the TG centrals 
is a member of 2 EGs, with a much lower probability of membership in the EG
that contains the central of the other TG compared to the probability of
membership in the other EG.

\section{Results}
\label{sec:results}

We now present the results of our tests on group fragmentation, galaxy
completeness and reliability, accuracies of group total masses, 
luminosities and stellar masses. We ran these tests on both {\sc
  maggie-l} and {\sc maggie-m}, using mocks without or with the inclusion
observational errors of 0.08 dex in luminosity and 0.2 dex in stellar masses.
We, however, defer the discussion of the impact of observational errors to Sect.~\ref{sec:obs_errs}.

\subsection{Fragmentation}

\begin{figure}
    \centering
    \includegraphics[width=\hsize]{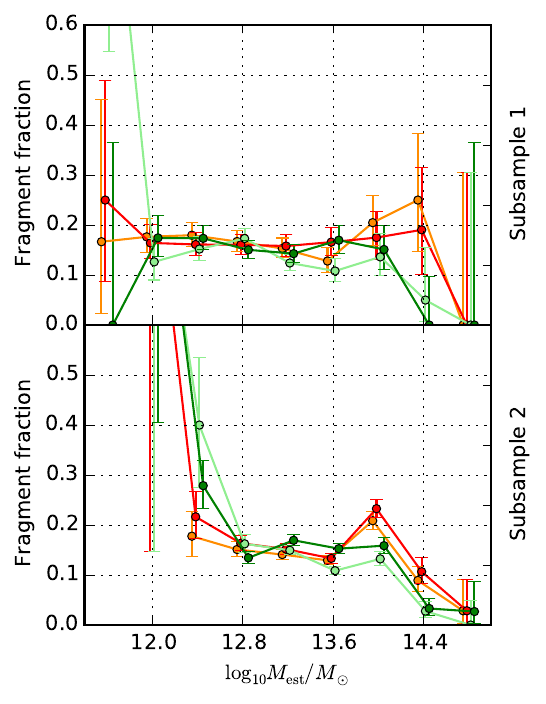}
    \caption{Fraction of extracted groups that are secondary fragments
        as a function of their estimated group mass, for unflagged groups of at
        least 3 members (for both the extracted and true groups), for both
        the nearby (\emph{top}) and distant (\emph{bottom}) subsamples.
        The line colours are  \emph{dark green} and \emph{red} for {\sc
          maggie-m} and {\sc maggie-l}, with respective 
        observational errors of 0.2 dex on stellar mass and 0.08 dex on
        luminosity, and
        \emph{light green} and \emph{orange} for
        {\sc maggie-m} and {\sc maggie-l}, with zero  observational errors.  
        The error bars are computed with the 
        Wilson (1927)
        formula (see text). The points have their abscissa slightly
          shifted for clarity. 
\label{fig:frag}}
\end{figure}
\nocite{Wilson27}
Figure~\ref{fig:frag} displays the fraction of extracted groups (EGs)
that are secondary fragments as a function of estimated group mass.
The error bars are obtained with the \cite{Wilson27} formula.\footnote{The
  \cite{Wilson27} formula avoids zero errors when the fraction is zero or
  unity. It is described in the 
Wikipedia entry {\sf Binomial proportion
          confidence interval}, http://en.wikipedia.org/wiki/Binomial\_proportion\_confidence\_interval.}
Both versions of {\sc maggie} lead to  fragmentation of typically
15\%, even when realistic errors on galaxy luminosities and stellar masses
are considered.
The fragmentation in {\sc maggie} is fairly independent of the
chosen doubly complete subsample, except that extracted groups with low
estimated  masses ($\log M_{\rm est}/{\rm M}_\odot < 12.5$)
are more likely to be secondary fragments when in the distant subsample,
where most of the groups lie too far to permit the detection of such low mass groups.
In the high-mass end, one finds that the {\sc maggie-m} EGs are less likely
to be secondary fragments than their {\sc maggie-l} counterparts.

\subsection{Completeness and reliability}

\begin{figure}
\centering
\includegraphics[width=\hsize]{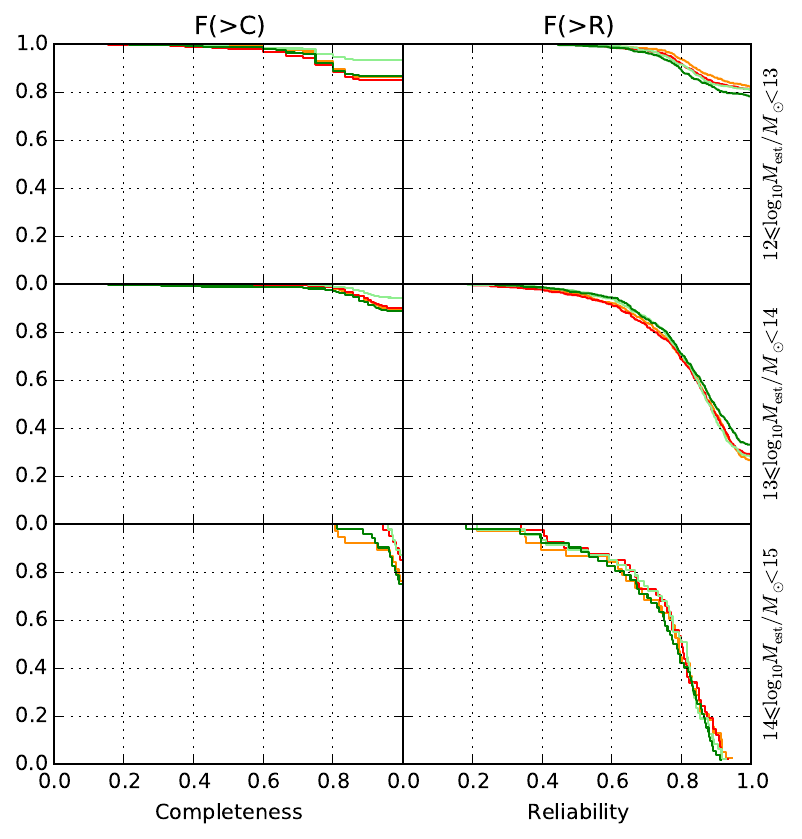}
\caption{Cumulative distribution functions  of the galaxy membership completeness
        (\emph{left}, computed with eq.~[\ref{Cdef2}])
and reliability (\emph{right},
both relative to the virial sphere of the true groups,
computed with
eq.~ [\ref{Rnew}])
in bins of estimated group
        masses, for the nearby subsample (unflagged galaxies in groups of 
        at least 3 true and 3 extracted members that
        are not secondary fragments). The colours are the same as in
        Figure~\ref{fig:frag}.
\label{fig:comp_rel_1_est}
}
\includegraphics[width=\hsize]{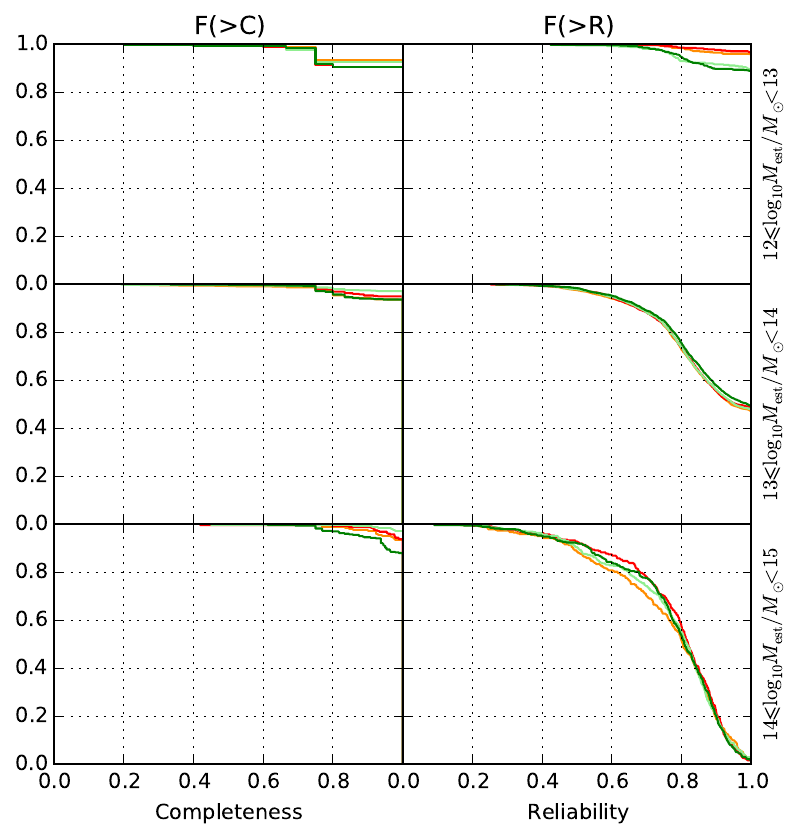}
\caption{Same as Figure~\ref{fig:comp_rel_1_est}, but for the distant subsample.
\label{fig:comp_rel_2_est}
}
\end{figure}

Figures~\ref{fig:comp_rel_1_est} and~\ref{fig:comp_rel_2_est} show 
that
the EGs from {\sc
  maggie-m} (dark green lines) and {\sc maggie-l} (red lines)
that are primary fragments are highly complete in galaxies. 
For the nearby subsample (Fig.~\ref{fig:comp_rel_1_est}), 
in the
worst performing among {\sc maggie-m} and {\sc maggie-l}, with observational errors,
100\% completeness is achieved for 
85\%, 89\% and 75\%
of the groups,
for the low,
intermediate and high mass bins, respectively, and 90\% completeness is
reached for 
90\%, 93\% and 95\%
of the groups in the same respective mass bins.
The completeness is even higher 
for the distant subsample
(Fig.~\ref{fig:comp_rel_2_est}).
The galaxy completeness values are roughly the same with {\sc maggie-l} and
{\sc maggie-m}, except for the high-mass end where {\sc maggie-l} shows higher completeness.

The galaxy reliability of {\sc maggie} decreases with increasing EG mass:
For the worst performing among {\sc maggie-m} and {\sc maggie-l} with
  observational errors, the fractions of 90\%-reliable 
  groups are respectively 
82\%, 43\%, and 4\% for the nearby subsample.
The median galaxy reliabilities for the high-mass bin are more similar: while
they are 
respectively
80\% and 82\% (78\% and 81\%)
with {\sc maggie-l} ({\sc maggie-m}), again with observational errors.
The galaxy reliabilities are very similar between {\sc maggie-l} and {\sc
  maggie-m}, except that {\sc maggie-l} shows higher reliability in the
low mass bin of the distant subsample.

\subsection{Accuracy in group total masses}

\begin{figure}
    \centering
    \includegraphics[width=\hsize]{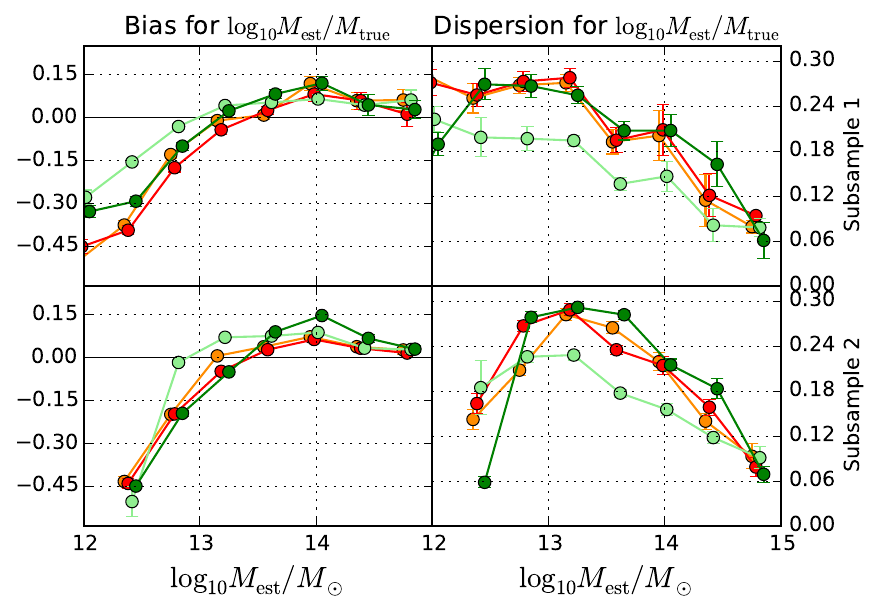} 
    \caption{Median bias and scatter (using 16th and 84th percentiles) of
      extracted group total mass within the 
        virial sphere (for unflagged primary groups of at least 
3 true and 3 extracted galaxies),
        for the nearby (\emph{top}) and distant (\emph{bottom}) subsamples.
The error bars for the bias and scatter are respectively $\sigma/\sqrt{N}$
and $\sigma/2\,[2/(N-1)+\kappa/N]^{1/2}$, where $\kappa$ is the kurtosis excess.
The points have their abscissa slightly shifted for clarity. 
Same colours as Figures~\ref{fig:frag} and~\ref{fig:masscomp}.
\label{fig:bias_disp_virial_mass}}
\end{figure}

Figure~\ref{fig:bias_disp_virial_mass} 
shows the bias and scatter in group mass. 
Both flavours of {\sc maggie}, without or with errors, have their group
masses biased low at low masses, by typically 0.4 dex at
 $\log M_{\rm est}/{\rm M}_\odot=12.5$.
The estimated masses are unbiased at $\log M_{\rm est}/{\rm M}_\odot \approx
13.5$ and are slightly positively biased at high mass, especially at 
$\log M_{\rm est}/{\rm M}_\odot \approx 14$, where the bias reaches $\approx
0.1$ dex.
The  bias is never more
than 0.1 dex in absolute value for groups with $\log M_{\rm est}/\rm M_\odot >
12.8$.
There are no significant differences in group total mass bias between {\sc
  maggie-l} and {\sc maggie-m}.

While bias can be corrected for, scatter is a more serious concern.
The dispersion in $M_{\rm est}/M_{\rm true}$
decreases with EG mass, from typically 0.2 dex at 
$\log M_{\rm est}/{\rm M}_\odot = 12$ to 13 
to better than 0.1 dex at 
$\log M_{\rm est}/{\rm M}_\odot = 14.8$.

\subsection{Accuracy in group luminosities and stellar masses}

\begin{figure}
    \centering
    \includegraphics[width=\hsize]{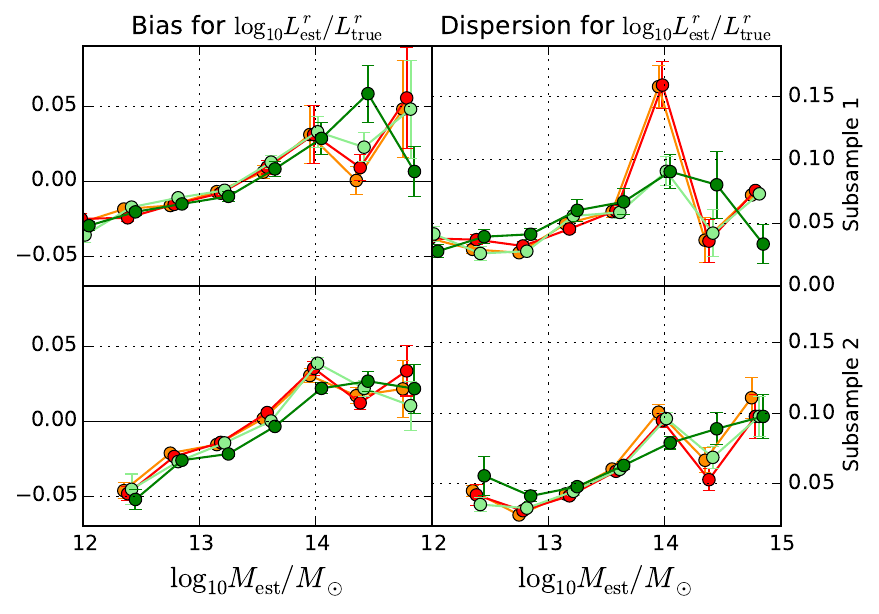}     
    \caption{Median bias and scatter (using 16th and 84th percentiles) 
 of extracted group
      $r$-band luminosity  within
      the virial sphere (for unflagged primary groups of at least 3 true and
      3 extracted
      galaxies).  
The points have their abscissa slightly shifted for clarity. 
Same colours as Figures~\ref{fig:frag}
      and~\ref{fig:masscomp}. 
\label{fig:Ltotbiassig}}
    \centering
    \includegraphics[width=\hsize]{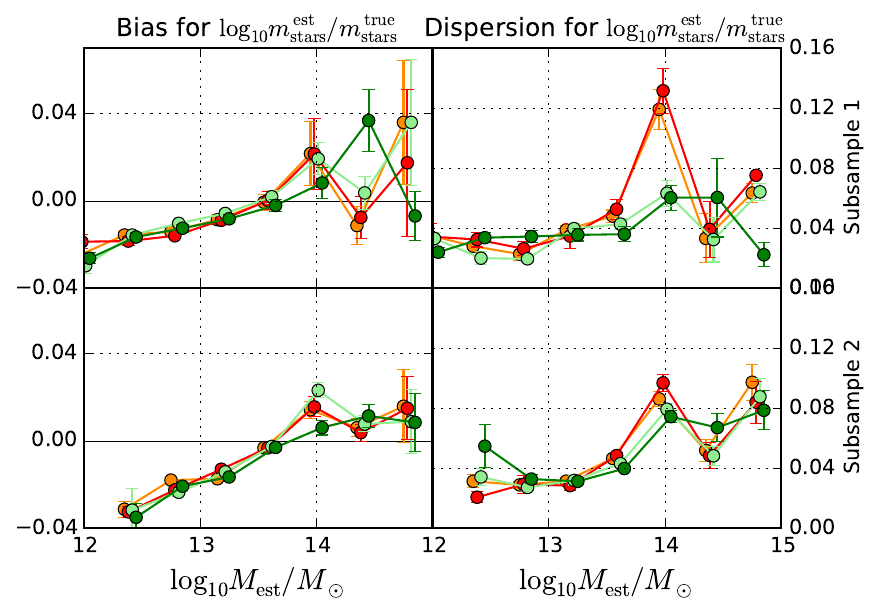} 
    \caption{Same as Figure~\ref{fig:Ltotbiassig}, but for group 
        stellar mass. 
\label{fig:mtotbiassig}}
\end{figure}

Figures~\ref{fig:Ltotbiassig} and~\ref{fig:mtotbiassig} show the bias and
inefficiency of the recovered group luminosities and stellar masses,
respectively (using eqs.~[\ref{Lgroup}] and [\ref{mgroup}]).
{\sc maggie}
groups with low estimated total masses are slightly
biased low, by up to typically 0.03 dex (7\%) at $\log M_{\rm
  est}/\rm M_\odot=12$ (but 0.05 dex for group luminosities in the distant subsample).
At high estimated group masses, the bias is less well measured (given the
lower number of groups in the higher mass bins), and some fluctuations
appear, but the overestimate of the group total masses is typically limited
to 0.04 dex (10\%).
There are virtually no differences in the luminosity and stellar mass biases 
 between {\sc maggie-l} and {\sc maggie-m}.
 
Surprisingly, the inefficiencies in group luminosity or stellar mass
estimation with both flavours of {\sc maggie} increase with EG mass,
typically from 0.03 or 0.04 dex in the low EG mass end to 0.05 dex at the
high end (0.08 dex for the distant subsample).
{\sc maggie-l} and {\sc maggie-m} have comparable dispersions in luminosity
and stellar mass, except that {\sc maggie-l} is worse in one mass bin 
($M_{\rm est}/{\rm M}_\odot \approx 14$), but this bad performance of {\sc
  maggie-l} in a single mass bin  might be of statistical nature.

\section{Discussion}
\label{sec:discus}

Our tests have been performed in an idealized situation for {\sc maggie}, 
with perfectly known
scaling relations, yet with realistic measurement errors on galaxy
luminosities and stellar masses.
We now discuss the general features of {\sc maggie}
(Sect.~\ref{sec:features}), and then the performance of {\sc maggie} in terms of its sensitivity to
observational errors (Sect.~\ref{sec:obs_errs}), and its robustness relative
to
the adopted initial mass -- luminosity scaling relation (Sect.~\ref{sec:prior_relation}),
the HMF (Sect.~\ref{sec:hmf_test}), cosmological parameters (Sect.~\ref{sec:cosmo}), and the
precise choice for the density of interlopers in PPS, $g_{\rm i}(R,v)$
(Sect.~\ref{sec:robust_gi}), before comparing its performances in
Sect.~\ref{sec:compalgos} 
to two popular group finders: FoF and the
halo-based group-finder of \cite{Yang+05,Yang+07}.

\subsection{General features of {\sc maggie}}
\label{sec:features}

Thanks to its probabilistic nature, {\sc maggie} generally performs well with
galaxy membership
reliability, since the least reliable galaxies along the LOS are assigned low
probabilities. 
In general, extracted global properties of groups should be less
biased and errors in membership are usually tempered by the low probabilities
assigned to uncertain members.

Since at high group (halo) mass, group luminosity
and stellar mass are less sensitive to group (halo) mass \citep{mh02},
the group mass that is obtained with the AM technique 
is more sensitive to the observable group luminosity or
stellar mass. Moreover, high-mass groups (clusters), which are known to have
a more prominent Red Sequence of galaxies, should have higher stellar mass to
luminosity ratios. This means that, measuring group mass with AM will be more
accurate using group stellar masses than with group luminosities. This is,
indeed, what is seen (Fig.~\ref{fig:bias_disp_virial_mass}) 
in our tests of the accuracies of {\sc maggie} group
total masses without observational errors, as {\sc maggie-m} (light green)
masses are significantly more accurate than those of {\sc maggie-l} (orange).
But the greater observational errors on stellar masses relative to
luminosities reverse this
hierarchy, making {\sc maggie-l} slightly more accurate than {\sc maggie-m}.
in extracting group
total masses.

{\sc maggie} assigns non-zero probabilities to all galaxies lying within the virial
cone, and only considers those for which $p > p_{\rm mem}=0.001$. If the
virial radius is overestimated, {\sc maggie} will be more prone
to group merging and the galaxy membership will be less reliable. 
The bloated sizes of high-mass groups, as witnessed by the mass bias going
from negative to positive for {\sc maggie} groups
(left panels of Fig.~\ref{fig:bias_disp_virial_mass}), explains 
the strong decrease in reliability with increasing estimated group mass
(Figs.~\ref{fig:comp_rel_1_est} and~\ref{fig:comp_rel_2_est}), especially
in {\sc maggie}. 

\subsection{Robustness of {\sc maggie} to observational errors}

\label{sec:obs_errs}

While geometric based algorithms such as FoF or Voronoi-Delaunay
methods are relatively immune to observational errors on luminosities and stellar
masses, such errors can affect group finders that derive group total masses
by AM with group luminosities or stellar masses, as those of
\cite{Yang+07}, \cite{MunozCuartas+12}, \cite{DominguezRomero+12} and {\sc
  maggie}.

We have run {\sc maggie} both with and without the observational errors on
galaxy luminosities and stellar masses.
Figures~\ref{fig:frag}--\ref{fig:mtotbiassig} show the effects of going from no observational errors 
(orange for {\sc maggie-l} and light green for {\sc maggie-m}) to
realistic observational errors
(red for {\sc maggie-l} and green for {\sc maggie-m}).

Including observational errors produces only small extra group
fragmentation, by typically 10 percent for {\sc maggie-l} and 20 percent {\sc
  maggie-m}, both in relative terms
(Fig.~\ref{fig:frag}).

Using the Kolmogorov-Smirnov (KS) test, we see (Figs.~\ref{fig:comp_rel_1_est} and
\ref{fig:comp_rel_2_est})
that the observational errors significantly 
worsen the galaxy completeness of {\sc maggie-m} for low mass EGs of the
nearby subsample (and marginally so for the intermediate and high mass EGs of
the distant subsample), while {\sc maggie-l} does not seem affected.
Neither {\sc maggie-l} nor {\sc maggie-m} see their galaxy reliability
affected by the observational errors.

The inefficiency of group mass estimation is significantly worsened by the
observational errors for {\sc maggie-m} (at virtually all EG masses), but not for {\sc maggie-l}
(Fig.~\ref{fig:bias_disp_virial_mass}).
While, with mocks that do not include observational errors, {\sc maggie-m}
produces significantly more accurate group total mass estimation than does
{\sc maggie-l}, the inclusion of observational errors inverses this
hierarchy, with {\sc maggie-l} producing slightly more accurate group masses.

On the other hand, observational errors do not worsen the inefficiency of the
estimation of group luminosities (Fig.~\ref{fig:Ltotbiassig}) and stellar
masses with either flavour of {\sc maggie}
(Fig.~\ref{fig:mtotbiassig}).

\subsection{Robustness of {\sc maggie} to details of the model}
\label{sec:robust}
\subsubsection{Initial halo mass -- central stellar mass relation}
\label{sec:prior_relation}

We tested how {\sc maggie} is affected by our initial relation between halo
mass and central stellar mass (item~2a of Sect.~\ref{algo}).  We found that
{\sc maggie} is insensitive to our adopted scheme of relating luminosity or
stellar mass to halo mass in its first pass: the final variation of group
$M/L_r$ vs. $L_r$ is precisely the same whether one adopts $M/L_r=300$ or 
the relation of $M/L_r$ vs. $L_r$ that \cite{BCW+10} derived from AM.
The same effect has been previously noticed by \cite{Yang+07}.

\subsubsection{Halo mass function model}
\label{sec:hmf_test}

\begin{figure}
    \includegraphics[width=\hsize]{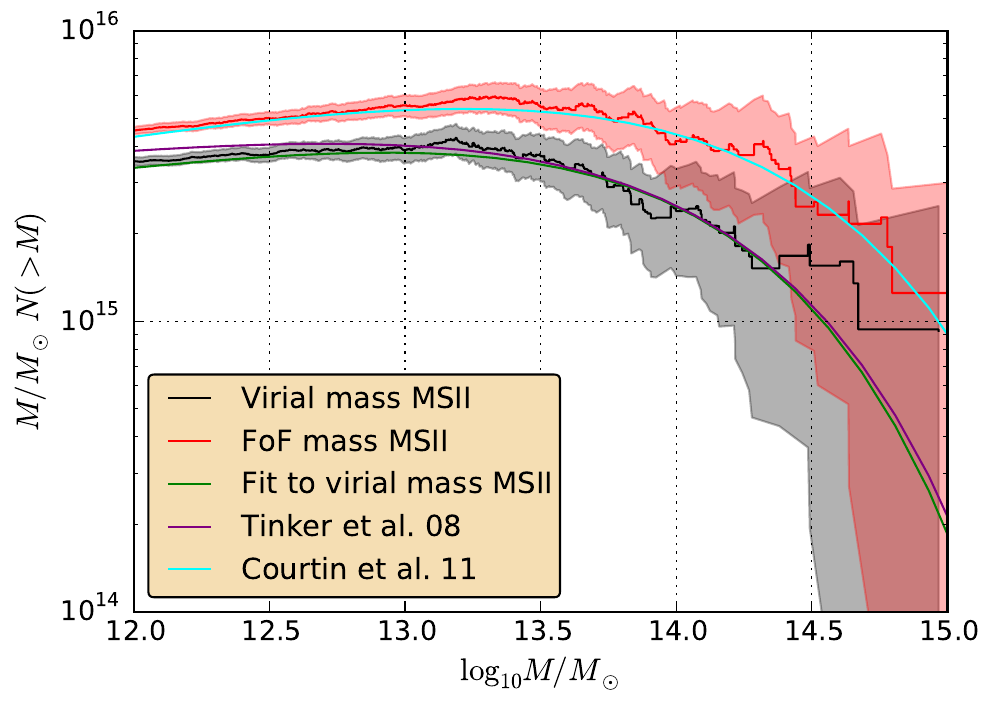}
    \caption{Cumulative $z$=0 halo mass functions (multiplied by halo 
mass for clarity) for the output of
    the
Millennium-II simulation for FoF masses (\emph{red}) and virial masses
($M_{200}$, i.e. computed in the sphere whose mean density is 200 times the
critical density of the Universe,
\emph{black}),
as well as the analytical forms from 
Courtin et al. (2011) for the halo FoF mass function (\emph{cyan})
and from
Tinker et al. (2008) for the halo mass function where the masses are in the
spheres of overdensity of 800 relative to the mean density of the Universe
(\emph{purple}, corresponding to 200 times the critical density for $\Omega_{\rm m}=0.25$
as in the Millennium Simulation).
Also shown is our maximum likelihood fit to the
virial mass function (\emph{green}).
\label{fig:hmf}}
\end{figure}

The estimation of the virial mass (or virial radius) is a crucial step
(item 2 in Sect.~\ref{algo}) of {\sc maggie} (and
of other methods that use priors such as \citealp{Yang+07} and
\citealp{DominguezRomero+12}). 
A biased estimate of group masses will affect the observed
trends of galaxy properties with the global environment.
The AM technique, used in {\sc maggie} (as well as by \citeauthor{Yang+07},
\citealp{MunozCuartas+12}
and \citeauthor{DominguezRomero+12}) appears to be a good way to estimate the virial mass of galaxy
group haloes.
There are, however, three issues that need to be considered. 

First, there may be haloes with no galaxies that may perturb the halo-group bijectivity
assumption of AM. We checked that no haloes above $10^{11}
\rm M_\odot$ in the Millennium-II simulation have zero galaxies assigned to them
in the SAM of \cite{Guo+11}.

Secondly, deriving group total masses by AM between an HMF 
and the inferred distribution of group galaxy luminosities or
stellar masses should cause  
inefficient estimation of group masses when these are in the high range 
($14 < \log_{10} M_{\rm est}/\rm M_\odot < 15$), because of the lower slope of
the high mass end of the group
luminosity (or stellar mass) as a function of halo mass
at high halo mass (e.g.,
\citealp{YMvdB08,Yang+09}).\footnote{This issue is much more
  severe when one uses central galaxy (instead of group) luminosity or
  stellar mass (e.g., \citealp{YMvdB08,Yang+09,Cattaneo+11,WM13}).}

\begin{figure*}
    \centering
\includegraphics[width=0.39\hsize]{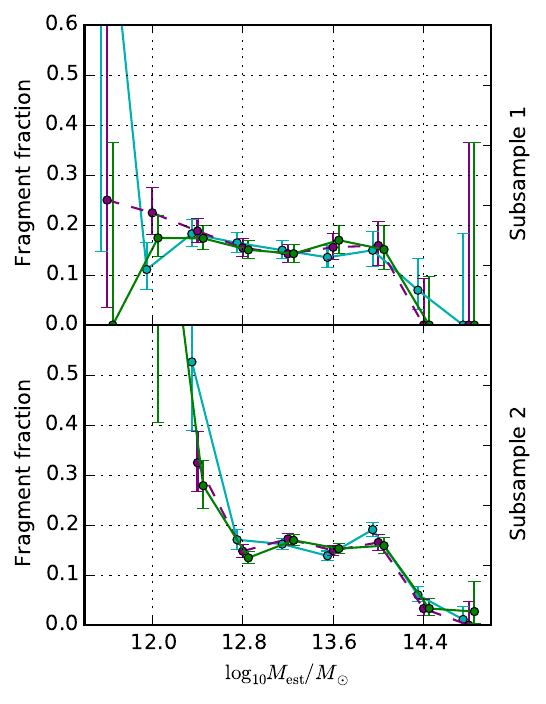}
\includegraphics[width=0.60\hsize]{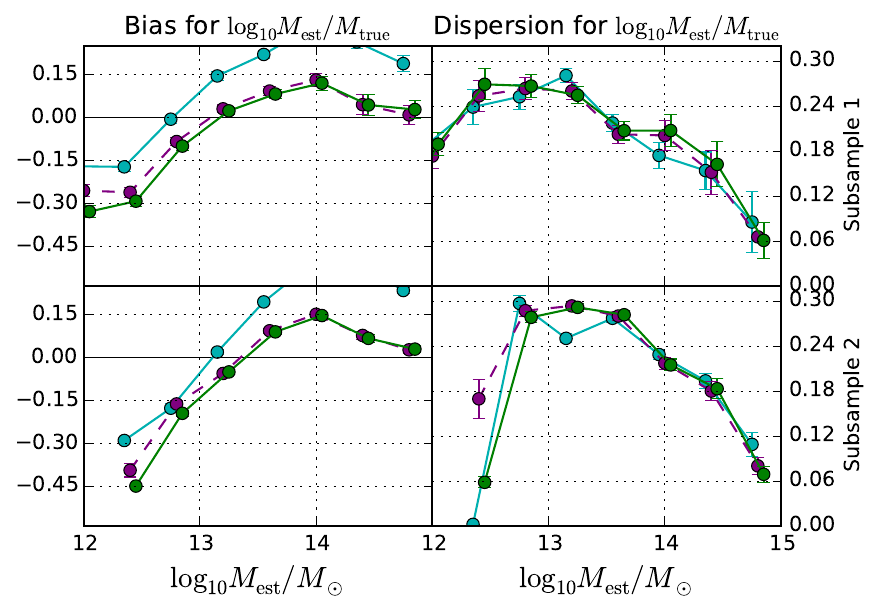}
\includegraphics[width=0.49\hsize]{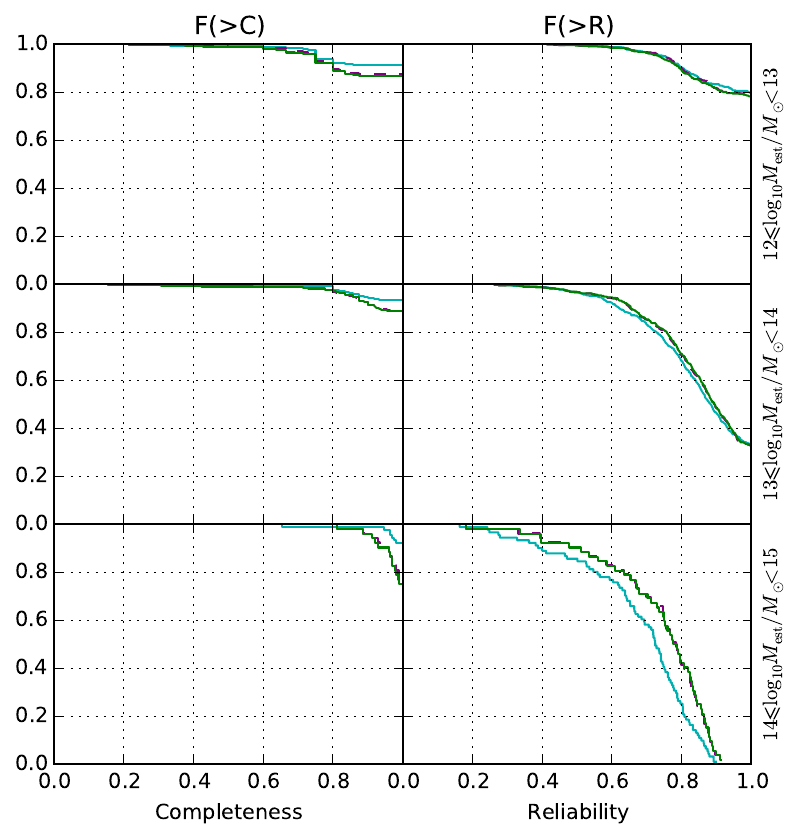}
\includegraphics[width=0.49\hsize]{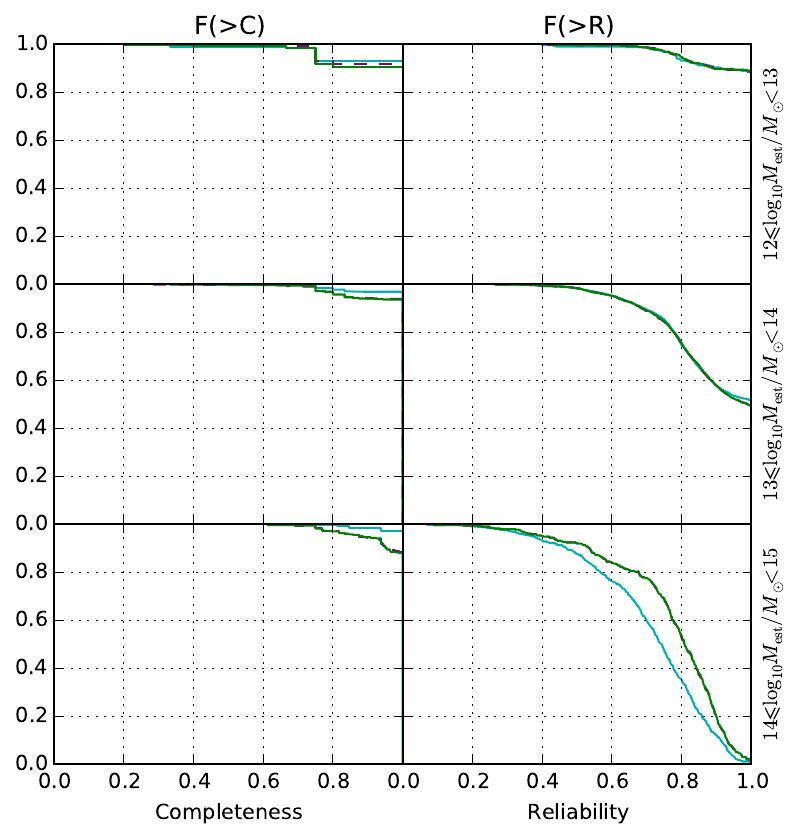}
        \caption{Effects of the choice of halo mass function on the
fraction of fragments (\emph{top left}),
the accuracy in the recovered halo mass (\emph{top right}),
and the          cumulative distribution functions of galaxy 
            completeness and reliability of the groups (\emph{bottom}) for
            {\sc maggie-m} (with 0.2 dex observational errors). The lines show the
            three adopted halo mass functions: our fit to
            the halo virial mass function  (\emph{green}, as used throughout
            article), and the 
            analytical halo mass functions of
Tinker et al. (2008, virial \emph{purple, dashed}, for overdensity 800 relative to the
mean, corresponding to 200 relative to critical for the cosmology of our mock) and 
Courtin et al. (2011, FoF, \emph{light blue}). 
The analysis is for unflagged groups of at least 3 true and 3 extracted
members.
The points in the upper plots have their abscissa slightly shifted for clarity. 
The Tinker et al. lines are usually indistinguishable from those obtained
with our fit.
\label{fig:tests_vs_hmf}
}
\end{figure*}

Thirdly, most analytical HMFs described in the literature are
based on fits to the FoF mass of the
haloes instead of the spherical over-density mass, which is how we defined the
virial mass of the halo. Since we used the galaxy catalogue from
\citet{Guo+11}, whose SAM was applied onto the Millennium-II
run, we fit the halo \emph{virial}  mass function directly on its output. 

Figure~\ref{fig:hmf} shows the cumulative HMF computed in various ways.
The figure clearly shows that the cumulative FoF HMF computed from the
Millennium-II Simulation is typically 0.2 dex
above the cumulative virial mass function computed from the same simulation.
While the analytical approximation of \citeauthor{Courtin+11}
(\citeyear{Courtin+11}, light blue) matches very well the halo FoF mass
function,\footnote{Other analytical forms of the halo FoF mass function by
  \cite{Jenkins+01}, \cite{Warren+06}, and \cite{Crocce+10} are virtually
  identical to that of \cite{Courtin+11}.}
Figure~\ref{fig:hmf} shows that the cumulative halo virial mass function obtained from the
analytical approximation of \citeauthor{Tinker+08} (\citeyear{Tinker+08},
purple) 
is slightly offset, at low masses, relative to
cumulative halo virial mass function extracted from the Millennium-II
simulation.

We therefore chose, in our present tests,  to fit ourselves the halo virial
  mass function of the Millennium-II simulation with the
  \cite{Tinker+08} model. Our maximum likelihood fit of the
  \citeauthor{Tinker+08} function to the set of halo masses (green curve)
 produces
  \citeauthor{Tinker+08} parameters $a=2.13$, $b=1.97$, and $c=1.75$, with
  normalization $A=0.188$ (instead of the corresponding values of $a=1.87$,
  $b=1.59$, $c=1.58$, and $A=0.248$ that \citeauthor{Tinker+08} found for
  $\Delta_{\rm m}=800$, i.e. $\Delta=200$ given $\Omega_{\rm m}=0.25$ of the
  Millennium-II simulation, purple curves).\footnote{Our fit
    is performed in the mass range $11 < \log M/\rm M_\odot < 15.5$. Following
    \cite{Jenkins+01}, \cite{Tinker+08} fit a simple form for 
$f(\sigma) = (M/\overline \rho)\,{\rm d}n/{\rm d}M\,/\,({\rm d}\ln
    \sigma^{-1}/{\rm d}\ln M)$, 
where $\sigma$ is the standard deviation of
  primordial perturbations of mass $M$ (linearly extrapolated to $z=0$). The
  expression $f(\sigma)$ turns out to be only very weakly sensitive to the cosmological parameters
  and
  fairly weakly sensitive to redshift \citep{Jenkins+01}.
However, the
  cumulative HMF of the Jenkins/Tinker model cannot be expressed in
  analytical form. Therefore, the normalisation of the probability density function of the halo
  mass distribution (required for the maximum likelihood estimate) requires
  in turn the
  numerical integration of the unnormalized HMF.}

Figure~\ref{fig:tests_vs_hmf}
shows how the
performance of {\sc maggie} is affected by the choice of HMF,
respectively for group fragmentation, as well as the mass accuracy,  galaxy
completeness and reliability of the primary groups.

All three HMFs lead to similar group fragmentation, group total mass
inefficiency and galaxy incompleteness.
However, adopting  the analytical approximation (by \citealp{Courtin+11}) 
to the halo FoF mass function (light blue
lines and symbols) leads to 
significantly positive mass bias (up to 0.3 dex) at high EG masses (upper right plot of 
Fig.~\ref{fig:tests_vs_hmf}, and
 lower reliability at high EG masses (bottom plots of
Fig.~\ref{fig:tests_vs_hmf},
only significant for the distant subsample according to our KS tests).

Substituting the HMF of \cite{Tinker+08}, for
overdensity 800 relative to the mean density of the Universe (corresponding
to 200 times the critical density for the value of the cosmological density
parameter used in the Millennium Simulation II, on which the galaxies of our
mock were modeled with the SAM of \citealp{Guo+11}) yields very
similar results to our standard HMF (which, we recall, is a fit of the
\citeauthor{Tinker+08} form to the halo mass distribution):
these two HMFs (purple lines and symbols for the \citeauthor{Tinker+08} HMF,
green ones for our HMF) often lead to indistinguishable positions in the various plots of 
Figure~\ref{fig:tests_vs_hmf} (recall that the points have their abscissa
slightly shifted for clarity).
This similarity reflects the similar HMFs seen in Figure~\ref{fig:hmf}
(purple vs. green curves). This suggests that one can use the \citeauthor{Tinker+08} analytical fits to
the halo virial mass function for the AM in {\sc maggie}.

\subsubsection{Cosmological parameters}
\label{sec:cosmo}
\begin{figure*}
\includegraphics[width=0.39\hsize]{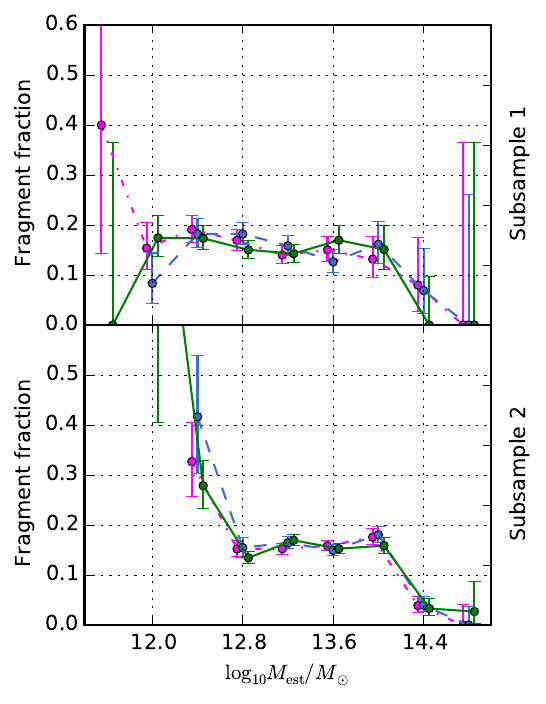}
\includegraphics[width=0.60\hsize]{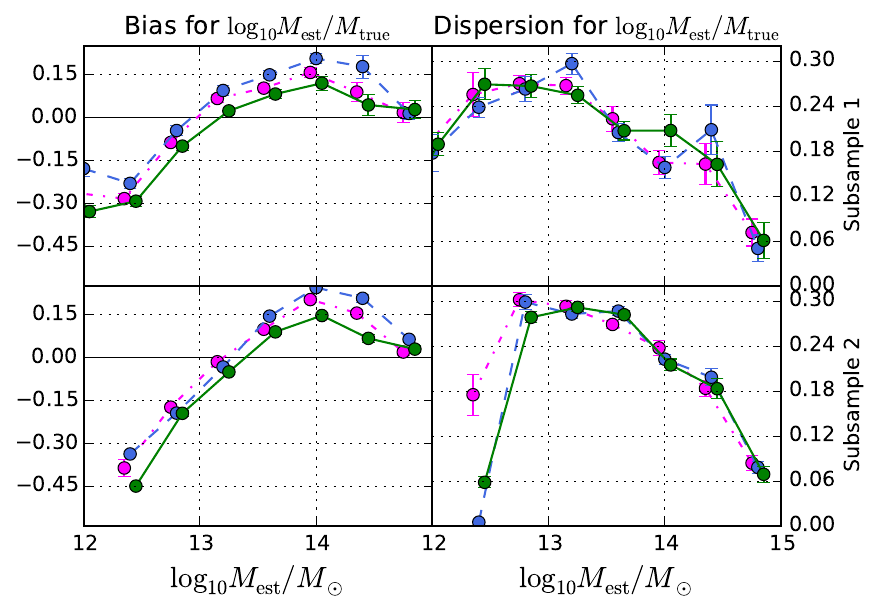}
\includegraphics[width=0.49\hsize]{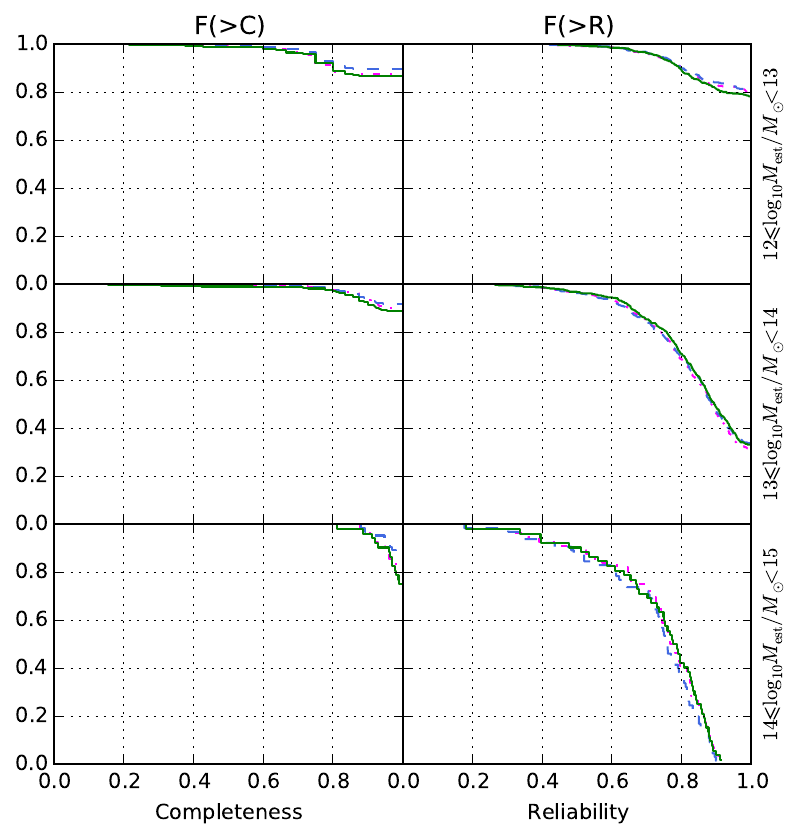}
\includegraphics[width=0.49\hsize]{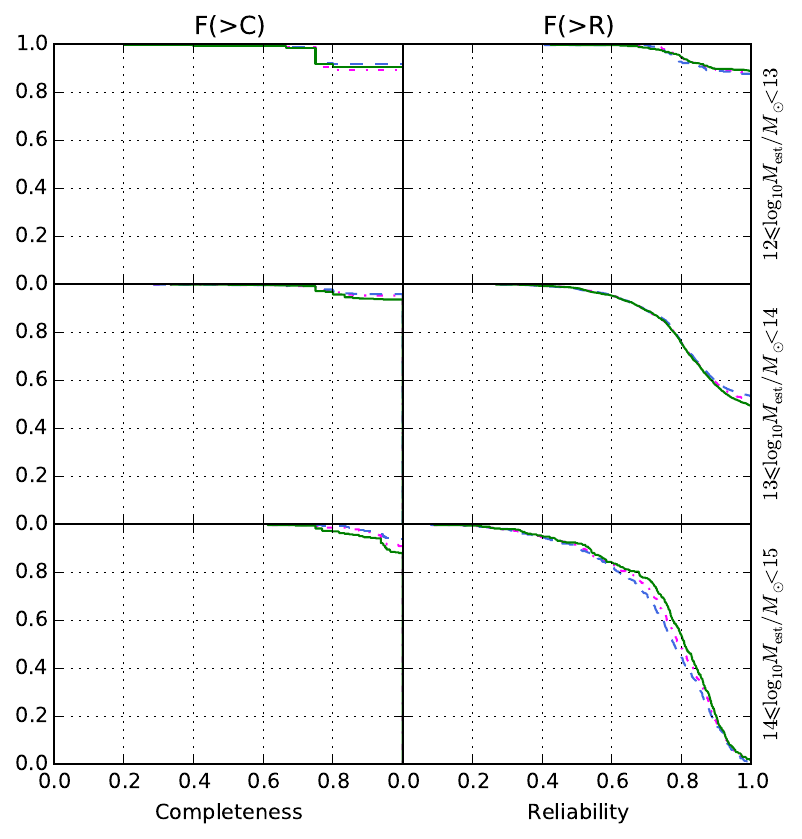}
        \caption{Effects of the choice of cosmological parameters on the
fraction of fragments (\emph{top left}),
the accuracy in the recovered halo mass (\emph{top right}),
and the          cumulative distribution functions of galaxy 
            completeness and reliability of the groups (\emph{bottom}) for
            {\sc maggie-m} (with 0.2 dex observational errors). 
The analysis is for unflagged groups of at least 3 true and 3 extracted members.
The colour
            code is 
\emph{green} for the Millennium Simulation cosmology 
($\Omega_{\rm m}=0.25$, $h=0.73$),
\emph{magenta} for WMAP9 
($\Omega_{\rm m}=0.279$, $h=0.70$),
and
\emph{royal blue} for Planck
($\Omega_{\rm m}=0.308$, $h=0.678$).
\label{fig:tests_vs_cosmo}
}
\end{figure*}

Groups extracted from group finders will depend
on the choice of cosmological parameters.
For example, when computing the projected radius of a galaxy at the redshift of
the group, we implicitly need to compute the cosmological angular distance,
$d_{\rm A}(z) = d_{\rm L}(z)/(1+z)^2$, 
hence the luminosity distance, $d_{\rm L}(z)$, which scales as the inverse
dimensionless Hubble parameter, $1/h= 100 \, {\rm km \,
  s^{-1}\,Mpc^{-1}}/H_0$, 
but also more subtly on $\Omega_{\rm
  m}$ and $\Omega_\Lambda$.
For our assumed flat $\Lambda$CDM Universe, the cosmological parameters
reduce to $h$ and $\Omega_{\rm m}$, and the luminosity distance 
is computed using elliptic integrals (\citealp{Liu+11}; see also
\citealp{Eisenstein97}).
Moreover, for all analytical HMFs tested in Sect.~\ref{sec:hmf_test}, we have
assumed the same cosmological parameters as in our mock, i.e. those of
the Millennium-II simulation, on which the galaxy SAM output was constructed,
which in turn was used to build our mock redshift space survey.
However, the observer may choose a different set
of cosmological parameters. 

We tested the sensitivity of {\sc maggie} to the choice of cosmological parameters,
by comparing the results of {\sc maggie-m} (with observational errors)
with the ``true'' cosmology ($\Omega_{\rm
  m}=0.25, h=0.73$ as assumed in the
Millennium-II simulation, on which our mock is based) with analogous runs of {\sc
  maggie-m} assuming instead
two ``false'' cosmologies (i.e. inconsistent with our mock): WMAP9, with
$\Omega_{\rm }=0.279, h=0.70$ \citep{Bennett+13} and Planck-2015 with
$\Omega_{\rm m}=0.308, h=0.678$ \citep{PlanckCollab+15_cosmo}. 

As seen in the plots of  Figure~\ref{fig:tests_vs_cosmo},
the
 choice of cosmological parameters affects very little the performance of
 {\sc maggie-m} on the fraction of groups that are secondary fragments (upper
 left plot), the inefficiency of group total mass estimation (right panels of
 upper right plot), as well as 
 galaxy completeness, and accuracy in total
 group mass, with no statistically significant trends with $\Omega_{\rm m}$.

However, with Planck (royal blue), the galaxy reliability is significantly less reliable
for the high EG mass bin of the distant sample (right panel of lower right
plot of Fig.~\ref{fig:tests_vs_cosmo},
again using a KS test)
than when the mock cosmology is assumed (green). 

Moreover, assuming WMAP9 (magenta) and especially Planck-2015 (royal blue) cosmologies leads to
increasingly positive biases
in group total mass  (left panels of upper right plot) at the high EG mass end up
to 0.15 dex higher (Planck) than with the cosmology of our mock (green).
Indeed, the lower Hubble constants of the WMAP and especially
Planck cosmologies, relative to the value used in the Millennium-II
simulation, hence in our mock, lead to halo masses that are larger by
the inverse ratio of Hubble constants in our AM
technique.\footnote{The group luminosities and stellar masses are also
  affected by the Hubble constant, but this does not affect the group total
  masses (halo masses) derived with AM, since the ranking of the group
  luminosities or stellar is independent of the Hubble constant, hence the
  first-rank group (by luminosity or stellar mass) will be assigned the
  highest halo mass, which scales as $1/h$, hence will be higher.}
This bias should thus be 
$\log (h_{\rm MS-II}/h_{\rm WMAP}) \simeq 0.02$ and 
$\log (h_{\rm MS-II}/h_{\rm Planck}) \simeq 0.03$, both independent of mass.
The left panels of the upper right plots confirm that the bias is roughly
independent of EG mass\footnote{We also find (not shown) that the bias in group total
  mass is even more independent of the true group mass.}, but is roughly 3
times greater than expected.

This general lack of sensitivity to cosmological parameters is expected, given the
low maximal redshift of our subsamples ($z=0.1$, see
Table~\ref{tab:subsamp}). One expects {\sc maggie} to be more
sensitive to the choice of cosmological parameters when applied to deeper
galaxy surveys.

\subsubsection{Effects of the form for the density of interlopers in projected
  phase space}
\label{sec:robust_gi}

We now probe the sensitivity of {\sc maggie-m} to the choice of the
functional forms that enter in the expression of the density of interlopers
in PPS, $g_{\rm i}(R,v)$ (see eqs.~[\ref{gilop}] and [\ref{gilophat}]),
namely the radial dependence of the normalization $A$ and standard deviation
$\sigma_{\rm i}$ of the Gaussian component of $g_{\rm i}$, as well as the
level of the constant component, $B$.

Figure~\ref{fig:compgi} in Appendix~\ref{sec:appgiGuo}
shows only small effects (note that the symbols in the upper plots have their
abscissa shifted for clarity) on the performance of {\sc
  maggie-l} and {\sc
  maggie-m} (both with observational errors) on group fragmentation, galaxy
completeness and reliability and group mass accuracy, when switching from
equations~(\ref{Ailop})-(\ref{Bilop}), derived by \cite{MBM10} on the dark
matter particles of a hydrodynamical cosmological simulation to
equations~(\ref{AilopGuo})-(\ref{BilopGuo}), that we fit to the mock that we
extracted from the SAM by \cite{Guo+11}.

\subsection{Comparison of {\sc maggie} to other group finders}
\label{sec:compalgos}

\subsubsection{Other grouping methods}

We now compare {\sc maggie} to other group finders. We first consider 
the FoF algorithm with the
 dimensionless linking lengths of $b_\perp=0.06$ and
        $b_\parallel=1.0$, which, in Paper~I, we had determined to be
 optimal for studies of environmental effects on galaxies (for the cosmology
 of the mock we are using here). These linking
 lengths are close to the values $b_\perp=0.06$, $b_\parallel=1.08$ optimized
 by \cite{Robotham+11}.
We consider two implementations of the FoF, where, for testing purposes, the
central galaxy is the most luminous (`FoF-L') or the most massive in
stars (`FoF-M').

\cite{Yang+07} provided extensive tests of their group finder, but these are
difficult to compare with ours, in particular, because 1) their mocks are
flux-limited, so they had to include uncertain corrections for
luminosity-incompleteness of groups in performing their AM, and 2) their
mocks do not include observational errors. 
We therefore additionally consider a simple implementation of the
\cite{Yang+05,Yang+07} group finder.

Yang et al. (2005, 2007) assigned a galaxy to a group according to a minimum density
in PPS, which they transformed into a number density contrast of galaxies in
redshift space:\footnote{Equation~(\ref{PM})
comes from writing the number of objects in a shell
($R,R$+${\rm d}R$) and redshift interval ($z,z$+$\Delta z$) as
$N(R,\Delta z)=2\pi R\,\Sigma_{\rm NFW}(R)\,p_{\rm Gauss}(\Delta z | R)$,
while the mean number 
of galaxies expected in the same volume $dV = 2 \pi R \,{\rm d}R
\,(c/H_0)\,{\rm d} \Delta z$ is
$\overline N = \overline n\,{\rm d}V = 2 \pi\,(c/H_0)\,\overline n\, R\,{\rm
  d}R\,{\rm d} \Delta z$.}
\begin{equation} 
P_M(R,\Delta z) = {H_0\over c}\,{\Sigma_{\rm NFW}(R)\over \overline n}\,p_{\rm Gauss}(\Delta z|R) >
{\cal B}  \ ,
\label{PM}
\end{equation} 
where $\overline n$ is the mean galaxy density at redshift $z_{\rm group}$,
$c$ is the speed of light, $H_0$ is the present-day Hubble constant,
$\Sigma_{\rm NFW}(R)$ is the NFW surface density profile,
$p_{\rm Gauss}$ is the Gaussian probability distribution
function, $\Delta z = z-z_{\rm group}$, and ${\cal B}$ is the dimensionless
local density contrast in redshift space.\footnote{Yang et al. (2005, 2007) assume that
  $p_{\rm Gauss}(\Delta z|R)$ is in fact independent of $R$, i.e. that
  $\sigma_{\rm LOS}$ is independent of $R$.}
Combining our notation with theirs, the total PPS density (what we would call
$g = g_{\rm h} + g_{\rm i}$ in {\sc maggie}) is
\begin{eqnarray} 
g(R,v) &=& \Sigma_{\rm NFW}(R)\,p_{\rm Gauss}(v|R) \nonumber \\
&=& \Sigma_{\rm NFW}(R)\,\left({1+z_{\rm group}\over c}\right)\,p_{\rm Gauss}(\Delta z|R) \\
&=& {(1+z_{\rm group})\over H_0}\,\overline n\,P_M(R,\Delta z) \ ,
\label{gYang1}
\end{eqnarray} 
where the second and third equalities are respectively obtained with
equations~(\ref{vofz})
and (\ref{PM}).
Equations~(\ref{PM}) and (\ref{gYang1}) lead to an expression of the Yang et
al. criterion in terms of the PPS density:
\begin{equation}
g(R,v) > {\cal B}\,{(1+z_{\rm group})\over H_0}\,\overline n \ .
\end{equation}

In our implementation of the Yang et al. algorithm, the probabilities of
membership of equation (\ref{prob1}) 
are replaced by
\begin{equation}
p(R,v) = \left\{
\begin{array}{ll}
1 & \hbox{ for } P_M(R,\Delta z) \geq {\cal B} \\
0 & \hbox { for } P_M(R,\Delta z) < {\cal B} \ .
\end{array}
\right. \ ,
\label{pYang}
\end{equation}
where equation~(\ref{vofz}) links $v$ with $\Delta z$.
In Appendix~\ref{sec:appYang}, we argue that
since we adopt different cosmological parameters (from the Millennium-II
simulation) than Yang et al. did (from their cosmological simulations), we
need to convert ${\cal B}=10$ to  ${\cal B}=29$.\footnote{In Appendix~\ref{sec:appYang}, we
  actually obtain ${\cal B}=6.1$ for the 
  cosmology adopted by Yang et al. (instead of ${\cal B}=10$), which translates
  to ${\cal B}=20$ in our cosmology.}
However,  Figure~\ref{fig:compYangB} in Appendix~\ref{sec:appYang}
indicates that the fraction of EGs that adopting ${\cal B}=10$ leads to  slightly
lower group fragmentation, much higher galaxy completeness, and less dispersed
total masses than with ${\cal B}=29$, but at the expense of much lower galaxy
reliability. 
We therefore adopt ${\cal B}=10$. 

We have implemented the Yang et al. group finder in this fashion using AM to
obtain halo masses with either the group luminosities (hereafter, Yang-L) or
the group stellar masses (hereafter, Yang-M), which we collectively refer to
as `Yang'.
Both of these implementations are a streamlined version
of the \cite{Yang+05,Yang+07} group finder (which also has two flavours according the
the observable used for the AM), as a full implementation is beyond the scope of the
present article.

\subsubsection{Fragmentation}

\begin{figure}
\centering
\includegraphics[width=0.8\hsize]{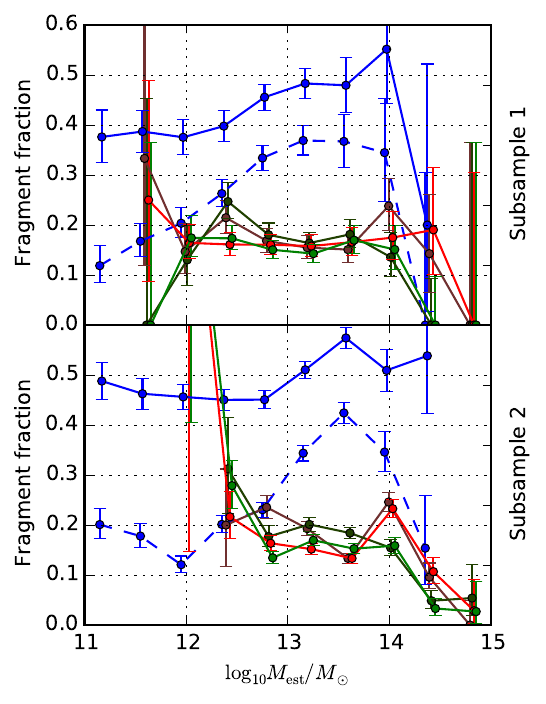} 
    \caption{Fraction of extracted groups with {\sc maggie-m}, {\sc
        maggie-l},
FoF and our
      slimmed-down version of Yang et al. that are secondary fragments,
        as a function of their estimated group mass, for unflagged groups of at
        least 3 members (for both the extracted and true groups), for both
        the nearby (\emph{top}) and distant (\emph{bottom}) subsamples.
        The line colours are  
\emph{green} for {\sc maggie-m},
\emph{red} for {\sc maggie-l},
\emph{blue} for FoF (with  \emph{dashed} and \emph{solid lines} for FoF-L and
FoF-M, respectively),
\emph{brown} for Yang-L
and
\emph{dark green} for Yang-M (the latter two with threshold ${\cal B}=10$).
All group finders were run on a catalogue with errors of 0.2 dex on stellar masses.
        The error bars are computed with the 
        Wilson (1927)
        formula. The points have their abscissa slightly
          shifted for clarity. 
\label{fig:others_fracfrag}}
\end{figure}

Figure~\ref{fig:others_fracfrag} shows that
the fraction of FoF-L groups
that are secondary fragments is over 30 percent for intermediate EG 
masses ($13 \leq \log M_{\rm est} \leq 14$), while the fraction of
FoF-M groups that are secondary fragments is over 38 percent at all masses.
In
comparison, the fraction of {\sc maggie} and Yang et al. 
groups that are secondary fragments
is typically less than 20\%, on average nearly three times less than for FoF-M
groups of the same estimated mass.
This very high contamination of FoF groups by secondary fragments had already
been noticed in Paper~I, where the fraction of groups that are secondary
fragments were found to be 49\%, 23\% and 46\% for the 3 bins of estimated
mass in the nearby subsample and 28\%, 29\% and 26\% for the corresponding
mass bins in the distant subsample. 
\textcolor{red}{The higher fractions of FoF EGs that are secondary fragments 
found in the present work are caused by
the inclusion of observational errors in our mock galaxy catalogue (in
contrast to how we proceeded in \citealp{DM+14a}), which sometimes
substitutes the wrong galaxy for the  central when matching the EG to the
TG. 
This effect is more important  
for stellar masses 
than for luminosities, 
since the assumed observational errors are higher for the
former (0.2 dex) than for the latter 
(0.08 dex).}

The fraction of secondary fragments is quite similar between {\sc maggie-l} and Yang-L
on one hand, and between {\sc maggie-m} and Yang-M on the other. On average,
{\sc maggie} performs slightly better than Yang by a few percent in the
absolute fraction of secondary fragments (4\% and 2.5\% for the L and M
flavors, respectively), in the range $12.5 < \log M_{\rm
  est}/\rm M_\odot < 14$, which is statistically significant in the distant
subsample (bottom panel of Fig.~\ref{fig:others_fracfrag}).

\begin{figure}
\centering
\includegraphics[width=0.8\hsize]{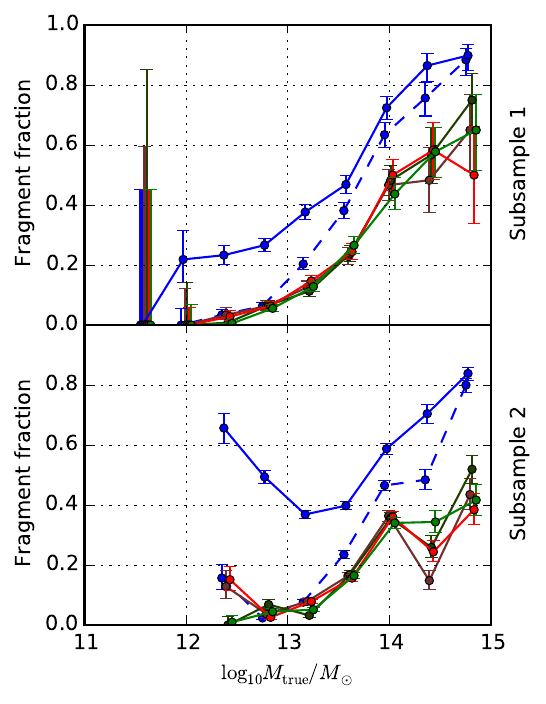} 
    \caption{Same as Figure~\ref{fig:others_fracfrag} but as a function of
      true group mass.
\label{fig:others_fracfragtrueM}
}
\end{figure}
The fraction of secondary fragments in Figure~\ref{fig:others_fracfrag} shows
that the FoF EG masses are limited to $\log M_{\rm est}/\rm M_\odot = 14.4$. This
is the consequence of the very high level of fragmentation at high TG masses,
as illustrated in Figure~\ref{fig:others_fracfragtrueM}. This high level of
fragmentation of FoF groups at high TG mass was also seen
in Figs. 4 and 6 of Paper~I. While {\sc maggie}
and Yang see their fraction of secondary fragments increase fairly moderately
from 0 to 40 to 60 percent (depending on the subsample) from low to high TG
masses, the FoF groups show much higher fractions of secondary fragments,
increasing to over 80 percent at high TG masses.

\begin{figure}
\centering
\includegraphics[width=0.75\hsize]{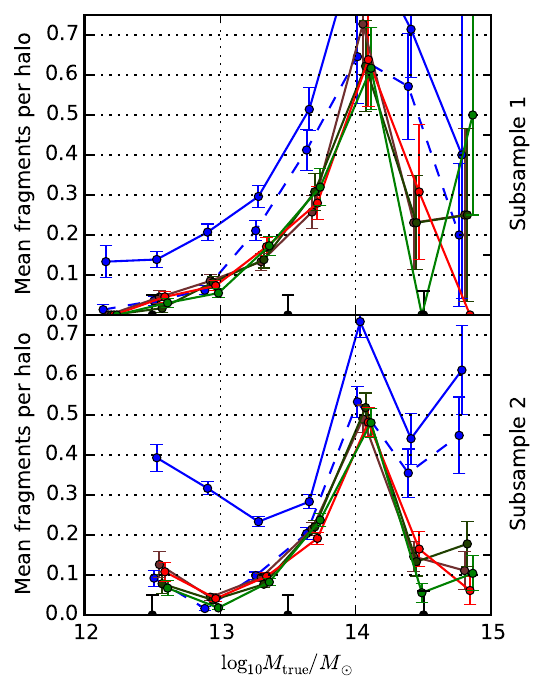}
\caption{Mean number of secondary fragments (estimated mass between 0.1 and 1
  times the true group mass) per true group as a function of true group mass
  for FoF (\emph{blue}), {\sc maggie-m} (\emph{green}), {\sc maggie-l}
  (\emph{red}), our implementation of Yang et al. (\emph{brown}, dashed and
  \emph{solid} for ${\cal B}=10$ and 29, respectively), and for Dom\'{\i}nguez
  Romero et al. (2012, \emph{black}). The \emph{upper} and \emph{lower}
  panels correspond to the nearby and distant subsamples, respectively, with
  the same Dom\'{\i}nguez Romero et al. values in both. All catalogues were
  given errors of 0.2 dex in stellar mass, except {\sc maggie-l}, which
  performed AM with luminosities that were given 0.08 dex
  errors, and Dom\'{\i}nguez
  Romero et al., which did not consider observational errors (and were
  measured on a flux-limited sample).
 The error
  bars indicate errors on means for all group finders except  Dom\'{\i}nguez
  Romero et al., for which standard deviations are used.
 The points have their abscissa slightly
  shifted for clarity.
\label{fig:fragDominguez}
}
\end{figure}

In the lower portion \footnote{The two panels
  of figure~1 of \cite{DominguezRomero+12} are confusing as they each mix two
 quantities.} of the right panel of their figure~1, \cite{DominguezRomero+12}
found that the mean number of secondary fragments was zero with less than
0.05 errors at all TG masses.
In contrast, {\sc maggie} and Yang lead to much higher mean numbers of
fragments, while FoF is even worse.

However, \citeauthor{DominguezRomero+12}  restricted their
secondary fragments to those accounting for at least 10\% of the TG mass, and
defined their primary fragments as the most massive, while our definition of
primary is the fragment containing the central galaxy. 

Figure~\ref{fig:masscomp}, further down, shows that a small minority of our secondary
fragments are more massive than the TG, so to make a clean comparison
with \citeauthor{DominguezRomero+12}, we show in
Figure~\ref{fig:fragDominguez} the mean number of secondary fragments with
mass between one-tenth and one times the TG mass.
Fragmentation worsens with increasing
TG mass, as we had found in Paper~I for FoF.
But Figure~\ref{fig:fragDominguez} also indicates that, using the measure of group fragmentation of
\citeauthor{DominguezRomero+12}, {\sc maggie}, Yang and FoF are unable to match
the zero mean number of secondary fragments per TG that
\citeauthor{DominguezRomero+12} found. 
This discrepancy would be even stronger had we used errors on the means
instead of standard deviations for
the points of \citeauthor{DominguezRomero+12}.

\subsubsection{Galaxy completeness}

\begin{figure}
\centering
\includegraphics[width=\hsize]{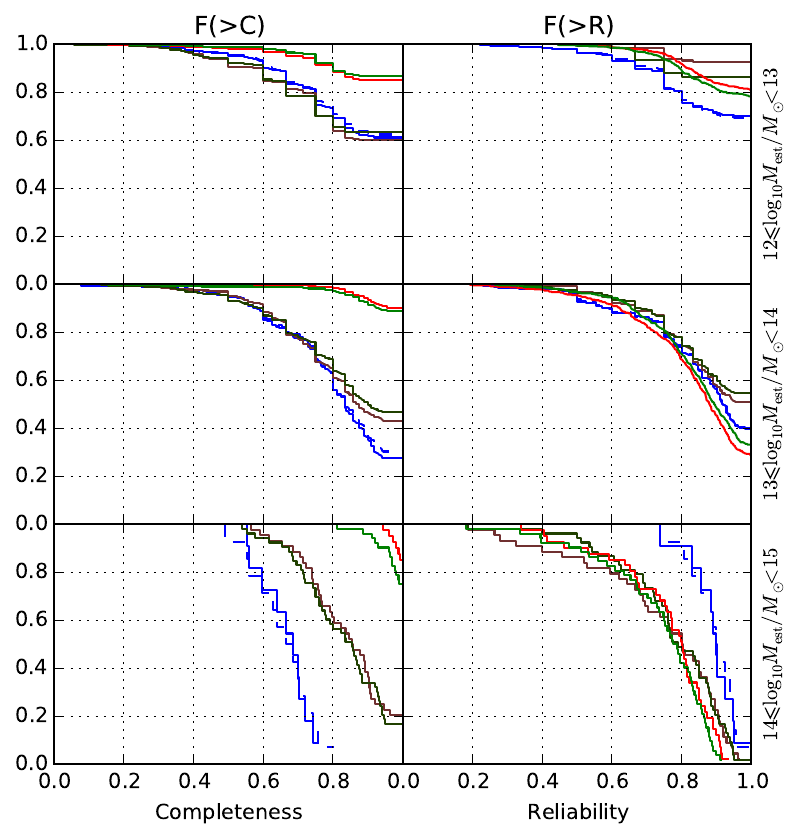}
\caption{Cumulative distribution functions  of the galaxy membership completeness
        (\emph{left}, eq.~[\ref{Cdef2}])
and reliability (\emph{right}, eq.~[\ref{Rnew}]), 
for groups extracted with {\sc maggie}, {\sc
  FoF}, and Yang for the nearby subsample
 (unflagged galaxies in groups of at
        least 3 members (for both the extracted and true groups) that
        are not secondary fragments .
Both completeness and reliability are relative to the virial sphere of the true groups,
and are derived in bins of estimated group
        masses.
Same colours and line-types as in Fig.~\ref{fig:others_fracfrag}.
\label{fig:CompRelia_vsOthers1}
}
\end{figure}

\begin{figure}
\centering
\includegraphics[width=\hsize]{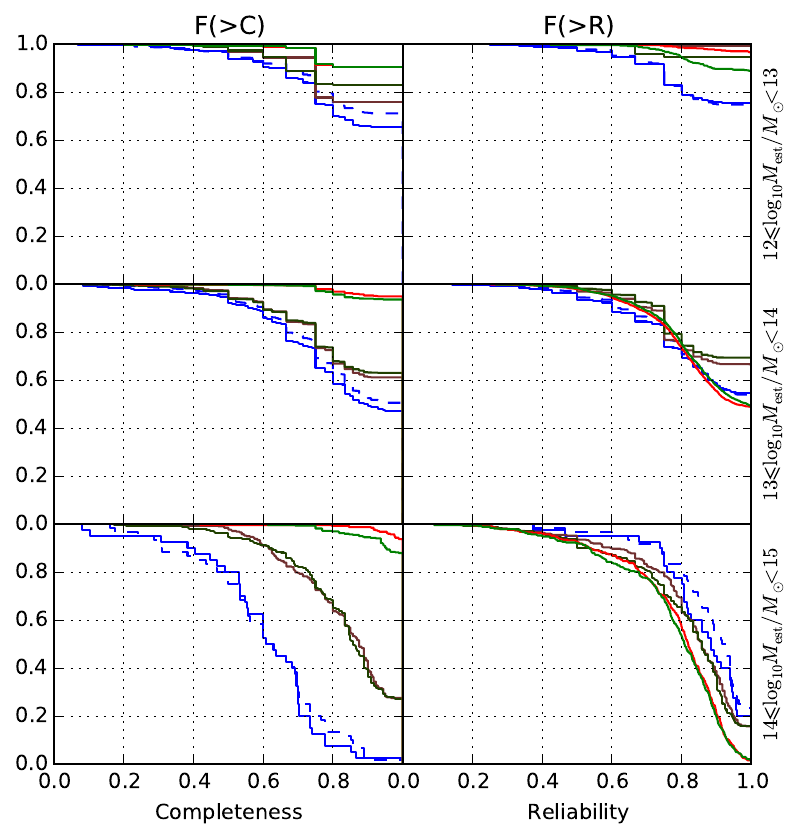}
\caption{Same as Fig.~\ref{fig:CompRelia_vsOthers1}, but for the distant
  sample.
In the lowest mass bin, the completeness of {\sc maggie-l} is virtually the
same as that of {\sc maggie-m} (which hides the former).
\label{fig:CompRelia_vsOthers2}
}
\end{figure}

The left panels of 
Figures~\ref{fig:CompRelia_vsOthers1} and \ref{fig:CompRelia_vsOthers2} show
that the galaxy completeness of {\sc maggie} is considerably higher than for
FoF and Yang, regardless of the flavour (L or M). 
For example, according to
Figure~\ref{fig:CompRelia_vsOthers2}, 
the fractions of EGs in the distant
subsample (which has better statistics)
with better than 80
percent incompleteness are 
91, 97 and 97
percent for {\sc maggie-m} and even higher for {\sc maggie-l},
compared to
80, 73 and 67 percent for Yang and
77, 63 and only 9 percent for FoF.
Similar values
are found for the nearby subsample (Fig.~\ref{fig:CompRelia_vsOthers1}).
The superior galaxy completeness of {\sc maggie} relative to Yang is
statistically significant (with the KS test) in all three EG mass bins and in
both subsamples. Similarly, Yang is significantly more complete than FoF in
all EG mass bins and in both subsamples, except for the low EG mass bin of
the nearby subsample, for which FoF is significantly more complete than Yang.

The decrease of galaxy completeness with increasing group mass was already
noticed by \cite{Yang+07} (upper panels of their figure~2), although this
decrease in completeness with mass is not as severe in their study, which
considers bins of TG mass rather than EG mass (and is also for a
flux-limited sample).

The left panel of figure~2 of  \citeauthor{DominguezRomero+12} indicates
median galaxy completeness of 96\%, 91\% and 92\% for the three mass bins we
used here (but for TG masses). In contrast, according to 
Figures~\ref{fig:CompRelia_vsOthers1} and \ref{fig:CompRelia_vsOthers2}, we
arrive at median galaxy completeness of 100 percent for {\sc maggie} in all
mass bins and in both subsamples, while our implementation of the Yang
algorithm leads to 100 percent galaxy completeness in the low mass bin, 85
percent in the high mass bin (somewhat worse in bins of true mass).
A more direct test involves a comparison to the flux-limited mock analyzed by
\cite{Yang+07}, whose upper left panel of figure~2 indicates 100 percent
completeness in all mass bins. 

The
FoF algorithm produces increasingly incomplete galaxy extractions for
increasing estimated group masses (see above).
The lower galaxy completeness of FoF at high estimated group masses is a consequence of the
very high fragmentation of high mass real-space groups (see
Fig.~\ref{fig:others_fracfragtrueM}).
If a TG has non-negligible secondary fragments, then its primary fragment will
tend to be incomplete.
Consider, for example, a TG with 5 galaxies that is fragmented into an EG of 3 galaxies
(containing the TG's central) and another EG of 2 galaxies; the EG of 3
galaxies will have a completeness of 3/5=0.6, a reliability of unity, and
will not be counted as a secondary fragment, while the EG of 2 galaxies
will be considered a secondary fragment, but will not have completeness and
reliability measured.

Finally, we compare the completeness levels to those that 
\cite{MunozCuartas+12} derived from the analysis of the groups of at least
two members that they extracted with their FoF-like group finder on a
doubly complete subsample with $M_r < -19$ (matching our nearby doubly complete
subsample). According to  the bottom panel of
their figure~13, the fraction of their EGs with over 90\% completeness is 82\%.
In contrast, in the nearby subsample, our Figure~\ref{fig:comp_rel_1_est} indicates
that the fractions of EGs with 90\%  galaxy completeness are higher in all
mass bins:
85\%, 94\%, and 100\% for {\sc maggie-l}
and 
87\%, 92\%, and 96\% for {\sc maggie-m}.

\begin{figure*}
    \centering
    \includegraphics[width=\hsize]{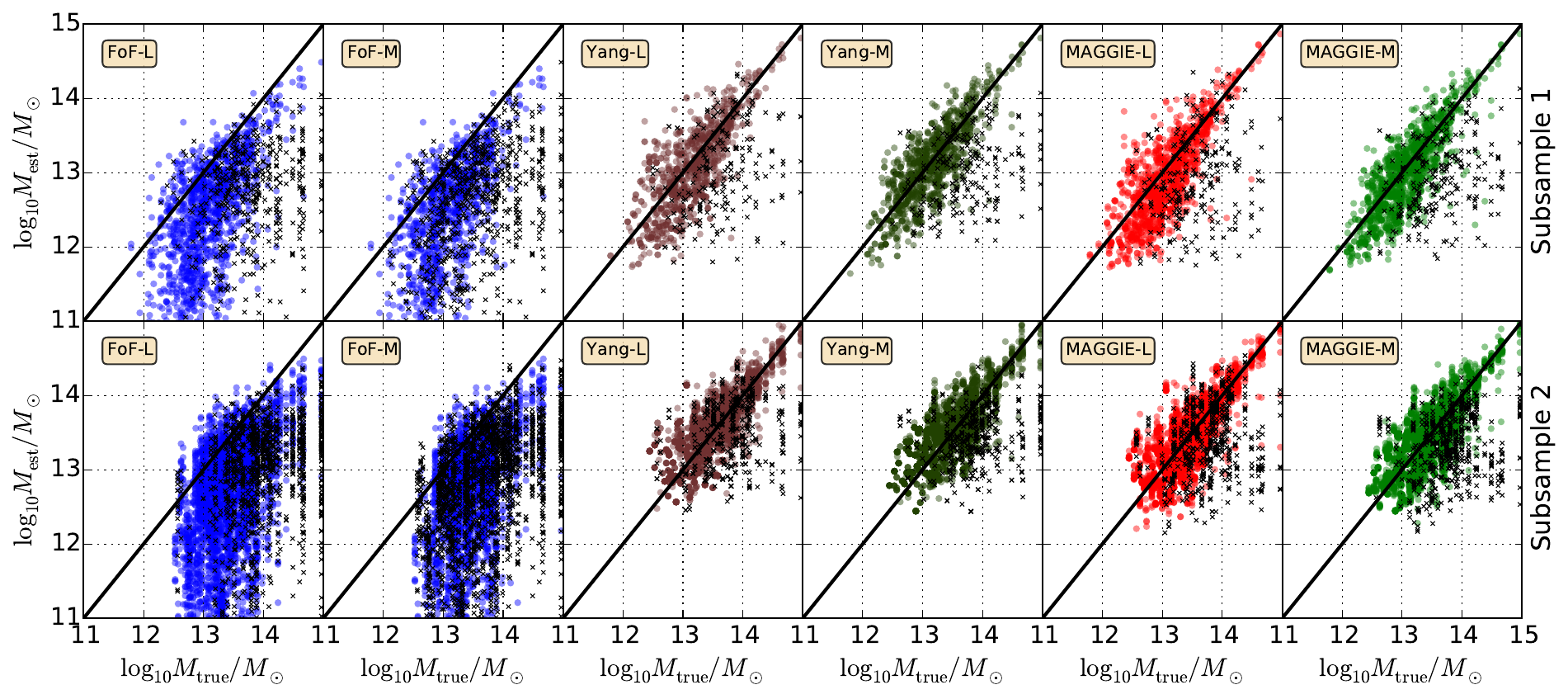}
    \caption{Estimated mass versus true mass for the unflagged non-secondary
      (\emph{filled coloured circles}) and secondary (\emph{black crosses}) EGs
        with at least 3 members (for both the extracted and true groups) 
        for six group finders (from left to right):
        FoF-L, FoF-M, Yang-L, Yang-M (both with ${\cal B}=10$), 
        {\sc maggie-l}, and
        {\sc maggie-m}.
The \emph{diagonal lines} indicate
        perfect mass recovery.
There are roughly as many {\sc maggie} groups
        as there are FoF groups, but the former occupy identical positions in
        the plots due to the replications of the simulation boxes
        causing identical stellar masses, hence identical group masses after
        the AM used to infer group masses. 
\label{fig:masscomp}}
\end{figure*}

\subsubsection{Galaxy reliability}
The right hand panels of
Figures~\ref{fig:CompRelia_vsOthers1} and \ref{fig:CompRelia_vsOthers2}
indicate that the rankings of the group finders in terms of galaxy
reliability depend on EG mass.
For the distant subsample (which has better statistics), at low EG mass (upper right
panel of Fig.~\ref{fig:CompRelia_vsOthers2}), 
the fraction of groups with 100 percent reliability is higher with
Yang (90 percent) than with {\sc maggie} (80 percent) or FoF (70 percent).
At 80 percent reliability, {\sc maggie} and Yang have similar reliabilities
(90 percent).
The overall differences in the CDFs, as quantified by the KS test, 
indicate that overall, Yang-M is
significantly more reliable than {\sc maggie-m}, while Yang-L and {\sc
  maggie-l} cannot be distinguished. In turn, {\sc maggie} is significantly
more reliable than FoF.

At intermediate EG mass (middle right panel of
Fig.~\ref{fig:CompRelia_vsOthers2}), all three group finders lead to similar
fractions of 
roughly three-quarters of
groups with over 80 percent reliability (80 percent for Yang-M).
Yang is significantly more reliable than either {\sc maggie} or FoF, while
{\sc maggie} is signficantly more reliable than FoF (although FoF has a
marginally significantly higher fraction of groups with 100 percent reliability).

At high EG mass (lower right panel of Fig.~\ref{fig:CompRelia_vsOthers2}),
there is a clear hierarchy, where FoF is significantly more reliable than Yang,
which in turn is significantly more reliable than {\sc maggie}. The fraction
of groups with 80 percent reliability is roughly 80 percent for FoF, roughly
two-thirds for Yang and slightly more than half for {\sc maggie}.

Our probabilistic method of measuring {\sc maggie}'s galaxy
reliability (eq.~[\ref{Rnew}]) leads to values that are rarely low or near
100 percent. This can be clearly seen in the intermediate-mass EG groups
 (middle right panel of
Fig.~\ref{fig:CompRelia_vsOthers2}), where for {\sc maggie} relative to FoF
or Yang, the decrease of the CDF is slower
for reliabilities below 80 percent and  faster at higher reliabilities.

\subsubsection{Mass accuracy}
Figure~\ref{fig:masscomp} shows how the estimated total masses (of the EGs)
compare with the total masses (within the virial sphere) of the TGs (for
clarity, we hereafter drop the term `total' before `mass' in this
subsection), both for the primary (large coloured circles) and secondary
(black crosses) fragments. 

The FoF
method, with the virial theorem to estimate masses (with the formula from \citealp{HTB85}),
leads to frequent strong underestimation of the mass for the primary
fragments of low-mass TGs. 
This is analogous to what is found by most group mass estimation
methods when the group center and its velocity are provided
(see \citealp{Old+14}, although \citealp{Old+15} find that a Bayesian fitted
slope of the estimated versus true mass relation is typically unity). 
Figure~\ref{fig:masscomp} shows that {\sc maggie} and Yang
do not underestimate the EG masses of primary fragments 
as frequently as does the FoF algorithm.
This better behaviour is likely to be the result of the use of AM 
to derive group masses. 

As expected, in secondary fragments, the estimated mass is usually lower than
the TG mass
(often by several dex). However, in some secondary
fragments, the estimated mass is higher than the TG mass. This can
occur when the group luminosity or stellar mass is higher in the group
with the lower central luminosity or stellar mass (which is how secondary fragments
are defined).
Figure~\ref{fig:masscomp} also shows that
the trend of estimated versus true mass for the secondary
fragments has a shallower slope than unity in contrast with the analogous
trend for primary fragments. 

\begin{figure}
\centering
\includegraphics[width=\hsize]{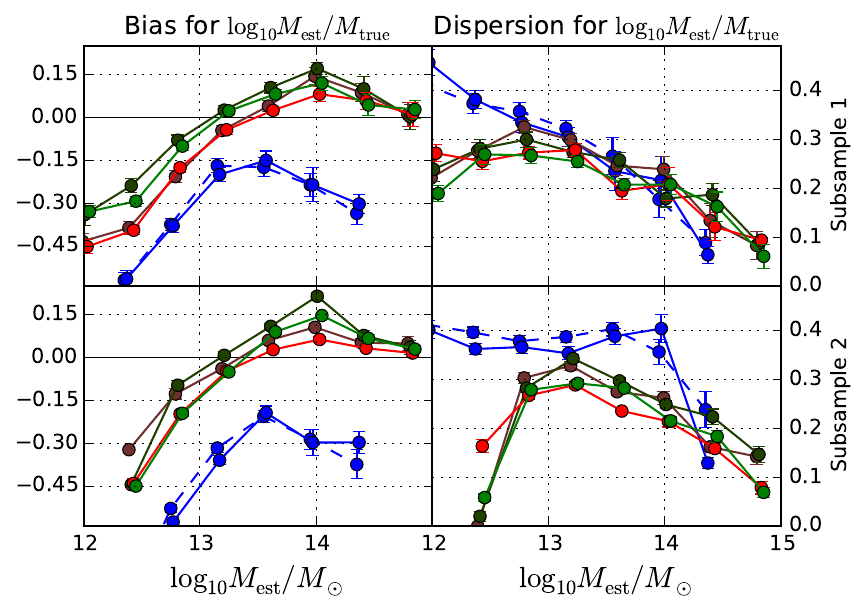}
\caption{Median bias and scatter (using 16th and 84th percentiles) of
      extracted group total mass within the 
        virial sphere, for groups
extracted with {\sc maggie}, {\sc
  FoF}, and our implementation of Yang et al..
The \emph{top} and \emph{bottom} panels are for  nearby and distant subsamples,
respectively
 (for unflagged primary groups of at least 3 true and 3 extracted members),
The error bars for the bias and scatter are respectively $\sigma/\sqrt{N}$
and $\sigma/2\,[2/(N-1)+\kappa/N]^{1/2}$, where $\kappa$ is the kurtosis excess.
Same colours as Figure~\ref{fig:others_fracfrag}.
\label{fig:others_accMhalo}
}
\end{figure}

The left panels
of Figures~\ref{fig:masscomp} and ~\ref{fig:others_accMhalo} show that 
 the virial-theorem masses of FoF EGs are biased low by a factor as great as
10 at the lowest estimated masses and by over 0.15 dex at high
estimated masses.
Similar trends of strong mass underestimation with FoF were found in Paper~I
for low-richness EGs. 
The mass biases of {\sc maggie} and Yang vary in a similar way as a function
of EG mass as for FoF, except that the biases are much lower (the mass ratio
is much closer to unity): while the FoF mass bias is greater than 15\% at all EG
masses,
the mass biases of {\sc maggie} and Yang are better than 15\% for 
$\log M_{\rm est}/\rm M_\odot > 13$.

Since bias can, in principle, be corrected for, 
it is more important to consider the scatter in the
mass estimation. 
The right hand panels of Figure~\ref{fig:others_accMhalo} indicate that, 
at low EG masses, FoF EGs have much greater mass dispersion than their {\sc
  maggie} and Yang counterparts. For example, at $\log M_{\rm est}/\rm M_\odot
\simeq 12.7$, the FoF mass scatter is 0.1 dex worse than that of {\sc maggie},
with Yang in between (nearby subsample) or as good as {\sc maggie} (distant
subsample).

At intermediate EG masses, 
{\sc maggie} remains the group finder with the lowest scatter in group masses
(roughly 0.24 and 0.26 dex in the nearby and distant subsamples,
respectively), 
with Yang slightly worse (0.25 dex and 0.30 dex scatter respectively), and
FoF slightly worse than Yang in the nearby subsample (0.28 dex scatter) and
much worse in the distant subsample (0.38 dex scatter).
Given the errors in the dispersions, these trends are statistically significant.

At high EG masses, the hierarchy between group finders is more difficult to
appraise. The factor of 2 typical
underestimate of group masses with FoF prevents high EG masses with FoF,
which explains why no points for FoF beyond
$\log M_{\rm est}/\rm M_\odot = 14.4$ in
Figure~\ref{fig:others_accMhalo}.
At this EG mass, FoF-M produces lower dispersion (0.06 and 0.12 dex for the
nearby and distant subsamples, respectively), while FoF-L is not as accurate
(0.09 and 0.24 for the two subsamples, respectively).
At this EG mass, the corresponding dispersions on EG masses are
0.12, 0.13, 0.17 and 0.19 for {\sc maggie-l}, Yang-L, {\sc maggie-m} and Yang-M
for the nearby subsample,
and
0.16, 0.16, 0.18 and 0.22 for the distant subsample (in the same order of
group finders). Averaged over the full bin of high EG masses, while {\sc
  maggie} and Yang are comparable in the nearby subsample, while {\sc maggie} is
more accurate than Yang by 0.04 dex in the distant subsample.

The scatter in group total mass  of the original Yang et al.
 algorithm was shown in figure~7
of \cite{Yang+07}. Considering the case where no correction for luminosity
incompleteness is required (their panel c), one finds that the scatter divided by
$\sqrt{2}$ is $\sigma_Q \approx 0.23$ for group estimated log masses in the range
12 to 14.4 (solar units),  
leading to a scatter of $\sqrt{2}$$\times$0.23=0.33.\footnote{We do not understand the factor of
   $\sqrt{2}$ in eq. (14) of \cite{Yang+07}, since the halo masses to which
   the group masses are compared are highly accurate as they are derived by
   summing over 1000 simulation particles, leading to 0.014 dex mass
   scatter from Poisson shot noise.}

The dispersion in group mass found with our implementation of Yang on our
doubly-complete subsamples (with observational errors)
are thus much lower, and diminish with increasing
EG mass from 0.3 dex to 0.1 dex (right panels of Fig.~\ref{fig:others_accMhalo}), 
contrary to the situation with the flux-limited sample (without
observational 
errors) analyzed by \cite{Yang+07}.

\subsubsection{Accuracy of group luminosity and stellar mass}

\begin{figure}
\centering
\includegraphics[width=\hsize]{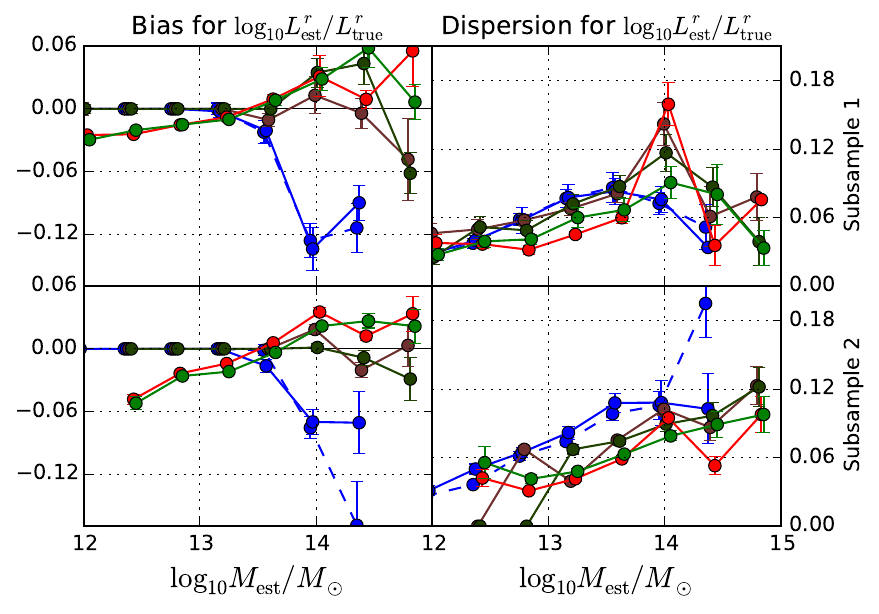}
\caption{Median bias and scatter (using 16th and 84th percentiles) of
      extracted group luminosity within the 
        virial sphere, for groups
extracted with {\sc maggie}, {\sc
  FoF}, and our implementation of Yang et al..
The \emph{top} and \emph{bottom} panels are for  nearby and distant subsamples,
respectively
 (for unflagged primary groups of at least 3 true and 3 extracted members),
The error bars for the bias and scatter are respectively $\sigma/\sqrt{N}$
and $\sigma/2\,[2/(N-1)+\kappa/N]^{1/2}$, where $\kappa$ is the kurtosis excess.
Same colours as Figure~\ref{fig:others_fracfrag}.
\label{fig:others_accL}
}
\end{figure}

\begin{figure}
\centering
\includegraphics[width=\hsize]{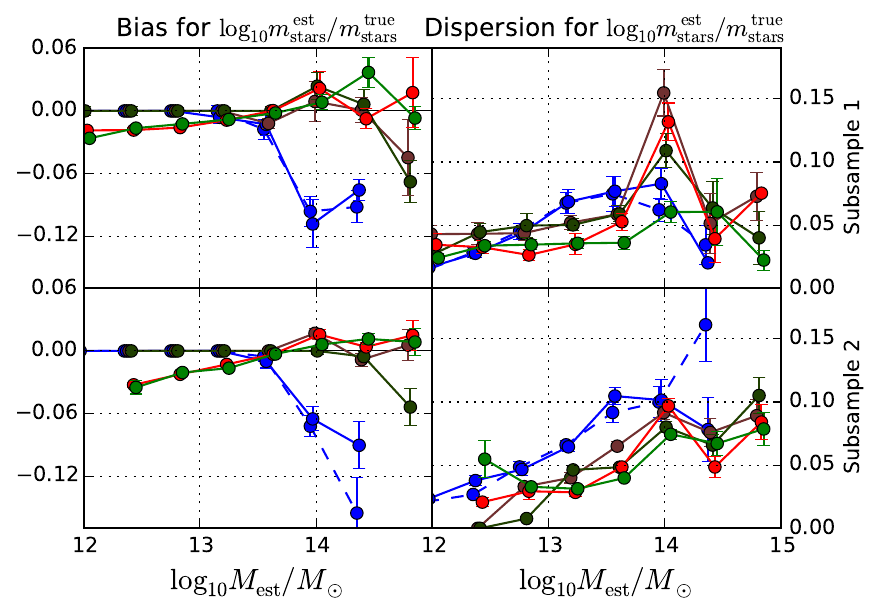}
\caption{Median bias and scatter (using 16th and 84th percentiles) of
      extracted group stellar mass within the 
        virial sphere, for groups
extracted with {\sc maggie}, {\sc
  FoF}, and our implementation of Yang et al..
The \emph{top} and \emph{bottom} panels are for  nearby and distant subsamples,
respectively
 (for unflagged primary groups of at least 3 true and 3 extracted members),
The error bars for the bias and scatter are respectively $\sigma/\sqrt{N}$
and $\sigma/2\,[2/(N-1)+\kappa/N]^{1/2}$, where $\kappa$ is the kurtosis excess.
Same colours as Figure~\ref{fig:others_fracfrag}.
\label{fig:others_accmstars}
}
\end{figure}

Figures~\ref{fig:others_accL} and \ref{fig:others_accmstars} respectively
show the accuracy on EG luminosities and stellar masses.  The left panels of
both figures show that, at low EG mass, FoF and Yang have zero median bias of
luminosity and stellar mass, while {\sc maggie} is biased low (always less
than 0.05 dex).
The median bias of zero for FoF and Yang is expected since the
 reliability of the primary EGs in the low EG mass bin is 100 percent for
 all group finders.
The negative median bias of {\sc maggie} at low EG mass must therefore be 
linked to its
probabilistic method of measuring luminosity (eq.~[\ref{Lgroup}]) and stellar
mass (eq.~[\ref{mgroup}]).

At intermediate EG mass, the biases in luminosity and stellar mass with
Yang-M remain at zero, 
those with Yang-L are within 0.01 dex of zero, those with {\sc maggie}
increase to typically 0.03 dex from zero, while those with FoF decrease
 to reach --0.07 (--0.12) dex at $\log M_{\rm est}/\rm M_\odot=14.0$ for the
 distant (nearby) subsample.

At high EG mass, Yang and {\sc maggie} have comparable biases in luminosity
and stellar mass, typically within 0.03 dex from zero (except Yang-M at the
highest EG mass bin for stellar masses, 
which is biased low by typically 0.06 dex). On the other hand,
the group luminosities and stellar masses are biased low by  FoF by typically
--0.08 dex for FoF-M and as much or worse than --0.10 dex for FoF-L (all with
fairly large uncertainties).

The FoF, Yang and {\sc maggie} group finders all lead to comparable 
inefficiencies in EG luminosity or stellar mass in the low EG mass bin of the
nearby subsample, 0.04 dex for {\sc maggie} and 0.045 dex for FoF, and 0.05
dex for Yang. In the distant subsample, while FoF remains at 0.05 
(luminosities) or 0.04 (stellar masses) dex inefficiency, {\sc maggie} has
lower dispersion of 0.03 dex, while Yang reaches even lower dispersions at
the lowest EG masses.

At intermediate EG mass, EGs extracted with FoF have the highest dispersion
(typically 0.09 dex), while
{\sc maggie} EGs have the lowest dispersion in luminosity and stellar mass
(typically 0.05 dex), and Yang EGs are in between.
However, {\sc maggie-l} and Yang-L (to a lesser extent) have a spike in
dispersion in luminosity and stellar mass at 
$\log M_{\rm est}/\rm M_\odot=14.0$ in the nearby subsample.

At high EG masses ($\log M_{\rm est}/\rm M_\odot>14.0$), FoF groups only reach
$\log M_{\rm est}/\rm M_\odot>14.4$. At that EG mass, FoF and {\sc maggie-l}
groups
 have the
lowest dispersions in luminosity and stellar mass in the nearby subsample
(typically 0.04 dex only),
whereas {\sc maggie-m} and Yang-M have the highest dispersions (0.06 dex in
luminosity and 0.08 dex in stellar mass), while Yang-L is in between.
In the distant subsample at the same EG mass, all group finders lead to the
same dispersions in luminosity (0.09 dex) and stellar mass (0.075 dex),
except that {\sc maggie-l} has significantly lower dispersions (0.05 dex),
while FoF-L has very high dispersions (over 0.16 dex).
At the highest EG mass, where $\log M_{\rm est}/\rm M_\odot=14.8$, {\sc maggie-m}
and Yang-M lead to the lowest dispersions in the nearby subsample (0.03 dex
only), while {\sc maggie-L} and Yang-L are marginally less efficient (0.075
dex).
In the distant subsample at this very high EG mass, both flavours of {\sc
  maggie} lead to the lowest dispersions in luminosity and stellar mass (0.10
dex in luminosity and 0.08 dex in stellar mass), while Yang-M is
marginally worse (0.12 dex in luminosity and 0.10 dex in stellar mass).

\subsubsection{Richnesss versus total mass}

\begin{figure}
\includegraphics[width=\hsize,bb=0 0 560 710]{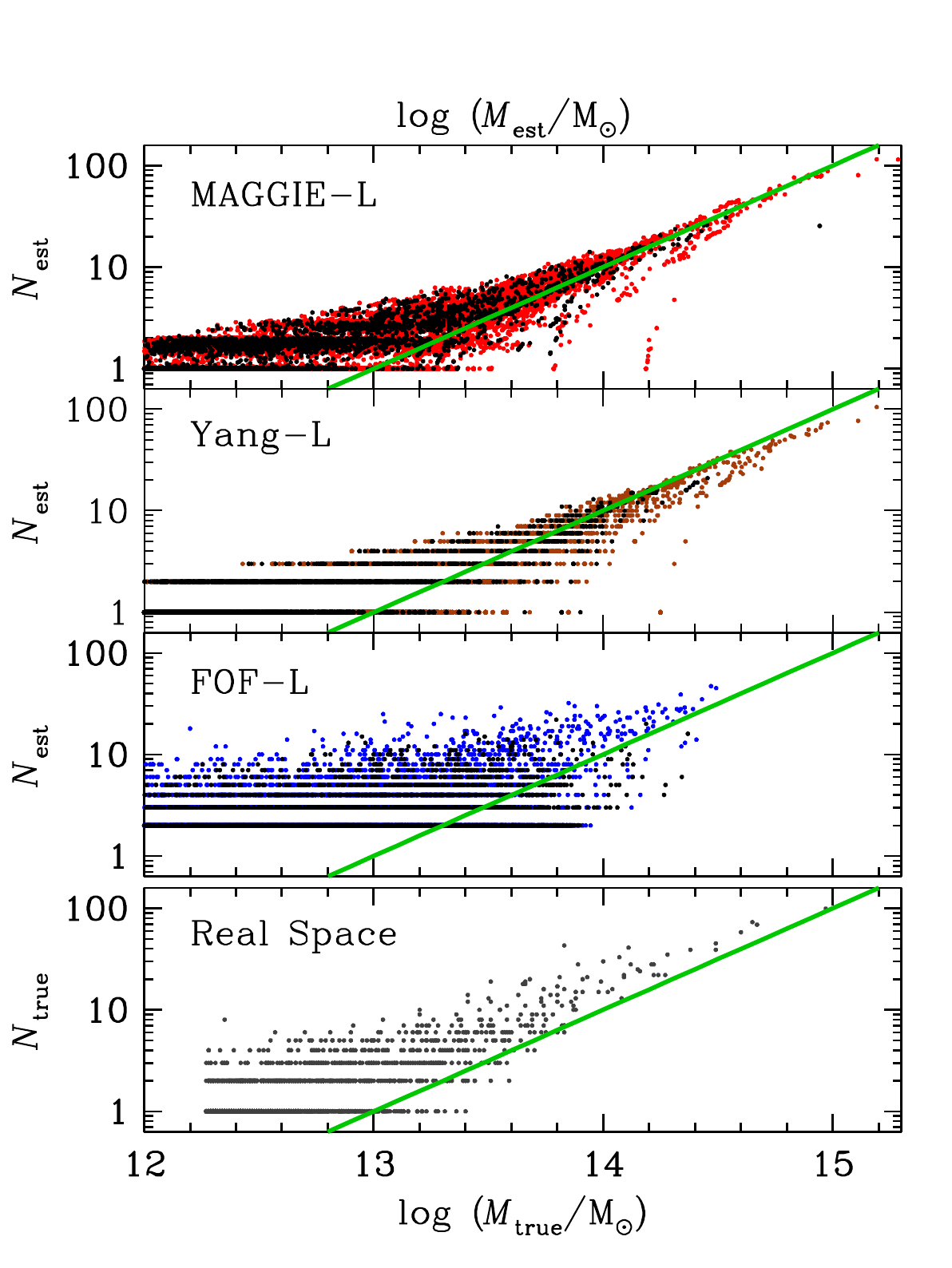}
\caption{Group richness versus total mass. 
\emph{Top three panels}: Estimated quantities for unflagged groups
  extracted with 
{\sc maggie-l} (\emph{top}),
Yang-L (\emph{middle}), and
FoF-L (\emph{bottom}), all for the distant subsample.
The primary fragments are shown in colour 
(\emph{red} for {\sc maggie-l},
\emph{brown} for Yang-L, and \emph{blue} for FoF-L), while the secondary
fragments are shown in \emph{black}.
\emph{Bottom panel}: True richness (galaxies more luminous than $M_r =-20.5$,
which corresponds to the luminosity limit of the distant subsample)
versus total mass of real space groups (\emph{dark grey}).
In all panels, the \emph{thick green line} shows $N = M/10^{13} \rm M_\odot$,
simply to guide the eye.
\label{fig:NvsM}}
\end{figure}

We complete the comparison of {\sc maggie} with Yang and FoF by comparing the
richness - mass relations of the three group finders.
Figure~\ref{fig:NvsM} shows the estimated richness versus the estimate mass
for {\sc maggie-l}, Yang-L and FoF-L, as well as the true richness versus
true mass (where the richness is measured for galaxies more luminous than
$M_r = -20.5$, conforming to the luminosity limit of the distant
subsample).\footnote{The lower number of groups in the Real Space panel is a
  consequence of the smaller volume of the Millennium-II simulation box
  in comparison to that of the distant doubly complete subsample.}
The richness-mass relations of {\sc maggie} and Yang resemble one another,
although the estimated galaxy numbers are integers for Yang (and FoF),
while they are floating numbers for {\sc maggie}, making the figure for {\sc
  maggie} more continuous. In particular the mean trend and scatter are very similar.
On the other hand, 
the poor mass accuracy of FoF causes scatter and the negative bias of mass
shows here as higher richness for given mass.
The richness mass relations are very similar for the nearby sample and for
the M flavour of the three group finders.
One also notices in Figure~\ref{fig:NvsM} that the overall trend of the
estimated richness-mass relations match fairly well that of the true
relation, with Yang, and to a lesser extend {\sc maggie} showing slight
negative bias in richness at given mass
relative to the real-space trend.

\subsubsection{Synthesis}

A comparison of {\sc maggie} with other group finders, performed on the same
mock catalogue, with the same observational errors, 
does not point to a
single superior group finder among the 6 tested group finders (3 algorithms:
{\sc maggie}, FoF and Yang; 2 definitions of central galaxy: most luminous or
most massive in stars).

Since most of our tests involve the primary fragments of the EGs, group
fragmentation appears to be the most serious issue. Our tests indicate that
{\sc maggie} has the least group fragmentation, with Yang a close second,
while FoF suffers tremendously from group fragmentation.
This hierarchy is repeated when we consider galaxy completeness of primary groups, which is
related to group fragmentation, since groups with important secondary
fragments will necessarily have lower galaxy completeness.

Galaxy reliability is generally the highest in the Yang group
finder, with {\sc maggie} second and FoF last, except at high EG masses,
where FoF is most reliable while {\sc maggie} is the least.

Group total mass is severely underestimated by the virial theorem mass of the
FoF group finder. This, again, is a consequence of the heavy fragmentation of
FoF groups. {\sc maggie} and Yang perform much better, with a slight
advantage for {\sc maggie}.
The ranking of the group finders concerning the dispersion in group total
mass depends on the interval of EG mass considered. At low EG mass, FoF
performs poorly and Yang performs slightly better than  {\sc maggie}.
At intermediate EG mass, {\sc maggie} performs slightly better than Yang
(starting at $\log M_{\rm est}/\rm M_\odot = 12.5$) and considerably better
than FoF.
At high EG mass, {\sc maggie} performs slightly better than Yang, with FoF
even better, but limited to half the high EG mass bin, given its negative bias.

Group luminosity and stellar mass are measured without bias with FoF and Yang
for low EG mass groups, while {\sc maggie} is slightly biased low (by typically less
than 0.05 dex). At intermediate EG mass, Yang EG luminosities and stellar
masses  remain unbiased, FoF is biased low and {\sc maggie} is only slightly
biased (in either direction).
At high EG mass, {\sc maggie} is the least biased and FoF the most.

Finally, the richness-mass relations of {\sc maggie} and Yang are much less
scattered than that of FoF, although 
 {\sc maggie} and Yang are biased low in richness (at given mass), 
by 0.2 dex ({\sc maggie}) and 0.3 dex (Yang), while FoF is not.

In summary, relative to FoF, {\sc maggie} suffers much less from
fragmentation, is much more complete, more reliable, except at high EG
masses, with much less biased group masses, considerably less dispersed group
masses, except for cluster-mass EGs, and produces more accurate group luminosities 
and stellar masses, except for low-mass EGs, where the median bias although
very small, is not zero as in FoF.

Relative to our implementation of Yang, {\sc maggie} is slightly less
affected by group fragmentation, considerably more complete but less reliable in its galaxy
membership, slightly more accurate in group total masses, slightly more
biased but slightly less dispersed in group luminosities and stellar masses.
 
These tests permit to assess how the characteristics of the group finder
influence its performance in reproducing the real space groups.
FoF is a group finder that makes no assumptions on the physics of groups and
of projection effects (once its linking lengths have been optimized).
The group finder by \cite{Yang+05,Yang+07} introduced priors on the galaxy
distribution in PPS as well as abundance matching to determine EG masses.
{\sc maggie} improved the adopted priors on the distribution of galaxies in
PPS and adopted a probabilistic approach for the membership of galaxies in
groups, following \cite{DominguezRomero+12}, but keeping the memberships
probabilistic instead of finishing with hard group assignments as done by
\citeauthor{DominguezRomero+12}.
Our tests indicate that, overall, there are fewer differences between {\sc
  maggie} and Yang, than between either and FoF. This suggests that the
probabilistic membership of {\sc maggie} and its more refined priors on the
distribution of galaxies in PPS play a smaller role than the use of priors,
as well as the AM used for measuring group masses.

\section{Conclusions}
\label{sec:maggie_discussion_conclusion}

We have introduced a new prior, halo-based and fully probabilistic group finder called
{\sc maggie}, where the total group/cluster masses are obtained by abundance matching
between the assumed known halo mass function and the derived group luminosity
({\sc maggie-l}) or stellar mass ({\sc maggie-m}) function.
This grouping algorithm is similar to that of \cite{Yang+05,Yang+07}, but
uses a more refined and probabilistic membership criterion, and is meant to be applied to
subsamples that are complete in both luminosity and distance, to avoid the
unavoidable luminosity incompleteness in flux limited samples, which are very
difficult to accurately correct for.

We extensively tested {\sc maggie}
as well as our implementations of the FoF group finder with the optimal linking
lengths derived by \citet{DM+14a} as well as a simplified version of 
the group finder of \cite{Yang+05,Yang+07}.  
For our tests, we used a mock SDSS Legacy spectroscopic survey derived from the
\citet{Guo+11} SAM, itself run on the
Millennium-II cosmological dark matter simulation.
We also compared the performances of {\sc maggie} with the similar published
tests by
\cite{Yang+07} and \cite{DominguezRomero+12} and
\cite{MunozCuartas+12} of their respective group finders, in all instances where this
could be done.

We find that both  flavours of {\sc maggie} perform better than FoF in all
our tests (fragmentation, galaxy completeness and reliability, accuracy in
group total mass, luminosity and stellar mass), except
for cluster-mass EGs, where {\sc maggie} produces less reliable galaxy
members and more dispersed total masses.
The superiority of {\sc maggie} relative to FoF appears to be linked with the
very high fraction of secondary fragments that FoF produces.

The performance of {\sc maggie} is much closer to that of our simple implementation
of the Yang et al. group finder: {\sc maggie} performs much better on galaxy
completeness,
 slightly better on
group fragmentation, and dispersion of group total masses, luminosities and
stellar masses, but slightly worse on bias in group luminosities and stellar
masses, and also worse on galaxy reliabilities.  

Given its  use of
realistic priors, abundance matching and probabilistic galaxy membership,
{\sc maggie-m} is an ideal grouping algorithm to be applied on large
galaxy spectroscopic 
surveys such as the Sloan Digital Sky Survey (SDSS) and the Galaxy And
Mass Assembly (GAMA), for several applications:
environmental effects on
galaxy properties such as SSFR, as well as mass/orbit modelling of groups and clusters
(possibly stacking the groups), for which {\sc maggie} will lead to more
realistic results compared to the Yang et al. group finder, given the more
realstic prirors of the former. Moreover, {\sc maggie} should in principle be
able to work for much deeper spectroscopic surveys, possibly including
surveys based upon photometric redshifts (since {\sc maggie} naturally
handles redshift errors), with applications to the evolution
of environmental effects, dark matter properties (normalization,
concentration), and velocity anisotropy (orbital shapes).

In particular, {\sc maggie} should be very useful for dark energy surveys
such as the Dark Energy Survey (DES), Euclid, and the Wide-Field Infrared
Survey Telescope (WFIRST, yet to be approved)
that will constrain dark energy parameters not only with cosmic shear and
baryonic acoustic oscillations, but also by measuring the mass function and
clustering of galaxy clusters. However, the abundance matching method -- used to
determine group masses -- involves an assumption on the halo mass function,
which is cosmology-dependent. This
implies that the current implementation of {\sc maggie} cannot be used as a
cosmographic tool to determine cosmological parameters from the derived halo
mass function. Nevertheless, {\sc maggie} should be an excellent tool to optimally detect and measure groups
and clusters in dark energy surveys, if a given cosmology is assumed. Moreover, by replacing  abundance
matching by other techniques, {\sc maggie} could be adapted into a powerful
cosmographic tool for such surveys.

\section*{Acknowledgments}

Paper I and this article were part of the doctoral thesis of MD.
We thank Radek Wojtak for encouraging us to avoid hard assignments of
galaxies to their final groups in our
probabilistic group finder, at a time when we were still hesitant on the approach.
We also warmly thank Reinaldo de Carvalho for a thorough reading of an
earlier draft with
constructive criticisms, Xiaohu Yang and Mariano Dom\'{\i}nguez Romero for
explaining details of their respective group
finders, as well as David Valls-Gabaud, Andrea Biviano, Bego\~{n}a Ascaso 
and Florence Durret for
useful comments.
\textcolor{darkgreen}{Thanks to Prajwal Kafle for spotting errors in eqs.~(\ref{nunfw})
  and (\ref{dilogapx}).} 
We finally thank the anonymous referee for his/her insightful and
constructive comments that significantly 
improved this work.
The Millennium-II Simulation database used in this paper and the web
application providing online access to them were constructed as part of the
activities of the German Astrophysical Virtual Observatory (GAVO).
We are grateful to Michael Boylan-Kolchin and Qi Guo for respectively allowing
the outputs of the  Millennium-II simulation and the Guo semi-analytical
model to be available to the public, and Gerard Lemson for maintaining the
GAVO database and for useful discussions.
We also made use of {\sc HMFCalc} \citep{Murray+13} to check our computations of halo mass functions.

\bibliography{references}

\onecolumn

\appendix

\section{Radial velocity dispersion for NFW model with ML velocity
anisotropy}
\label{sec:sigr}

The expression for the radial velocity dispersion can be obtained from the
Jeans equation~(\ref{Jeans}), yielding \citep{vanderMarel94,ML+05}
\begin{equation}
    \sigma_r^2(r) = {G\over \nu(r)}
    \int_r^\infty \!\!\exp \left [2\int_r^s \beta(t) {{\rm d}t\over t}\right]
    \nu(s) {M(s) \over s^2} \,{\rm d}s\,,
    \label{sigmaofr}
\end{equation}
where the term in brackets is expressed in analytical form for simple
anisotropy models in an appendix of~\cite{MBB13}.\footnote{Even if the halo
  component is limited to the virial radius $r_{200}$, the upper integration
  limit  in equation~(\ref{sigmaofr}) must be infinity.}
 With the anisotropy  model
of equation~(\ref{betaML}), the exponential in equation~(\ref{sigmaofr})
becomes $(s+r_{200}/c)/(r+r_{200}/c_{\textcolor{darkgreen}{200}})$. The solution of
equation~(\ref{sigmaofr}) for a pure NFW model (eq.~[\ref{nunfw}]) with the \cite{ML+05}
velocity anisotropy (eq.~[\ref{betaML}]) is then
\textcolor{darkgreen}{(replacing for clarity $c_{200}$ by $c$)}
\begin{eqnarray}
 {\sigma_r^2(r)\over GM_{200}/ r_{200}} &\!\!\!\!=\!\!\!\!&
    {c/[6\,y\,(y+b)]\over \ln (c+1)-c/(c+1)} \nonumber \\
    &\!\!\!\!\mbox{}\!\!\!\!& \times \left \{
    6\,(3\,b-2) y^2 (y+1)^2  {\rm Li}_2(-y)
    + 6\,b\,y^4\,\coth^{-1}(2y+1)
    - 3\,b\,y^2(2y+1) \ln y
   \right.
    \nonumber \\
    &\mbox{}&
    \quad\quad
     +3\,\left[2y\,(y+1)\,(2y+1)-b\,\left(4y^3+8y^2+2y-1\right)\right]\,\ln(y+1)
    \nonumber \\
    &\mbox{}&
    \quad\quad
    \left.
    + (3\,b-2)\,y^2(y+1)^2 \left[\pi^2+3\,\ln^2(y+1) \right]
    +3\,y\,\left[(4-7b)\,y^2+(5-9\,b)\,y-b\right]
    \right \} \ ,
    \label{sigmaofrML}
\end{eqnarray}
where $y = c\,r/r_{200}$, $b = c\,r_\beta/r_{200}$, while ${\rm Li}_2$ is
the dilogarithm or Spence function:
\begin{equation}
    {\rm Li}_2(x) = - \int_0^x \ln(1-u)\,{{\rm d}u\over u}
    = \sum_{i=1}^\infty
    {x^i\over i^2} \ .
\end{equation}
For our choice of $r_\beta=r_{-2}$, i.e. $b=1$, equation~(\ref{sigmaofrML})
simplifies to 
\begin{eqnarray}
 {\sigma_r^2(r)\over GM_{200}/ r_{200}} &\!\!\!\!=\!\!\!\!&
    {c/[6\,y\,(y+1)]\over \ln (c+1)-c/(c+1)} \nonumber \\
    &\!\!\!\!\mbox{}\!\!\!\!& \times \left \{
    6\, y^2 (y+1)^2  {\rm Li}_2(-y)
    + 6\,y^4\,\coth^{-1}(2y+1)
    - 3\,y^2(2y+1) \ln y
     +y^2(y+1)^2 \left[\pi^2+3\,\ln^2(y+1) \right]
   \right.
    \nonumber \\
    &\mbox{}&
    \left.
    \quad\quad
     -3\,(2y^2-1)\,\ln(y+1)
    -3\,y\,(y+1)\,(3\,y+1)
    \right \} \ .
    \label{sigmaofrMLb1}
\end{eqnarray}

In equations~(\ref{sigmaofrML}) and (\ref{sigmaofrMLb1}), 
the dilogarithm of negative argument, ${\rm
  Li}_2(-x)$ can be approximated using series 
expansions around $x=0$, $x=1$, and $x\to\infty$, yielding
\begin{equation}
{\rm Li}_2(-x) \simeq \left \{
\begin{array}{ll}
\displaystyle  \sum_{i=1}^{10} (-1)^i {x^i\over i^2} & \qquad x < 0.35 \\
\displaystyle -{\pi^2\over 12} +\sum_{i=1}^{10} 
\left ({\ln 2\over i}-{a_i\over b_i}\right)\,(1-x)^i & \qquad 0.35 \leq x < 1.95\\
\displaystyle  -{\pi^2\over 6}-\textcolor{darkgreen}{ {1\over2}\,}\ln^2(x) + \sum_{i=1}^{10} (-1)^i {x^{-i}\over
  i^2} & \qquad x \geq
  1.95
\end{array}
\right . \ ,
\label{dilogapx}
\end{equation}
where the coefficients $a_i$ and $b_i$ given in Table~\ref{coeffsLi2}. 
Equation~(\ref{dilogapx}) has relative accuracy better than $\textcolor{darkgreen}{
2.7}\times10^{-6}$ for
all $x$.
With the approximation of equation~(\ref{dilogapx}) for ${\rm Li}_2(-x)$, the
radial velocity dispersion  $\sigma_r$ in equation~(\ref{sigmaofrMLb1}) has
relative accuracy 
better than $10^{-4}$ for all $r$.

\begin{table}
    \centering
    \caption{\label{coeffsLi2} Coefficients for the approximation of the
      dilogarithm (eq.~[\ref{dilogapx}])}
    \begin{tabular}{lrrrrrrrrrr}
        \toprule
$i$   & 1 & 2 &  3 & 4 &   5 &    6 &     7 &     8 &               9 & 10 \\
\hline
$a_i$ & 0 & 1 &  5 & 1 & 131 &  661 &  1$\,$327 &  1$\,$163 &      148$\,$969 &
        447$\,$047 \\
$b_i$ & 1 & 4 & 24 & 6 & 960 & 5$\,$760 & 13$\,$440 & 13$\,$440 & 1$\,$935$\,$360 &
        6$\,$451$\,$200 \\
        \bottomrule
    \end{tabular}
\end{table}

\section{Projected phase space density of interloping galaxies}
\label{sec:giGuo}
We estimate the projected phase space density of interloping galaxies
following \cite{MBM10}, this time using the galaxies from the $z$=0 output of
the SAM of \cite{Guo+11} instead of the dark matter
particles of the hydrodynamical cosmological simulation of \cite{Borgani+04}.

\begin{figure}
\centering
\includegraphics[width=0.5\hsize,trim=0 120 0 100]{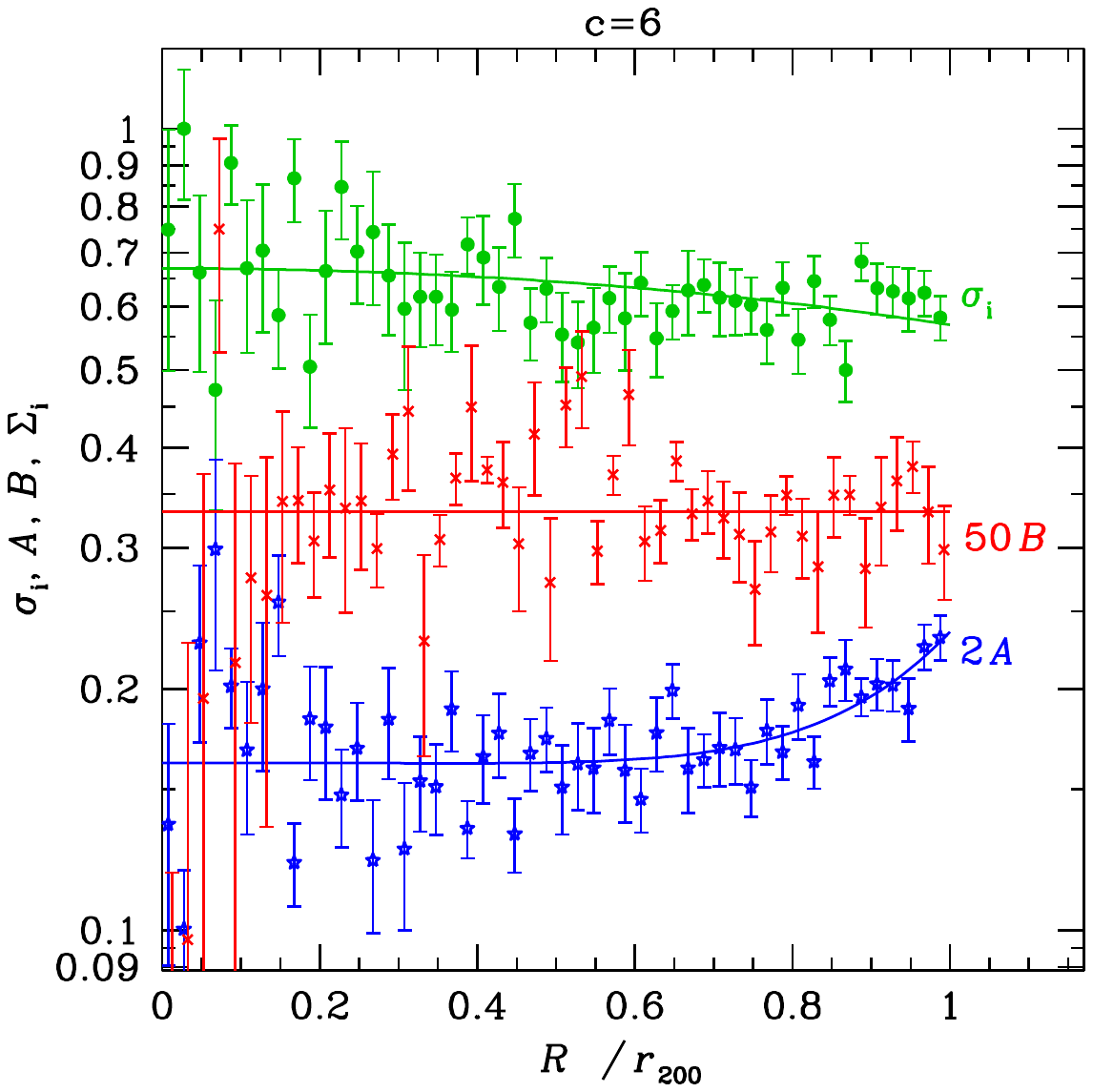}
\caption{Parameters of the density of interlopers in projected phase space
  vs. projected radius, taken from the $z$=0 galaxies of the semi-analytical
  model of Guo et al. (2011). The values are from maximum likelihood fits,
  assuming that the scale radius is 1/6th the virial radius,
  and the error bars represent the uncertainties on these values.
the curves are the best fits given by equations~(\ref{giGuo_A}),
(\ref{giGuo_sigmai}), and (\ref{giGuo_B}).
\label{fig:avzstats}
}
\end{figure}

Figure~\ref{fig:avzstats} shows the variations with projected radius of the 
interloper PPS density parameters $A$,
$\sigma_{\rm i}$ and $B$ (defined in eq.~[\ref{gilophat}]).
The best fitting parameters are
\begin{eqnarray}
\log_{10} A(X) &=&  -1.092 - \textcolor{red}{ 0.01922}\,X^3 + 0.1829\,X^6 \ , 
\label{giGuo_A} 
\\
\sigma_{\rm i}(X) &=& 0.6695 - 0.1004\, X^2 \ , 
\label{giGuo_sigmai}\\
B(X) &=& 0.0067 \ ,
\label{giGuo_B}
\end{eqnarray}
where $X = R/r_{200}$.

\section{SDSS errors on galaxy luminosity and stellar mass}
\label{app:sdsserrors}
Despite its very high quality, the SDSS survey is not immune to errors on
galaxy stellar mass and  luminosity. We estimate these errors below. 

\subsection{SDSS errors on stellar mass}
\label{app:errm}
The SDSS-DR10 database contains 8 measures of stellar mass for the primary
spectroscopic sample.
\begin{figure}
    \centering
    \includegraphics[width=\hsize]{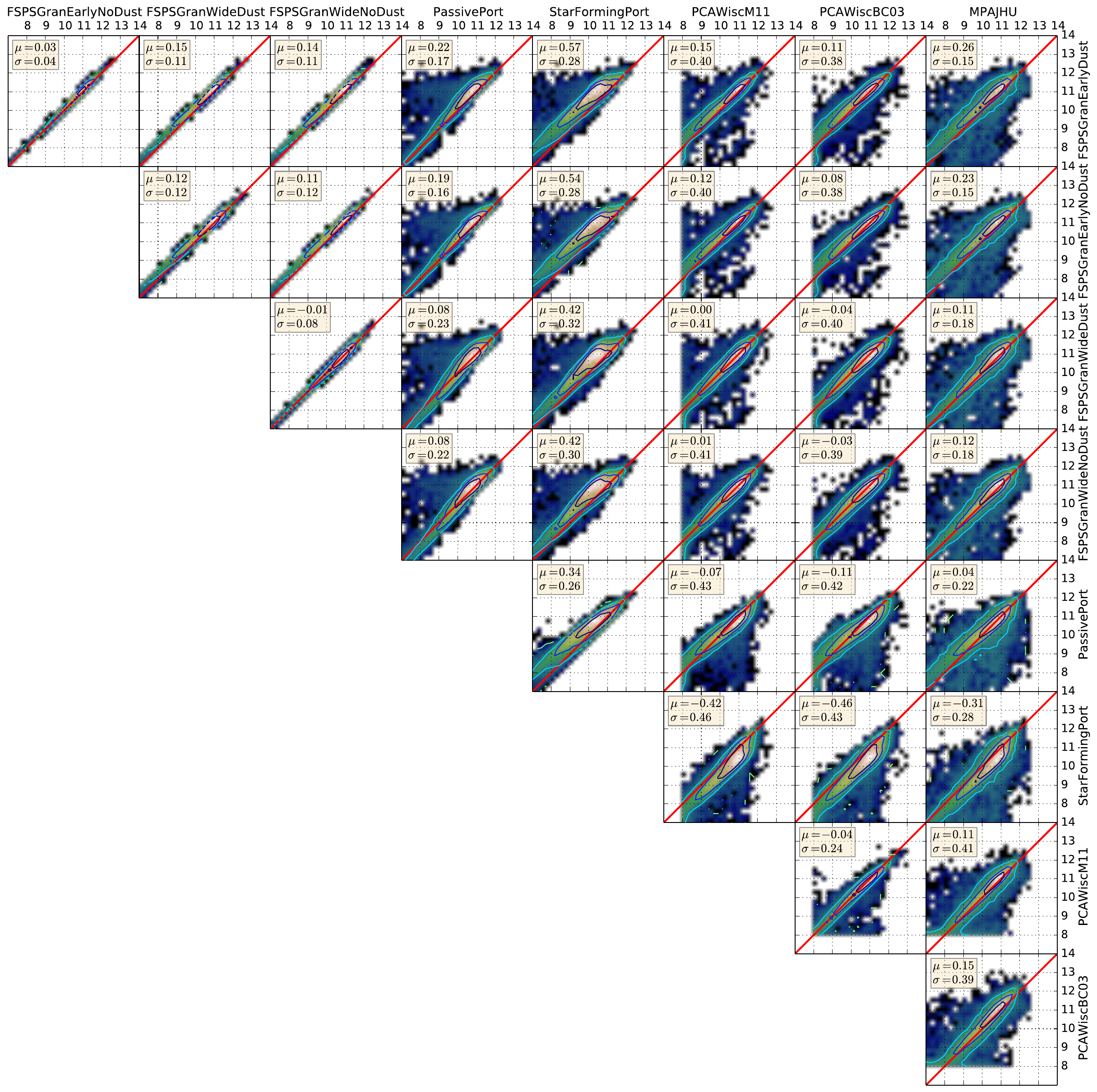}
    \caption{Comparison of the 8 measures of log stellar mass (solar units)
    in the SDSS-DR10 database. The biases ($\mu$) and bias-corrected
differences ($\sigma$) are highlighted. 
These measures are from the following models.
FSPSGranEarlyDust,
FSPSGranEarlyNoDust,
FSPSGranWideDust, and
FSPSGranWideNoDust: {\tt logMass} respectively from 
{\tt stellarMassFSPSGranEarlyDust},
{\tt stellarMassFSPSGranEarlyNoDust},
{\tt stellarMassFSPSGranWideDust},
and
{\tt stellarMassFSPSGranWideNoDust} 
(Conroy et al. 2009);
PassivePort and StarFormingPort: {\tt logMass} 
respectively from {\tt stellarMassPassivePort} and  
{\tt stellarMassStarFormingPort}
(Maraston et al. 2009);
PCAWiscM11 and PCAWiscBC03: {\tt mstellar\_median} respectively from {\tt
  stellarMassPCAWiscM11} and
 {\tt stellarMassPCAWiscBC03} 
(Chen et al. 2012), respectively using 
the Maraston \& Stromback (2011) and Bruzual \& Charlot (2003)
stellar population sythesis models;
MPAJHU: {\tt lgm\_tot\_p50} from {\tt GalSpecExtra} (Brinchmann et al. 2004) using
the Bruzual \& Charlot (2003) stellar population sythesis model. 
\label{fig:stellar_mass_comp}}
\end{figure}
\nocite{Conroy+09}
\nocite{Maraston+09}
\nocite{Chen+12}
\nocite{MarastonStromback11}
\nocite{BC03}
\nocite{Brinchmann+04} 
Figure~\ref{fig:stellar_mass_comp} compares these 8 different measures. Apart
from those from the Wisconsin group, the models generally agree to better than
0.3 dex, i.e.\ the errors on individual masses are of order $0.3/\sqrt{2} = 0.2$
dex. In particular, the MPA/JHU masses agree with all others to typically
better than 0.2 dex for $\sigma$ and 0.3 dex for the rms
($\sqrt{\mu^2+\sigma^2}$). We therefore adopt an error of 0.2 dex on stellar
mass.

\subsection{SDSS errors on galaxy luminosity}
\label{app:errL}
Writing the $r$-band absolute magnitude
of a galaxy as
\begin{equation}
    M_r = r - \mu(z) - k_r(z) - A_r^{\rm Gal} - A_r^{\rm int}
\end{equation}
where $\mu$ is the distance modulus, while $r$, $k_r$, $A_r^{\rm Gal}$, and
$A_r^{\rm int}$ are respectively the
apparent magnitude, k-correction, Galactic extinction and internal extinction, all in the $r$ band. The
photometric errors are expected to be less than 0.05 mag, i.e.\ less than
0.02 dex on luminosity.  The error caused by the uncertain distance can be
written as the quadratic sum of the error on redshift (as a distance
indicator) and the neglect of group peculiar velocities relative to the observer. We dot not consider here
the galaxy peculiar velocities within a group, as the group finders handle this.
\[
    \epsilon(\log_{10} L_r) = {1\over \ln 10}\,{\left[{\left({\epsilon(v)\over c
            z}\right)}^2+
    {\left({\sigma(v_{\rm p})\over cz}\right)}^2\right]}^{1/2} \la 0.056\,\rm dex
\]
for $\epsilon(v)\simeq 30 \, \rm km \, s^{-1}$, $\sigma(v_{\rm p}) \simeq
200 \, \rm km \, s^{-1}$, and $z>0.01$ (where the assumption of zero
difference in peculiar velocity between the galaxy and the observer dominates
the error).
According to Figure 2 of \cite{Chilingarian+10}, the intrinsic scatter in the k-correction
is of order 0.015 mag, i.e. 0.006 dex. Admittedly, the k-correction of
\citeauthor{Chilingarian+10} suffers from some catastrophic errors, but since 99.9\% of
the galaxies with $z < 0.12$ have k-corrections between $-0.15$ and 0.25, it
suffices to impose these limits to $k_r$.
Finally, since SDSS spans high galactic latitudes, the uncertainty on the
Galactic extinction should be $\simeq 0.075$ mag (the median $r$-band extinction of SDSS/Legacy
galaxies), i.e. 0.03 dex.
The uncertainty on internal extinction is more difficult to measure, but can
be estimated to be 0.1 mag, i.e. 0.04 dex.
Combining these 6 errors (photometry, redshift, assumption of no peculiar
velocity, k-correction, Galactic extinction and internal extinction)  in quadrature, we deduce that the error on luminosity
is of order of 0.08 dex.

\section{Graphical representation of the effects of the interloper density in
  projected phase space on the performance of {\sc maggie}}
\label{sec:appgiGuo}

\begin{figure*}
\centering
\includegraphics[width=0.39\hsize]{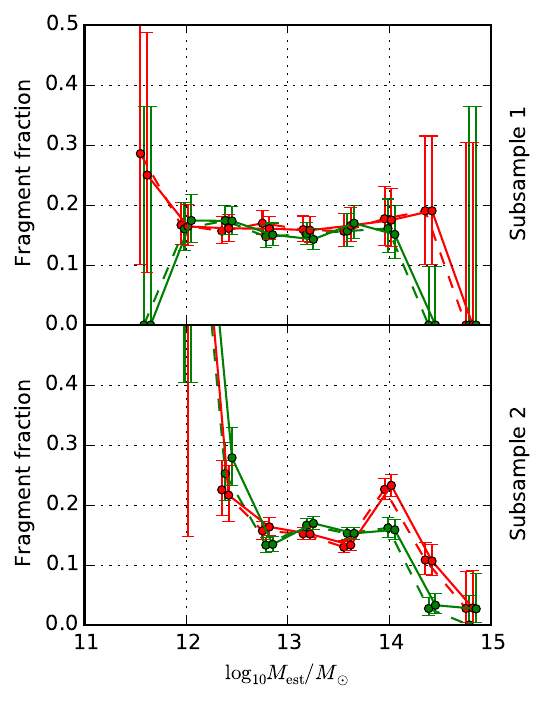}
\includegraphics[width=0.60\hsize]{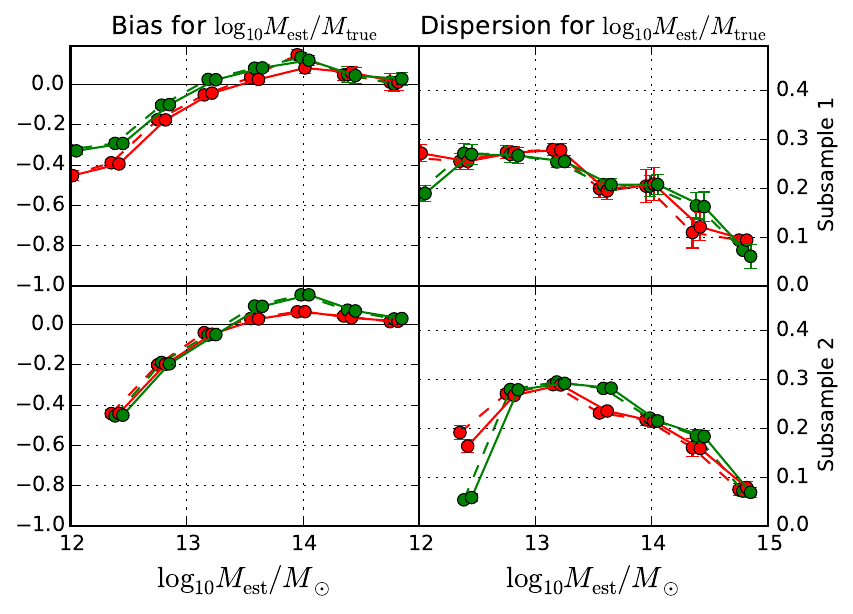}
\includegraphics[width=0.49\hsize]{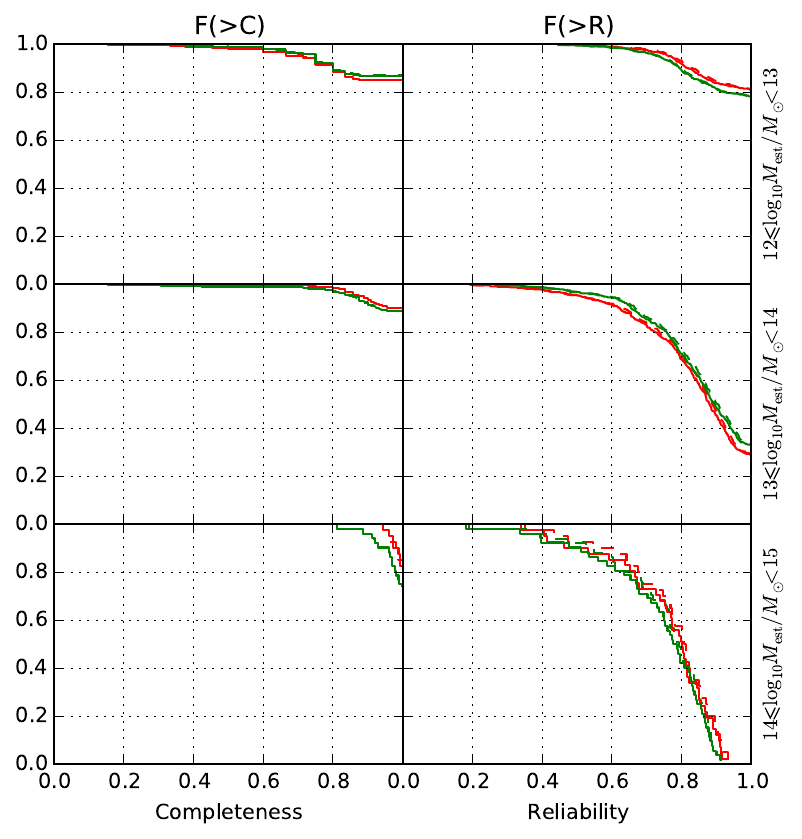}
\includegraphics[width=0.49\hsize]{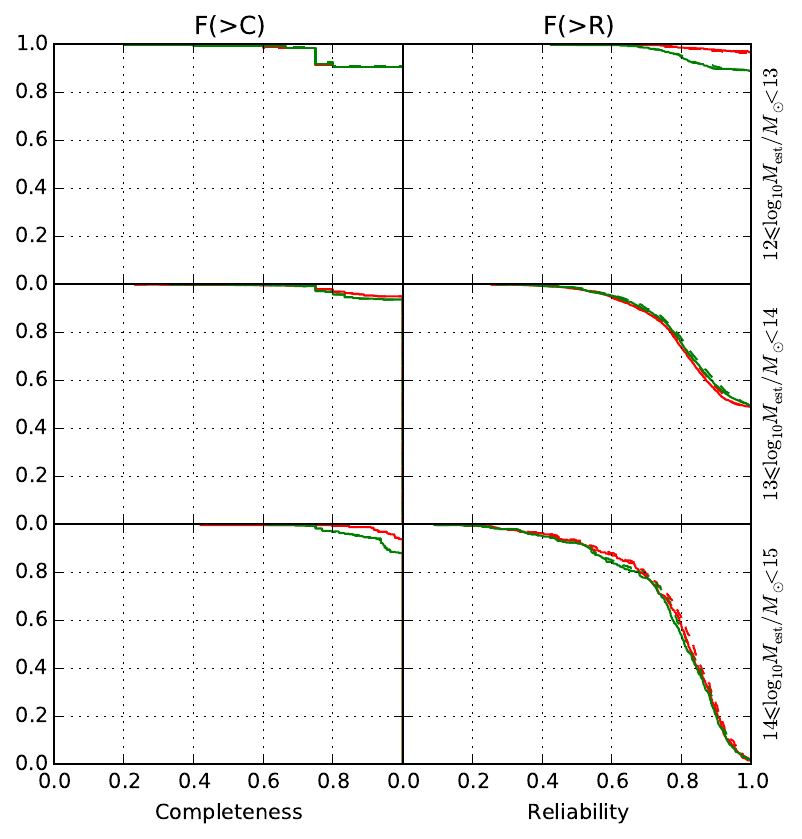}
\caption{
Effects of the choice of the $A(X)$, $\sigma_{\rm i}(X)$ and $B$ entering the
expression of the density of interlopers in projected phase space
(eq.~[\ref{gilophat}]) on the performance of {\sc maggie-l} (red) and {\sc
  maggie-m} (green), both with
observational errors, comparing the expression for $\widehat
g_{\rm i}(X,u)$ derived by Mamon et al. (2010) on the dark matter particles of a
cosmological hydrodynamical simulation (eqs.~[\ref{Ailop}]-[\ref{Bilop}], \emph{dashed}) with that derived in
Appendix~\ref{sec:giGuo} on the galaxies of the semi-analytical of Guo et
al. (2011) at $z=0$ (eqs.~[\ref{AilopGuo}]-[\ref{BilopGuo}], \emph{solid}). 
The analysis is for unflagged groups of at least 3 true and 3 extracted
members.
The points in the upper plots have their abscissa slightly shifted for clarity. 
\label{fig:compgi}
}
\end{figure*}

In this appendix, we illustrate, in Figure~\ref{fig:compgi}, the effects of 2 choices for the
dimensionless density of interlopers in projected phase space, $\widehat
g_{\rm i}(X,u)$ (eq.~[\ref{gilophat}]) on the performance of {\sc maggie-l}
and {\sc maggie-m}.
See the discussion in Sect.~\ref{sec:robust_gi}.

\section{Dimensionless surface density threshold for the Yang et al. group finder}
\label{sec:appYang}

\cite{Yang+05} defined the redshift space local density contrast (relative to the mean
density of the Universe) at the edge of the group as
\begin{equation} 
{\cal B} =  {V_{\rm sph}(r_{\rm v})\over V_{\rm cyl}(r_{\rm v})}\,{\rho(r_{\rm v})\over \rho_{\rm U}}
\ ,
\label{Bdef}
\end{equation}
where the first term represents the ratio of volumes of the virial sphere in
real space to the virial cylinder in redshift space, the second term is
the local overdensity at the surface of the virial sphere, and the galaxy
number density field is assumed to trace the mass density field.
With $V_{\rm sph} = 4\pi/3 r_{\rm v}^3$ and $V_{\rm cyl} =
2\,(\sigma_v/H_0)\,\pi r_{\rm v}^2$ (assuming that the cylinder's half-length
is $\sigma_v/H_0$),\footnote{Equation~(\ref{Bdef}) with these
  two formulae for the volumes are analogous to
  equation (11) of Yang et al. (2005), who seemed to have forgotten the factor 2
for $V_{\rm cyl}$.} and assuming an NFW density profile (eq.~[16]), the local
mass density at the
virial sphere is 
\begin{equation} 
\rho_{\rm NFW}(r_{\rm v}) = {\Delta\,\rho_{\rm U}\over 3\,\Omega_{\rm m}}\,{[c_\Delta/(c_\Delta+1)]^2\over
  \ln(c_\Delta+1)-c_\Delta/(c_\Delta+1)} \ ,
\label{rholocalNFW}
\end{equation} 
where $\Delta$ is the overdensity of the group relative to the critical density of the
Universe, $\rho_{\rm U}$ is the mean density of the Universe, 
$\Omega_{\rm m}$ is the cosmological density parameter, and
$c_\Delta$ is the concentration of the group.
Combining equations~(\ref{Bdef}) and (\ref{rholocalNFW}), one finds
\begin{eqnarray} 
{\cal B} &=& {\sqrt{8\,\Delta} \over 9\,\eta\,\Omega_{\rm m}}\,f_{\cal B}(c_\Delta) \ , 
\label{BYang}
\\
f_{\cal B}(c) &=& {[c/(c+1)]^2\over
  \ln(c+1)-c/(c+1)} \ ,
\label{fBofc}
\end{eqnarray} 
where $\eta =
\sigma_v/v_{\rm v}$ is the ratio of velocity dispersion to virial velocity
for the group.


Yang et al. (2005, 2007) adopted an overdensity of 180 relative to the mean
density of the Universe, and use a cosmological $N$-body simulation with
$\Omega_{\rm m} = 0.3$ to calibrate their group finder. This corresponds to
an overdensity relative to critical of 
$\Delta = 0.3\times 180 = 54$.
We can solve for the concentration $c_{54}$ relevant for the median Yang et
al. halos, $\log h M_{54,\rm median} \simeq 13.5$ as follows. We loop over values
of $\log M_{200}$, for which we extract $c_{200}$ from the relation of
Maccio et al. (2008), which also considers $h$. Assuming the NFW density model, 
we then solve for $c_{54}$ given $c_{200}$, as well as $M_{54}$ given
$M_{200}$.
Given our derived relation between $c_{54}$ and $M_{54}$, we solve for the median
$c_{54}$, which yields $c_{54,\rm median} = 8.72$. Then, through
equations~(\ref{BYang}) and (\ref{fBofc}), we obtain ${\cal B}=6.14$, which can be
contrasted to ${\cal B}=10$
estimated by Yang et al. (2005) and also adopted by \cite{Yang+07}.

We can also apply equations~(\ref{BYang}) and (\ref{fBofc}) to our case, where $\Delta=200$,
$\Omega_{\rm m}=0.25$ (the Millennium-II simulation on which the mock catalog
was built) and a median halo mass of $\log M_{200,\rm median}\simeq 13.5$ (without the $h$
term). This yields $c_{200,\rm median} = 4.97$ and ${\cal B}=19.9$.

Given that Yang et al. (2005) checked that ${\cal B}=10$ gave them the best
results for the groups they extracted from their mock galaxy catalogue (given their groups in real space), we need to rescale their ${\cal B}=10$ to our values of $\Delta$,
$\Omega_{\rm m}$ and $c$. With equations~(\ref{BYang}) and (\ref{fBofc}), we find
${\cal B}=10\times [\sqrt{200/54} / (0.25/0.3)] \times [f_B(4.97)/f_B(8.72)]=29$ (with a factor
2.3 from the first term in brackets and a factor 1.24 from the 2nd term).

\begin{figure*}
\centering
\includegraphics[width=0.39\hsize]{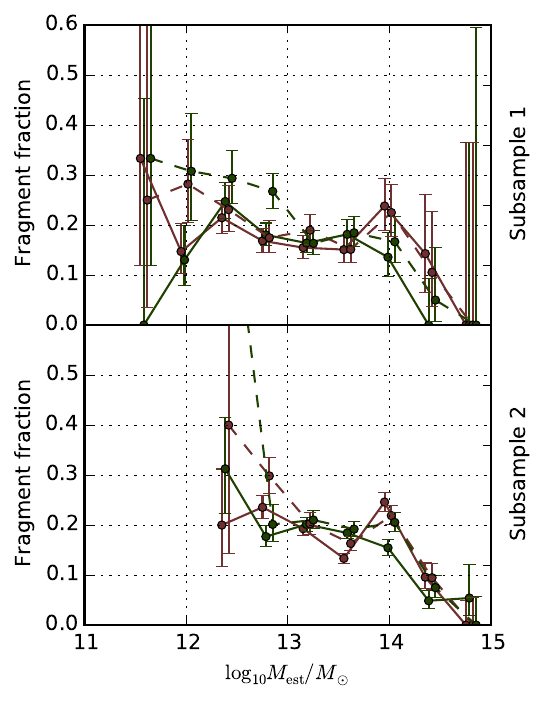}
\includegraphics[width=0.60\hsize]{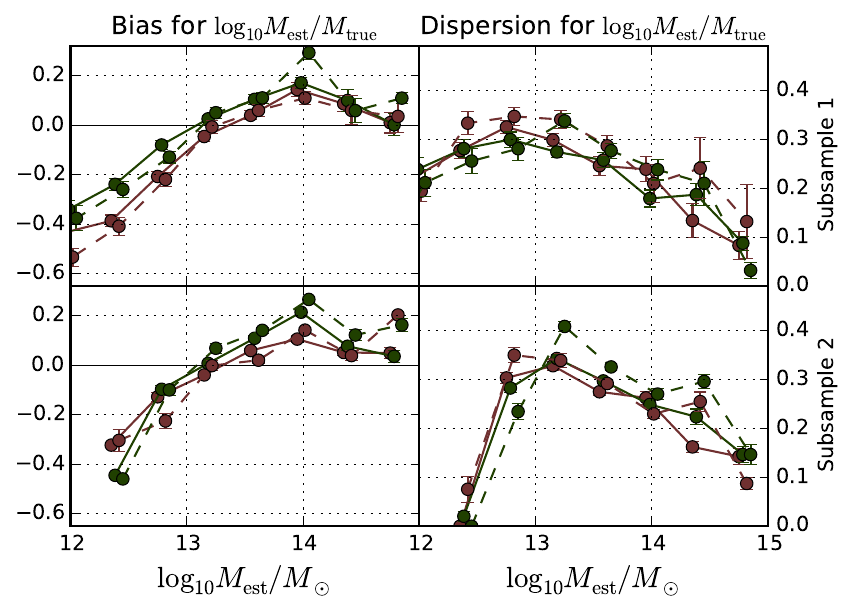}
\includegraphics[width=0.49\hsize]{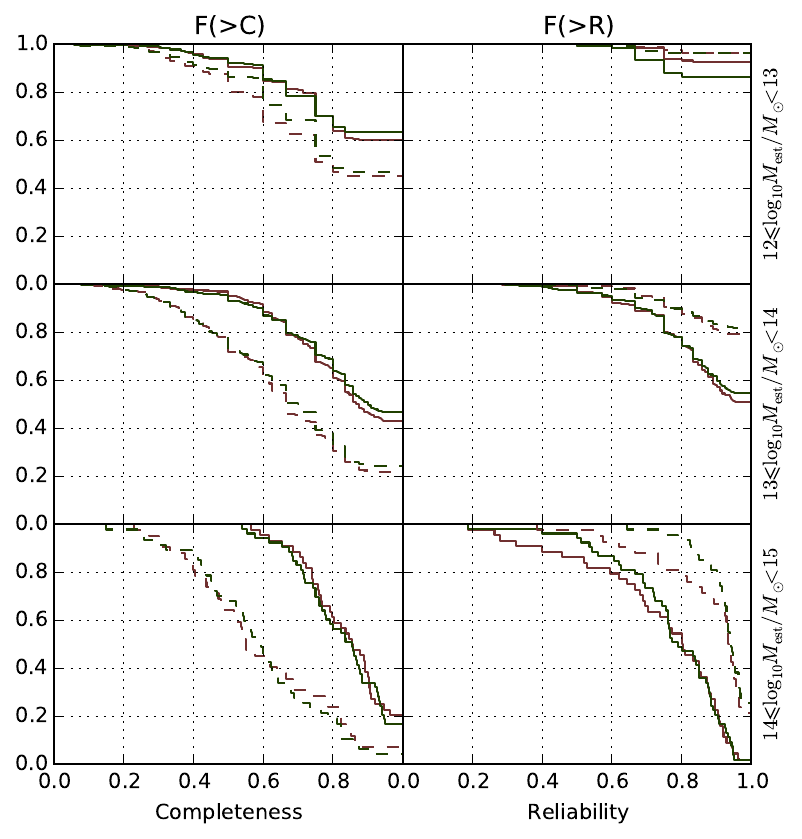}
\includegraphics[width=0.49\hsize]{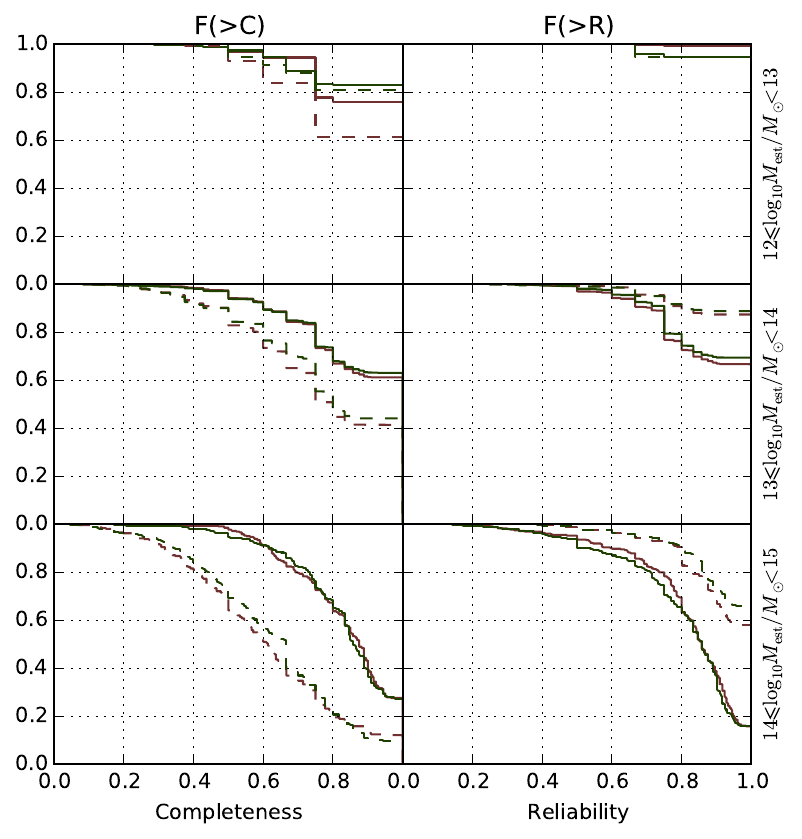}
\caption{
Effects of the choice of the density contrast threshold ${\cal B}$ on the
performance of our implementation of the Yang et al. group finder,
for Yang-L (\emph{brown}), Yang-M (\emph{dark green}), with ${\cal B}=10$
(\emph{solid lines}) and ${\cal B}=29$ (\emph{dashed lines}),
both with
observational errors.
The analysis is for unflagged groups of at least 3 true and 3 extracted
members.
The points in the upper plots have their abscissa slightly shifted for clarity. 
\label{fig:compYangB}
}
\end{figure*}

We test the effects of the choice of ${\cal B}$ in Figure~\ref{fig:compYangB}.
The higher threshold of ${\cal B}=29$ leads to slighter higher group fragmentation
(upper left panel), much lower galaxy completeness, but much higher galaxy reliability
(lower panels), yet slightly higher dispersion on group total masses (upper
right panel). 

\label{lastpage}

\end{document}